\documentclass[12pt]{article}
\usepackage{epsfig}
\usepackage{amsmath}
\usepackage{hhline}
\usepackage{amssymb}
\usepackage{times}
\usepackage{cite}
\usepackage{multirow}
\usepackage{lscape}
\usepackage{xtab}
\usepackage{url}
\newlength{\dinwidth}
\newlength{\dinmargin}
\begin{document}  
%%%%%%%%%%%%%%%% Pre-defined commands, you can use for the most obvious notations
\newcommand{\pom}{{I\!\!P}}
\newcommand{\reg}{{I\!\!R}}
\newcommand{\slowpi}{\pi_{\mathit{slow}}}
\newcommand{\fiidiii}{F_2^{D(3)}}
\newcommand{\fiidiiiarg}{\fiidiii\,(\beta,\,Q^2,\,x)}
\newcommand{\n}{1.19\pm 0.06 (stat.) \pm0.07 (syst.)}
\newcommand{\nz}{1.30\pm 0.08 (stat.)^{+0.08}_{-0.14} (syst.)}
\newcommand{\fiidiiiful}{F_2^{D(4)}\,(\beta,\,Q^2,\,x,\,t)}
\newcommand{\fiipom}{\tilde F_2^D}
\newcommand{\ALPHA}{1.10\pm0.03 (stat.) \pm0.04 (syst.)}
\newcommand{\ALPHAZ}{1.15\pm0.04 (stat.)^{+0.04}_{-0.07} (syst.)}
\newcommand{\fiipomarg}{\fiipom\,(\beta,\,Q^2)}
\newcommand{\pomflux}{f_{\pom / p}}
\newcommand{\nxpom}{1.19\pm 0.06 (stat.) \pm0.07 (syst.)}
\newcommand {\gapprox}
   {\raisebox{-0.7ex}{$\stackrel {\textstyle>}{\sim}$}}
\newcommand {\lapprox}
   {\raisebox{-0.7ex}{$\stackrel {\textstyle<}{\sim}$}}
\def\gsim{\,\lower.25ex\hbox{$\scriptstyle\sim$}\kern-1.30ex%
\raise 0.55ex\hbox{$\scriptstyle >$}\,}
\def\lsim{\,\lower.25ex\hbox{$\scriptstyle\sim$}\kern-1.30ex%
\raise 0.55ex\hbox{$\scriptstyle <$}\,}
\newcommand{\pomfluxarg}{f_{\pom / p}\,(x_\pom)}
\newcommand{\dsf}{\mbox{$F_2^{D(3)}$}}
\newcommand{\dsfva}{\mbox{$F_2^{D(3)}(\beta,Q^2,x_{I\!\!P})$}}
\newcommand{\dsfvb}{\mbox{$F_2^{D(3)}(\beta,Q^2,x)$}}
\newcommand{\dsfpom}{$F_2^{I\!\!P}$}
\newcommand{\gap}{\stackrel{>}{\sim}}
\newcommand{\lap}{\stackrel{<}{\sim}}
\newcommand{\fem}{$F_2^{em}$}
\newcommand{\tsnmp}{$\tilde{\sigma}_{NC}(e^{\mp})$}
\newcommand{\tsnm}{$\tilde{\sigma}_{NC}(e^-)$}
\newcommand{\tsnp}{$\tilde{\sigma}_{NC}(e^+)$}
\newcommand{\st}{$\star$}
\newcommand{\sst}{$\star \star$}
\newcommand{\ssst}{$\star \star \star$}
\newcommand{\sssst}{$\star \star \star \star$}
\newcommand{\tw}{\theta_W}
\newcommand{\sw}{\sin{\theta_W}}
\newcommand{\cw}{\cos{\theta_W}}
\newcommand{\sww}{\sin^2{\theta_W}}
\newcommand{\cww}{\cos^2{\theta_W}}
\newcommand{\trm}{m_{\perp}}
\newcommand{\trp}{p_{\perp}}
\newcommand{\trmm}{m_{\perp}^2}
\newcommand{\trpp}{p_{\perp}^2}
\newcommand{\alp}{\alpha_s}

\newcommand{\alps}{\alpha_s}
\newcommand{\sqrts}{$\sqrt{s}$}
\newcommand{\LO}{$O(\alpha_s^0)$}
\newcommand{\Oa}{$O(\alpha_s)$}
\newcommand{\Oaa}{$O(\alpha_s^2)$}
\newcommand{\PT}{p_{\perp}}
\newcommand{\JPSI}{J/\psi}
\newcommand{\sh}{\hat{s}}
\newcommand{\uh}{\hat{u}}
\newcommand{\MP}{m_{J/\psi}}
\newcommand{\PO}{I\!\!P}
\newcommand{\xbj}{x}
\newcommand{\xpom}{x_{\PO}}
\newcommand{\ttbs}{\char'134}
\newcommand{\xpomlo}{3\times10^{-4}}  
\newcommand{\xpomup}{0.05}  
\newcommand{\dgr}{^\circ}
\newcommand{\pbarnt}{\,\mbox{{\rm pb$^{-1}$}}}
\newcommand{\gev}{\,\mbox{GeV}}
\newcommand{\WBoson}{\mbox{$W$}}
\newcommand{\fbarn}{\,\mbox{{\rm fb}}}
\newcommand{\fbarnt}{\,\mbox{{\rm fb$^{-1}$}}}
\newcommand{\dsdx}[1]{$d\sigma\!/\!d #1\,$}
\newcommand{\eV}{\mbox{e\hspace{-0.08em}V}}
%
% Some useful tex commands
%
\newcommand{\qsq}{\ensuremath{Q^2} }
\newcommand{\gevsq}{\ensuremath{\mathrm{GeV}^2} }
\newcommand{\et}{\ensuremath{E_t^*} }
\newcommand{\rap}{\ensuremath{\eta^*} }
\newcommand{\gp}{\ensuremath{\gamma^*}p }
\newcommand{\dsiget}{\ensuremath{{\rm d}\sigma_{ep}/{\rm d}E_t^*} }
\newcommand{\dsigrap}{\ensuremath{{\rm d}\sigma_{ep}/{\rm d}\eta^*} }

%%% Dstar stuff
\newcommand{\dstar}{\ensuremath{D^*}}
\newcommand{\dstarp}{\ensuremath{D^{*+}}}
\newcommand{\dstarm}{\ensuremath{D^{*-}}}
\newcommand{\dstarpm}{\ensuremath{D^{*\pm}}}
\newcommand{\zDs}{\ensuremath{z(\dstar )}}
\newcommand{\Wgp}{\ensuremath{W_{\gamma p}}}
\newcommand{\ptds}{\ensuremath{p_t(\dstar )}}
\newcommand{\etads}{\ensuremath{\eta(\dstar )}}
\newcommand{\ptj}{\ensuremath{p_t(\mbox{jet})}}
\newcommand{\ptjn}[1]{\ensuremath{p_t(\mbox{jet$_{#1}$})}}
\newcommand{\etaj}{\ensuremath{\eta(\mbox{jet})}}
\newcommand{\detadsj}{\ensuremath{\eta(\dstar )\, \mbox{-}\, \etaj}}

%%% Cross sections expressions
\newcommand{\xsw}{\ensuremath{\sigma_{W}}}
\newcommand{\xsisolep}{\ensuremath{\sigma_{\ell+{P}_{T}^{\rm miss}}}}

%%% WWgamma coupling expressions
\newcommand{\dkl}{\ensuremath{\left(\Delta\kappa,\lambda\right)}}
\newcommand{\dkldk}{\ensuremath{\left(\Delta\kappa,\lambda=0\right)}}
\newcommand{\dkll}{\ensuremath{\left(\Delta\kappa=0,\lambda\right)}}
\newcommand{\dklp}{\ensuremath{\left(\Delta\kappa^{\prime},\lambda^{\prime}\right)}}

%% W polarisartion fraction expressions
\newcommand{\cosths}{\ensuremath{\cos\,\theta^{*}}}
\newcommand{\qcosths}{\ensuremath{q_{\ell}\cdot\cosths}}

% Journal macro
\def\Journal#1#2#3#4{{#1} {\bf #2} (#3) #4}
\def\NCA{\em Nuovo Cimento}
\def\NIM{\em Nucl. Instrum. Methods}
\def\NIMA{{\em Nucl. Instrum. Methods} {\bf A}}
\def\NPB{{\em Nucl. Phys.}   {\bf B}}
\def\PLB{{\em Phys. Lett.}   {\bf B}}
\def\PRL{\em Phys. Rev. Lett.}
\def\PRD{{\em Phys. Rev.}    {\bf D}}
\def\ZPC{{\em Z. Phys.}      {\bf C}}
\def\EJC{{\em Eur. Phys. J.} {\bf C}}
\def\CPC{\em Comp. Phys. Commun.}

%%%%%%%%%%%%%%%%%%%%%%%%%%%%%%%%%%%%%%%%%%%%%%%%%%%%%%%%%%%%%%%%%%%%%%%
\begin{titlepage}

\noindent
\begin{flushleft}
{\tt DESY 08-170    \hfill    ISSN 0418-9833} \\
{\tt September 2009}                  \\
\end{flushleft}

\vspace{2cm}
\begin{center}
\begin{Large}

{\bf Events with Isolated Leptons and Missing Transverse Momentum and
Measurement of {\boldmath $W$} Production at HERA}

\vspace{2cm}

H1 Collaboration

\end{Large}
\end{center}

\vspace{2cm}

\begin{abstract}
\noindent

Events with high energy isolated electrons, muons or tau leptons and
missing transverse momentum are studied using the full $e^{\pm}p$ data
sample collected by the H1~experiment at HERA, corresponding to an
integrated luminosity of $474$~pb$^{-1}$.
Within the Standard Model, events with isolated leptons and missing
transverse momentum mainly originate from the production of single $W$
bosons.
The total single $W$ boson production cross section is measured as
$1.14 \pm 0.25~({\rm stat.}) \pm 0.14~({\rm sys.})$~pb, in agreement
with the Standard Model expectation.
The data are also used to establish limits on the $WW\gamma$ gauge
couplings and for a measurement of the $W$ boson polarisation.

\end{abstract}

\vspace{1.5cm}

\begin{center}
Accepted by \EJC
\end{center}

\end{titlepage}

%%%%%%%%%%%%%%%%%%%%%%%%%%%%%%%%%%%%%%%%%%%%%%%%%%%%%%%%%%%%%%%%%%%%%%%
\begin{flushleft}
%-- H1AUTS Author list by names 
%-- Status: Tue Nov  4 16:26:59 CET 2008  Number of authors = 256 
F.D.~Aaron$^{5,49}$,           %BUCH-PD        11/06           Aaron               
C.~Alexa$^{5}$,                %BUCH-PD        06/06           Alexa               
V.~Andreev$^{25}$,             %LPI -PD        8/88            Andreev             
B.~Antunovic$^{11}$,           %DESY-PD        05/07           Antunovic           
S.~Aplin$^{11}$,               %DESY-LEFT      01/08           Aplin               
A.~Asmone$^{33}$,              %ROME-ST        07/2            Asmone              
A.~Astvatsatourov$^{4}$,       %BRUX-LEFT      01/08           Astvatsatourov      
S.~Backovic$^{30}$,            %PODG-PD        03/2            Backovic            
A.~Baghdasaryan$^{38}$,        %YERE-PD        09/03           Baghdasaryana       
E.~Barrelet$^{29}$
% ...

%-- H1 Institutes 
\bigskip{\it
 $ ^{1}$ I. Physikalisches Institut der RWTH, Aachen, Germany$^{ a}$ \\
% ...
$ ^{4}$ Inter-University Institute for High Energies ULB-VUB, Brussels;
          Universiteit Antwerpen, Antwerpen; Belgium$^{ c}$ \\
 $ ^{5}$ National Institute for Physics and Nuclear Engineering (NIPNE) ,
          Bucharest, Romania \\
% ...
$ ^{11}$ DESY, Hamburg, Germany \\
% ...
$ ^{25}$ Lebedev Physical Institute, Moscow, Russia$^{ e}$ \\
% ...
$ ^{29}$ LPNHE, Universit\'{e}s Paris VI and VII, IN2P3-CNRS,
          Paris, France \\
 $ ^{30}$ Faculty of Science, University of Montenegro,
          Podgorica, Montenegro$^{ e}$ \\
% ...
$ ^{33}$ Dipartimento di Fisica Universit\`a di Roma Tre
          and INFN Roma~3, Roma, Italy \\
% ...
$ ^{38}$ Yerevan Physics Institute, Yerevan, Armenia \\
% ...
$ ^{49}$ Also at Faculty of Physics, University of Bucharest,
          Bucharest, Romania \\
% ...
$ ^a$ Supported by the Bundesministerium f\"ur Bildung und Forschung, FRG,
      under contract numbers 05 H1 1GUA /1, 05 H1 1PAA /1, 05 H1 1PAB /9,
      05 H1 1PEA /6, 05 H1 1VHA /7 and 05 H1 1VHB /5 \\
% ...
$ ^c$ Supported by FNRS-FWO-Vlaanderen, IISN-IIKW and IWT
      and  by Interuniversity
Attraction Poles Programme,
      Belgian Science Policy \\
% ...
$ ^e$ Supported by the Deutsche Forschungsgemeinschaft \\
% ...
}
\end{flushleft}
%%%%%%%%%%%%%%%%%%%%%%%%%%%%%%%%%%%%%%%%%%%%%%%%%%%%%%%%%%%%%%%%%%%%%%%

\newpage

%%%%%%%%%%%%%%%%%%%%%%%%%%%%%%%%%%%%%%%%%%%%%%%%%%%%%%%%%%%%%%%%%%%%%%%
\section{Introduction}
\label{ch:introduction}
%%%%%%%%%%%%%%%%%%%%%%%%%%%%%%%%%%%%%%%%%%%%%%%%%%%%%%%%%%%%%%%%%%%%%%%

Events containing high energy leptons and missing transverse momentum
produced in high energy particle collisions are interesting as they
may be a signature of physics beyond the Standard Model~(SM).
Such events have been observed by the H1 Collaboration in $ep$
collisions at HERA~\cite{Ahmed:1994,Adloff:1998aw}.
In the SM, the production of single $W$ bosons with subsequent
leptonic decay gives rise to this topology.
An excess of electron\footnote{In this paper the term ``electron'' is
used generically to refer to both electrons and positrons, if not
otherwise stated.} and muon events with large missing transverse
momentum containing a hadronic final state at high transverse momentum
$P_{T}^{X}$ was previously reported by H1 using $105$~pb$^{-1}$ of
$e^{+}p$ data~\cite{Andreev:2003pm}.
In the region $P_{T}^{X}>25$~GeV ten events were observed compared to
a SM prediction of $2.9\pm0.5$.
This observation inspired searches for anomalous single top
production~\cite{Aktas:2003yd,Chekanov:2003yt} and bosonic stop decays
in $R$--parity violating SUSY~\cite{Aktas:2004tm} and has motivated
further possible
interpretations~\cite{Diaconu:2004mj,Carli:2004up,Choi:2006ms,Choi:2007qx}.
The observed excess of events over the SM prediction was not confirmed
by the ZEUS Collaboration~\cite{Chekanov:2003yt,Chekanov:2008new}.
In this paper the complete H1 data sample, collected in the period
$1994$--$2007$, is analysed.

%%%

The search for isolated tau leptons complements the analysis of the
electron and muon channels.
If lepton universality holds, the same rate of tau leptons is expected
from SM processes.
Moreover, an enhanced rate of tau leptons is expected in many new
physics scenarios~\cite{Carli:2004up}.
The search for events containing high energy tau leptons and missing
transverse momentum, where the tau is identified by its hadronic
decay, has previously been performed by the ZEUS and H1
experiments~\cite{Chekanov:2003bf,Aktas:2006fc}.
ZEUS reported two tau candidates at high $P_{T}^{X}>25$~GeV, where
only $0.2\pm0.05$ were predicted from the SM.
No events with large hadronic transverse momentum were observed in the
H1 data.

%%%

As the SM expectation in the cleaner electron and muon channels is
dominated by single $W$ production, this cross section is determined.
This measurement is also used to constrain coupling parameters of the
$WW\gamma$ vertex.
The polarisation fractions of the $W$ boson are also measured for the
first time at HERA.

%%%%%%%%%%%%%%%%%%%%%%%%%%%%%%%%%%%%%%%%%%%%%%%%%%%%%%%%%%%%%%%%%%%%%%%
\section{Standard Model Processes and their Simulation}
\label{ch:sm}
%%%%%%%%%%%%%%%%%%%%%%%%%%%%%%%%%%%%%%%%%%%%%%%%%%%%%%%%%%%%%%%%%%%%%%%

In the search for events with isolated leptons and missing transverse
momentum, processes are considered signal if they produce events
containing a genuine isolated lepton and genuine missing transverse
momentum in the final state.
All other processes are defined as background and contribute to the
selected sample through misidentification or mismeasurement.
Studies on how background processes enter the sample are discussed
in appendix~\ref{ch:bgstudies}.

%%%

\begin{figure}[]
  \begin{center}
    \includegraphics[width=0.45\textwidth]{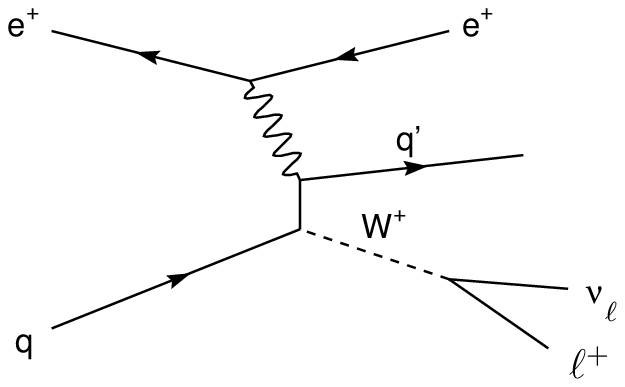}
    \hfill
    \includegraphics[width=0.45\textwidth]{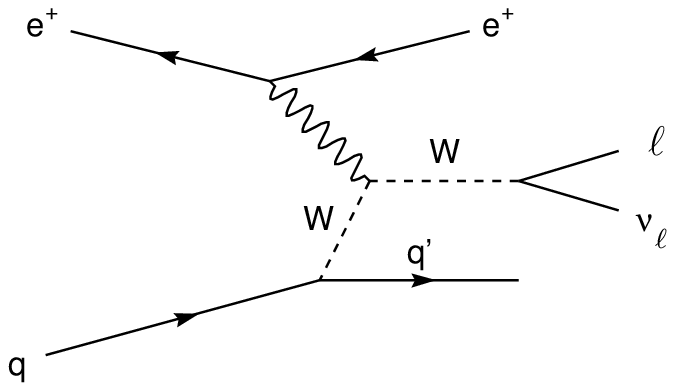}\\
      \vspace{1.8cm}
    \includegraphics[width=0.45\textwidth]{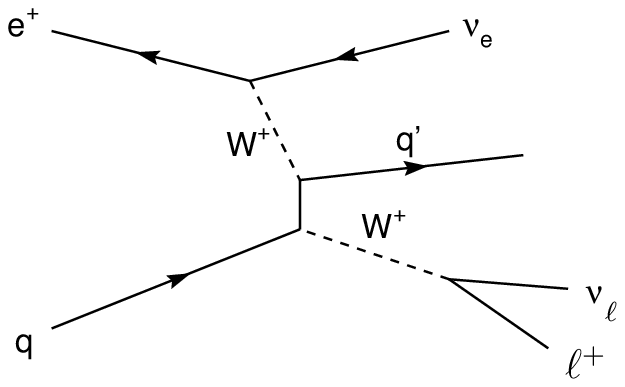}
    \hfill
    \includegraphics[width=0.45\textwidth]{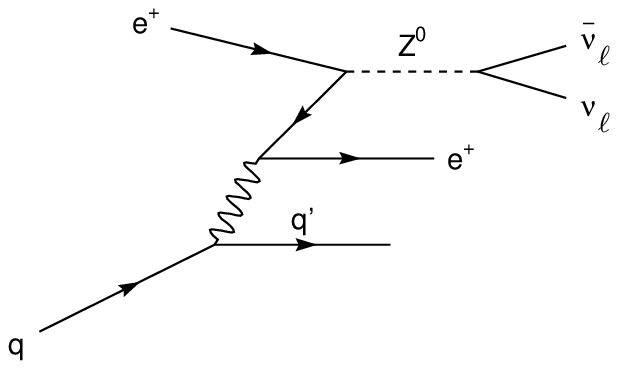}
      \vspace{1.8cm}
  \end{center} 
  \begin{picture} (0.,0.) 
    \setlength{\unitlength}{1.0cm}
    \put ( 5.3,8.6){(a)} 
    \put (13.2,8.6){(b)} 
    \put ( 5.3,2.3){(c)} 
    \put (13.2,2.3){(d)} 
  \end{picture} 
  \caption{Diagrams of processes at HERA with an isolated lepton and
  missing transverse momentum in the final state:
  (a)~$ep~\rightarrow~eW^{\pm}(\rightarrow\ell\nu)X$; (b)~$W$
  production via the $WW\gamma$ triple gauge boson coupling;
  (c)~$ep~\rightarrow~\nu~W^{\pm}(\rightarrow\ell\nu)X$;
  (d)~$ep~\rightarrow~eZ(\rightarrow\nu\bar{\nu})X$. The diagrams are
  shown for $e^{+}p$ collisions.}
  \label{fig:feynman}
\end{figure}

Single $W$ boson production in $ep$ collisions with subsequent
leptonic decay $W\rightarrow \ell\nu$, as illustrated in
figures~\ref{fig:feynman}~(a)--(c), is the main SM process that
produces events with high energy isolated leptons and missing
transverse momentum.
The SM prediction for $W$ production via $ep\rightarrow eW^{\pm}X$ is
calculated in the framework of the EPVEC event
generator~\cite{Baur:1991pp}, which employs the full set of LO
diagrams, including $W$ production via the $WW\gamma$ triple gauge
boson coupling as illustrated in figure
\ref{fig:feynman}~(b).
Each event generated by EPVEC according to its LO cross section is
weighted by a factor dependent on the transverse momentum and rapidity
of the $W$, such that the resulting cross section corresponds to the
NLO calculation~\cite{Diener:2003df,Diener:2002if}.
The ACFGP~\cite{Aurenche:1992sb} parameterisation is used for the
photon structure and the CTEQ4M~\cite{Lai:1996mg} parton distribution
functions are used for the proton.
The renormalisation scale is taken to be equal to the factorisation
scale and is fixed to the $W$ mass.
Final state parton showers are simulated using the PYTHIA
framework~\cite{Diaconu:1998kr}.
The NLO corrections are found to be of the order of $30\%$ at low $W$
transverse momentum (resolved photon interactions) and typically
$10\%$ at high $W$ transverse momentum (direct photon interactions)
\cite{Diener:2003df,Diener:2002if}.
The NLO calculation reduces the theory error to $15\%$, compared to
$30\%$ at leading order.
The charged current $W$ production process $ep\rightarrow \nu
W^\pm X$, illustrated in figure~\ref{fig:feynman}~(c), is calculated
at LO with EPVEC and found to contribute only $7\%$ of the predicted
signal cross section.
The total $W$ production cross section, calculated in this way,
amounts to $1.1$ pb for an electron--proton centre of mass energy of
$\sqrt{s}=301$~GeV and $1.3$ pb for $\sqrt{s}=319$~GeV.

%%%

Signal events may also arise from $Z$ production with subsequent decay
to neutrinos.
The outgoing electron from this reaction is the isolated lepton in the
event, while genuine missing transverse momentum is produced by the
neutrinos.
This process, illustrated in figure~\ref{fig:feynman}~(d), is also
calculated with the EPVEC generator and found to contribute less than
$3\%$ of the predicted signal cross section.

%%%

The SM background processes that may mimic the signature through
misidentification or mismeasurement are neutral current (NC) and
charged current (CC) deep--inelastic scattering (DIS), 
photoproduction, lepton pair production and photons from wide
angle bremsstrahlung.
The RAPGAP~\cite{Jung:1993gf} event generator, which implements the
Born level, QCD Compton and boson gluon fusion matrix elements, is
used to model NC DIS events.
The QED radiative effects arising from real photon emission from both
the incoming and outgoing electrons are simulated using the
HERACLES~\cite{Kwiatkowski:1990es} program.
Contributions from elastic and quasi--elastic QED Compton scattering
are simulated with the WABGEN~\cite{Berger:1998kp} generator.
Direct and resolved photoproduction of jets and prompt photon
production are simulated using the PYTHIA~\cite{Sjostrand:2006za}
event generator.
The simulation is based on Born level hard scattering matrix elements
with radiative QED corrections.
In RAPGAP and PYTHIA, jet production from higher order QCD radiation
is simulated using leading logarithmic parton showers and
hadronisation is modelled with Lund string fragmentation.
The leading order prediction for NC DIS and photoproduction processes
with two or more high transverse momentum jets is scaled by a factor
of $1.2$, to account for higher order QCD corrections to the Monte
Carlo (MC) generators~\cite{Adloff:2002au,Aktas:2004pz}.
Charged current DIS events are simulated using the
DJANGO~\cite{Schuler:yg} generator, which includes first order
leptonic QED radiative corrections based on HERACLES.
The production of two or more jets in DJANGO is accounted for using
the colour--dipole~model~\cite{Lonnblad:1992tz}.
The contribution to the SM background from lepton pair production $ep
\rightarrow e~\ell^{+}\ell^{-}X$ is calculated using the
GRAPE~\cite{Abe:2000cv} generator, based on a full set of electroweak
diagrams.
The production mechanisms include $\gamma \gamma$, $\gamma Z$, $ZZ$
interactions, internal photon conversion and the decay of virtual or
real $Z$ bosons.

%%%

Generated events are passed through the full GEANT~\cite{Brun:1987ma}
based simulation of the H1 detector, which takes into account the
running conditions of the different data taking periods, and are
reconstructed and analysed using the same program chain as for the
data.

%%%%%%%%%%%%%%%%%%%%%%%%%%%%%%%%%%%%%%%%%%%%%%%%%%%%%%%%%%%%%%%%%%%%%%%
\section{Experimental Conditions}
\label{ch:experiment}
%%%%%%%%%%%%%%%%%%%%%%%%%%%%%%%%%%%%%%%%%%%%%%%%%%%%%%%%%%%%%%%%%%%%%%%

The data were recorded at electron and proton beam energies of
$27.6$~GeV and $820$~GeV or $920$~GeV, corresponding to
centre--of--mass energies $\sqrt{s}$ of $301$~GeV or $319$~GeV,
respectively.
The total integrated luminosity of the analysed data is
$474$~pb$^{-1}$, which represents a factor of four increase with
respect to the previous published results.
The data set is made up of $183$~pb$^{-1}$ recorded in $e^-p$
collisions ($1998$--$2006$) and $291$~pb$^{-1}$ in $e^+p$ collisions
($1994$--$2007$), of which $36$~pb$^{-1}$ were recorded at $\sqrt{s} =
301$~GeV ($1994$--$1997$).
Data collected from $2003$ onwards were taken with a longitudinally
polarised lepton beam, with polarisation typically at a level of
$35\%$.
The residual polarisation of the combined left--handed and
right--handed data periods is about $2\%$ left--handed.
While previous measurements were performed using mainly $e^{+}p$ data,
an $e^{-}p$ data set with more than a ten--fold increase in integrated
luminosity is now analysed.

%%%

A detailed description of the H1 experiment can be found
in~\cite{Abt:1997}. Only the detector components relevant to the
present analysis are briefly described here.
The origin of the H1 coordinate system is the nominal $ep$ interaction
point, with the direction of the proton beam defining the positive
$z$--axis (forward region). Transverse momentum is measured in the
$x-y$ plane. The pseudorapidity $\eta$ is related to the polar angle
$\theta$ by $\eta = -\ln \, \tan (\theta/2)$.

%%%

The Liquid Argon (LAr) calorimeter~\cite{Andrieu:1993kh} covers the
polar angle range $4^\circ < \theta < 154^\circ$ with full azimuthal
acceptance.
Electromagnetic shower energies are measured with a precision of
$\sigma (E)/E = 12\%/ \sqrt{E/\mbox{GeV}} \oplus 1\%$ and hadronic
energies with $\sigma (E)/E = 50\%/\sqrt{E/\mbox{GeV}} \oplus 2\%$, as
measured in test beams~\cite{Andrieu:1993tz,Andrieu:1994yn}.
In the backward region, energy measurements are provided by a
lead/scintillating--fibre calorimeter\footnote{This device was
installed in 1995, replacing a lead-scintillator sandwich calorimeter
\cite{Abt:1997}.} (SpaCal)~\cite{Appuhn:1996na} covering the range
$155^\circ < \theta < 178^\circ$.

%%%

The central ($20^\circ < \theta < 160^\circ$) and forward ($7^\circ <
\theta < 25^\circ$) inner tracking detectors are used to measure
charged particle trajectories, to reconstruct the interaction vertex
and to complement the measurement of hadronic energies.
In each event the tracks are used in a common fit procedure to
determine the $ep$ interaction vertex.
Tracks not fitted to the event vertex are referred to as NV~tracks in
the following.
The LAr calorimeter and inner tracking detectors are enclosed in a
super--conducting magnetic coil with a field strength of $1.16$~T.
From the curvature of charged particle trajectories in the magnetic
field, the central tracking system provides transverse momentum
measurements with a resolution of $\sigma_{P_{T}}/P_{T} = 0.005 P_{T}
/ \rm{GeV} \oplus 0.015$~\cite{Kleinwort:2006zz}.
The return yoke of the magnetic coil is the outermost part of the
detector and is equipped with streamer tubes forming the central muon
detector and tail catcher ($4^\circ < \theta < 171^\circ$).
In the forward region ($3^\circ < \theta < 17^\circ$) a set of drift
chamber layers forms the forward muon detector, which together with an
iron toroidal magnet allows a momentum measurement.

%%%

The luminosity is determined from the rate of the Bethe--Heitler
process $ep \rightarrow ep \gamma$, measured using a photon detector
located close to the beam pipe at $z=-103~{\rm m}$, in the backward
direction.

%%%

The LAr calorimeter provides the main trigger for events in this
analysis \cite{Adloff:2003uh}.
The trigger efficiency is almost~$100\%$ for events containing an
electron with a transverse momentum above $10$~GeV.
For events with a transverse momentum imbalance of $12$~GeV measured
in the calorimeter the trigger efficiency is about~$60\%$, rising to
about~$98\%$ for an imbalance greater than $25$~GeV.
In addition to the calorimeter trigger, muon events may also be
triggered by a pattern consistent with a minimum ionising particle in
the muon system in coincidence with tracks in the tracking detectors.

%%%

In order to remove events induced by cosmic rays and other non--$ep$
background, the event vertex is required to be reconstructed within
$\pm 35$~cm in $z$ of the average nominal interaction point.
In addition, topological filters and timing vetoes are applied.

%%%%%%%%%%%%%%%%%%%%%%%%%%%%%%%%%%%%%%%%%%%%%%%%%%%%%%%%%%%%%%%%%%%%%%%
\section{Particle Identification and Event Kinematics}
\label{ch:reco}
%%%%%%%%%%%%%%%%%%%%%%%%%%%%%%%%%%%%%%%%%%%%%%%%%%%%%%%%%%%%%%%%%%%%%%%

The identification of electrons is based on the measurement
\cite{pbruel} of a compact and isolated electromagnetic cluster in the
LAr calorimeter or SpaCal.
The energy measured within a cone in pseudorapidity--azimuth $(\eta -
\phi)$ of radius $R=\sqrt{\Delta \eta^2 + \Delta \phi^2}=0.5$ around the
electron is required to be less than $3\%$ of the total cluster
energy.
For $\theta_{e}>20^{\circ}$, electrons are also required to have an
associated track with an extrapolated distance of closest approach
(DCA) to the cluster of less than $12$~cm.
The distance from the first measured track point in the central drift
chambers to the beam axis is required to be below $30$~cm in order to
reject photons that convert late in the central tracker material.
For $\theta_{e}<20^{\circ}$, no track conditions are imposed, but an
additional calorimetric isolation criterion is used, where the energy
in a cone of radius $R=0.75$ is required to be less than $5\%$ of the
total cluster energy.
The charge of the electron, measured from the associated track, is
used for events with $\theta_{e}>20^{\circ}$ and if the charge is
measured with a significance $\sigma_{q} = |\kappa|/
\delta\kappa>1.0$, where the curvature $|\kappa|$ of the track is
proportional to $1/P_{T}^{\rm track}$ and $\delta\kappa$ is the
associated error on the curvature measurement.
Electrons found in regions between calorimeter modules containing
large amounts of inactive material are excluded
\cite{Adloff:1999ah}.
The energy and polar angle of the electron are measured from the
calorimeter cluster.
The azimuthal angle is taken from the associated track.
The calibration of the electromagnetic part of the LAr calorimeter is
performed using NC events with the method described
in~\cite{Adloff:1999ah}.

%%%

Muon identification is based on the measurement of a track in the
inner tracking systems associated with a track or an energy deposit in
the central muon detector or forward muon
detector~\cite{Aktas:2003sz,Aaron:2008jh}.
Muons which do not reach the muon detector, or enter inefficient
regions of the muon detector, may be identified by a central track
linked to a signature of a minimal ionising particle in the LAr
calorimeter and a hit in the tail catcher.
The muon momentum is measured from the track curvature in the
solenoidal or toroidal magnetic field.
For very high energy muons, the curvature may be compatible with zero
within two standard deviations of its error $\delta\kappa$, and in
such cases is re--evaluated at $|\kappa| + \delta\kappa$, allowing a
lower limit to be placed on the transverse momentum.
A muon may have no more than $5$~GeV deposited in a cylinder, centred
on the muon track direction, of radius $25$~cm and $50$~cm in the
electromagnetic and hadronic sections of the LAr calorimeter,
respectively.

%%%

Calorimeter energy deposits and tracks not previously identified as
electron or muons are used to form combined cluster--track objects,
from which the hadronic final state is
reconstructed~\cite{matti,benji}.
Jets with a minimum transverse momentum of $4$~GeV are reconstructed
from these combined cluster--track objects using an inclusive $k_T$
algorithm~\cite{Ellis:1993tq,Catani:1993hr}.

%%%

The isolation of leptons with respect to jets and tracks in the event
is quantified using the distances of separation in $(\eta-\phi)$ space
$D(\ell;{\rm jet})$ and $D(\ell;{\rm track})$, respectively.

%%%

Hadronic one--prong decays of tau leptons are identified using
isolated narrow jets with a track multiplicity of one.
Such tau candidate jets are only identified in the dedicated search
described in section~\ref{sec:isotausel}.

%%%

The following kinematic quantities are employed in the analysis, some
of which are sensitive to the presence of undetected high energy
particles and/or can be used to reduce the main background
contributions:
\begin{itemize}

\item $P^{X}_{T}$, the transverse momentum of the inclusive hadronic final state.

\item $P^{\rm miss}_{T}$, the total missing transverse momentum
  calculated from all reconstructed particles. In events with large
  $P_{T}^{\rm{miss}}$, the only non--detected particle in the event is
  assumed to be a neutrino.

\item $P^{\rm calo}_{T}$, the net transverse momentum calculated
  from all reconstructed particles, where for muons only the energy
  deposited in the calorimeter is included. For events containing high
  energy muons $P_T^{\rm calo} \simeq P_T^{X}$ (the hadronic
  transverse momentum), otherwise $P_{T}^{\rm calo}~=~P_{T}^{\rm
  miss}$.

\item $V_{\rm ap}/V_{\rm p}$, a measure of the azimuthal
  balance of the event.  It is defined as the ratio of the
  anti--parallel to parallel components of all measured calorimetric
  clusters, with respect to the direction of the calorimetric
  transverse momentum \cite{Adloff:1999ah}.

\item $\Delta \phi_{\ell-X}$, the difference in azimuthal angle between
  the lepton and the direction of $P^X_T$. For events with low
  hadronic transverse momentum $P^X_T<1.0$~GeV, $\Delta \phi_{\ell-X}$
  is set to zero.

\item $\delta_{\rm miss}=2E^{0}_{e}- \sum_i (E^{i} - P_{z}^{i})$,
  where the sum runs over all detected particles, $P_{z}$ is the
  momentum along the proton beam axis and $E^{0}_{e}$ is the electron
  beam energy. For an event where only momentum in the proton
  direction is undetected, $\delta_{\rm miss}$ is zero.
  
\item ${\zeta}^{2}_{e}=4 E_{e}E^{0}_{e} \cos^2 \theta_e/2$, where
  $E_{e}$ is the energy of the final state electron. For NC events,
  where the scattered electron is identified as the isolated high
  transverse momentum electron, ${\zeta}^{2}_{e}$ is equal to the four
  momentum transfer squared $Q^{2}_{e}$, as measured by the electron
  method~\cite{JacquetBlondel}.

\end{itemize}

%%%%%%%%%%%%%%%%%%%%%%%%%%%%%%%%%%%%%%%%%%%%%%%%%%%%%%%%%%%%%
\section{Event Selection}
\label{ch:sel}
%%%%%%%%%%%%%%%%%%%%%%%%%%%%%%%%%%%%%%%%%%%%%%%%%%%%%%%%%%%%%

%%%%%%%%%%%%%%%%%%%%%%%%%%%%%%%%%%%%%%%%%%%%%%%%%%%%%%%%%%%%%
\subsection{Events with Isolated Electrons and Muons}
\label{sec:isolepsel}
%%%%%%%%%%%%%%%%%%%%%%%%%%%%%%%%%%%%%%%%%%%%%%%%%%%%%%%%%%%%%

\begin{table}[t]
  \renewcommand{\arraystretch}{1.4}
\begin{center}
\begin{tabular}{|c||cc|} \hline
\multicolumn{3}{|c|}{\bf H1 Isolated Lepton {\boldmath $+ P_{T}^{\rm miss}$} Event Selection}\\
\hline
\multicolumn{1}{|c}{}
& \multicolumn{1}{c}{Electron}
& \multicolumn{1}{c|}{Muon} \\
\hline
\hline
{\bf Basic Event} &  \multicolumn{2}{c|}{$5^\circ<\theta_\ell<140^\circ$}\\

{\bf Selection} & \multicolumn{2}{c|}{$P_{T}^{\ell}>10$ GeV}\\

& \multicolumn{2}{c|}{$P_{T}^{\rm miss}>12$ GeV}\\ 

& \multicolumn{2}{c|}{$P_{T}^{\rm calo}>12$ GeV}\\

\hline
{\bf Lepton Isolation}

& \multicolumn{2}{c|}{$D(\ell;{\rm jet})>1.0$}\\

& $D(e;{\rm track})>$ $0.5$ for $\theta_{e}>45^\circ$ & $D(\rm{\mu;track})>0.5$ \\

\hline

{\bf Background} & \multicolumn{2}{c|}{$V_{\rm ap}/V_{\rm p}<$ $0.5$}\\

{\bf Rejection} & $V_{\rm ap}/V_{\rm p}<0.2$ for $P_{T}^{e}<25$ GeV & $V_{\rm ap}/V_{\rm p}<0.2$ for $P_{T}^{\rm calo} <25$ GeV\\

& $\Delta\phi_{e-X}<160^\circ$ & $\Delta\phi_{\mu-X}<170^\circ$\\

& $\delta_{\rm miss}>5$ GeV & -- \\

& $\zeta^{2}_{e}>5000$ GeV$^{2}$ for $P_{T}^{\rm calo} <25$ GeV & -- \\

& \multicolumn{2}{c|}{$M_{T}^{\ell\nu}>10$ GeV}\\

& -- & {$P_{T}^{X}>12$ GeV}\\

\hline
\end{tabular}
\end{center}
  \caption{Summary of selection requirements for the electron and muon
  channels in the search for events with isolated leptons and missing
  transverse momentum.}
\label{tab:isolepcutstable}
\end{table}

%%%

The event selection for isolated electrons and muons is based on the
analysis described in~\cite{Andreev:2003pm,South:2003se,Brandt:2007}
and is summarised in table~\ref{tab:isolepcutstable}.
The lepton is required to have a high transverse momentum
$P_{T}^{\ell}>10$~GeV, be observed in the central or forward region of
the detector, $5^\circ<\theta_{\ell}<140^\circ$, and be isolated with
respect to jets and other tracks in the event.
A large transverse momentum imbalance $P_{T}^{\rm miss}$ is required
and the cut on $P_{T}^{\rm calo}$ is imposed to ensure a high trigger
efficiency.

%%%

In order to reject the NC background contribution in the electron
channel, further cuts are applied on the longitudinal momentum
imbalance $\delta_{\rm miss}$ and the azimuthal balance of the event,
using $V_{\rm ap}/V_{\rm p}$ and $\Delta\phi_{e-X}$.
The cut on $\delta_{\rm miss}$ is only performed if the event contains
exactly one electron, which has the same charge as the beam lepton.
Electron events with low values of $P_{T}^{\rm calo}$ are additionally
required to have large $\zeta^{2}_{e}$ in order to further reduce NC
background, which is predominantly at low $Q^{2}_{e}$.
Lepton pair background is removed from the muon channel by rejecting
azimuthally balanced events, using $V_{\rm ap}/V_{\rm p}$ and
$\Delta\phi_{\mu-X}$, and by rejecting events with two or more
isolated muons.
To ensure that the two lepton channels are exclusive, electron events
must contain no isolated muons.

%%% 

Some changes have been made to the selection with respect to that used
in the previous publication~\cite{Andreev:2003pm}.
The cut on $V_{\rm ap}/V_{\rm p}$ at low transverse momentum in both
the electron and muon channels has been relaxed to $0.2$, leading
to an improved signal efficiency.
As described in section~\ref{ch:reco}, in the very forward region
$\theta_{e}<20^{\circ}$ the electron DCA requirement is replaced by a
stricter calorimetric isolation criterion, which results in a better
CC rejection.
Finally, the lepton--neutrino transverse mass:

\[
 M_{T}^{\ell\nu} = \sqrt{(P_T^{\rm miss} + P_T^{\ell})^2 - (\vec{P}_T^{\rm miss} + \vec{P}_T^{\ell})^2}
\]

\noindent
is required to be larger than $10$~GeV in order to further reject NC
(lepton pair) background in the electron (muon) channel.

%%%

Following the selection criteria described above, the overall
efficiency to select SM $W \to e \nu$ events is $44\%$ and to select
SM $W \to \mu \nu$ events is $17\%$, calculated using the EPVEC
simulation.
The main difference in efficiency between the two channels is due to
the cut on $P_{T}^{\rm calo}$, which for muon events effectively acts
as a cut on $P_{T}^{X}$.
As the efficiency for muons at low $P_{T}^{X}$ is small, an additional
requirement is made of $P_{T}^{X}>12$~GeV.
For events with $P_{T}^{X}>25$ GeV the selection efficiencies of the
two channels are comparable at $\sim 42\%$.

%%%%%%%%%%%%%%%%%%%%%%%%%%%%%%%%%%%%%%%%%%%%%%%%%%%%%%%%%%%%%
\subsection{Events with Isolated Tau Leptons}
\label{sec:isotausel}
%%%%%%%%%%%%%%%%%%%%%%%%%%%%%%%%%%%%%%%%%%%%%%%%%%%%%%%%%%%%%

The selection of events with isolated tau leptons and missing
transverse momentum is based on the analysis described
in~\cite{Aktas:2006fc} and is summarised in
table~\ref{tab:isotaucutstable}.
The tau identification algorithm selects narrow (pencil--like), low
multiplicity jets typical for hadronic tau decays.
Only hadronic decays with one charged hadron (one--prong) are considered.
Tau decays to electrons and muons enter the electron and muon channels
described in section~\ref{sec:isolepsel}.

%%%

To ensure the presence of neutrinos in the event, large $P_{T}^{\rm
miss}$ and $P_{T}^{\rm calo}$, significant $\delta_{\rm miss}$ and low
$V_{\rm ap}/V_{\rm p}$ are required.
The event should also exhibit large inclusive hadronic transverse
momentum $P_{T}^{X}$.
At this stage the selection contains $96\%$ CC events and is denoted
CC--like in the following.

%%%

Tau jet candidates are based on jets found in the hadronic final
state.
The tau lepton four--vector is approximated using the four--vector of
the tau jet candidate.
The tau lepton four--vector is subtracted from the inclusive hadronic
final state $X$ to obtain the transverse momentum of the remaining
hadronic system $X'$ as:

\[
\vec{P}_{T}^{X'}=\vec{P}{}_{T}^{X}-\vec{P}_{T}^{\tau}.
\]

%%%

Jets with transverse momentum $P_{T}^{\rm jet}>7$~GeV found in the
central region $20^{\circ}<\theta_{\rm jet}<120^{\circ}$ of the
detector are considered as tau jet candidates.
The angular region is reduced with respect to that used in the
electron and muon channels since in the forward region the track
multiplicity cannot be reliably measured and in the backward region
the rate from NC DIS is high, leading to a correspondingly high
expectation of narrow jets from falsely identified electrons.
The jet radius is used as a measure for the collimation of the jet and
is calculated as:

\[
R_{\rm jet}=\frac{1}{E_{\rm jet}}\sum_{h}E_{h}\sqrt{\Delta\eta({\rm jet},h)^{2}+\Delta\phi({\rm jet},h)^{2}},
\]

\noindent
where $E_{\rm jet}$ is the total jet energy and the sum runs over all
jet daughter hadronic final state particles of energy $E_{h}$.
Narrow jets are selected by requiring $R_{\rm jet}<0.12$.
At least one track measured in the central tracking detector with
transverse momentum $P_{T}^{\rm track}>5$ GeV is required to be
associated to the jet.
Jets meeting the above criteria are denoted tau--like jets in the
following and are investigated in control samples as described in
appendix~\ref{sec:isotaucontrol}.

%%%

\begin{table}[h]
  \renewcommand{\arraystretch}{1.4}
  \begin{center}
    \begin{tabular}{|c||c|} \hline
    \multicolumn{2}{|c|}{\bf H1 Isolated Tau Lepton {\boldmath $+ P_{T}^{\rm miss}$} Event Selection}\\
    \hline
    \hline
    \bf{CC--like Sample}&
      $P_{T}^{\rm miss}>12$~GeV\\
      &
      $P_{T}^{\rm calo}>12$~GeV\\
      &
      $P_{T}^{X}>12$~GeV\\
      &
      $\delta_{\rm miss}>5$~GeV\\
      &
      $V_{\rm ap}/V_{\rm p}<0.5$\\
      &
      $V_{\rm ap}/V_{\rm p}<0.15$ for $P_{T}^{\rm{miss}}<25$~GeV\\
    \hline
    \hline 
    \bf{Tau--like Jets}&
      $P_{T}^{\rm jet}>7$~GeV\\
      &
      $20^{\circ}<\theta_{\rm jet}<120^{\circ}$\\
      &
      $R_{\rm jet}<0.12$\\
      &
      $N_{\rm tracks}^{\rm jet}\geq1$ for $P_{T}^{\mathrm{track}}>5$ GeV\\
    \hline     
    \hline     
      \bf{Isolation}&
      $D(\tau; e,\mu,\rm{jet})>1.0$\\
    \hline
      {\bf Acoplanarity}&
      $\Delta\phi_{\tau-X'}<170^{\circ}$ for $P_{T}^{X'}>5$~GeV\\
    \hline
      {\bf One--prong}&
      $N_{\rm tracks}^{D{\rm jet}<1.0}=1$\\
      &
      $N_{\rm NV tracks}^{D{\rm track}<0.3}=1$\\
    \hline
  \end{tabular}
  \end{center}

  \caption{Summary of selection requirements for isolated tau jets in
  events with missing transverse momentum. $N_{\rm tracks}^{D{\rm
  jet}<1.0}$ denotes the number of tracks in a cone of $1.0$ in the
  $(\eta-\phi)$~plane around the jet--axis. $N_{\rm NV tracks}^{D{\rm
  track}<0.3}$ denotes the number of NV tracks in a cone of $0.3$ in
  the $(\eta-\phi)$~plane around the one--prong track.}

  \label{tab:isotaucutstable}
\end{table}

%%%

For the final selection, the tau jet should be isolated from
electrons, muons and other jets in the event by requiring a
minimum separation $D(\tau;e, \mu, {\rm jet})>1.0$.
To avoid the misidentification of jets or electrons as tau jets due to
distortion of the shower shapes, jets pointing to regions between
calorimeter modules containing large amounts of inactive material are
not considered~\cite{Adloff:1999ah}.
Acoplanarity between the tau jet and the remaining hadronic system
$X'$ in the transverse plane $\Delta\phi_{\tau-X'}$ is required to
suppress events with back--to--back topologies, primarily NC events
and photoproduction events with jets.
Finally, a track multiplicity of one is required in a cone of radius
$R=1.0$ around the jet axis.
To improve the rejection of low multiplicity hadronic jets with
additional charged hadrons, no further NV~tracks are allowed in a cone
of radius $R=0.3$ around the one--prong track.

%%%%%%%%%%%%%%%%%%%%%%%%%%%%%%%%%%%%%%%%%%%%%%%%%%%%%%%%%%%%%%%%%%%%%%%
\section{Systematic Uncertainties}
\label{ch:systematics}
%%%%%%%%%%%%%%%%%%%%%%%%%%%%%%%%%%%%%%%%%%%%%%%%%%%%%%%%%%%%%%%%%%%%%%%

The following experimental systematic uncertainties are considered:
\begin{itemize}

\item The uncertainty on the electromagnetic energy scale varies
  depending on the polar angle from $0.7\%$ in the backward region up
  to $2\%$ in the forward region. The uncertainties on the $\theta_e$
  and $\phi_e$ measurements are $3$~mrad and $1$~mrad,
  respectively. The identification efficiency of electrons is known
  with an uncertainty of $2\%$.

\item The scale uncertainty on the transverse momentum of high $P_T$
  muons is $2.5\%$. The uncertainties on the $\theta_\mu$ and
  $\phi_\mu$ measurements are $3$ mrad and $1$ mrad, respectively. The
  muon identification efficiency has an error of $5\%$ in the region
  $\theta_{\mu}>12.5^\circ$ and $15\%$ in the region
  $\theta_{\mu}<12.5^\circ$.

\item The hadronic energy scale is known within $2\%$ for events with
  $P_{T}^{X}>8$~GeV and $5\%$ for events with $P_{T}^{X}<8$~GeV. The
  uncertainties on the $\theta$ and $\phi$ measurements of the
  hadronic final state are both $10$~mrad. The error on the
  measurement of $V_{\rm ap}/V_{\rm p}$ is $\pm$~$0.02$.

\item The uncertainty on the track reconstruction efficiency is $3\%$.

\item The uncertainty on the trigger efficiency for the muon channel
  varies from $3.4\%$ for events with $P_{T}^{X}>12$~GeV~to $1.3\%$
  for events with $P_{T}^{X}>25$~GeV.

\item The luminosity measurement has an uncertainty of $3\%$.

\end{itemize}

%%%

The effect of the above systematic uncertainties on the SM expectation
is determined by varying the experimental quantities by $\pm 1$
standard deviation in the MC samples and propagating these variations
through the whole analysis.
The contribution to the total error in the combined electron and muon
channels from the experimental systematic uncertainties is $4.5\%$ at low $P_{T}^{X}$,
rising to $8.1\%$ for events with $P_{T}^{X}>25$~GeV.

%%%

A theoretical uncertainty of $15\%$ is quoted for the predicted
contributions from signal processes.
This is mainly due to uncertainties in the parton distribution
functions and the scales at which the calculation is performed
\cite{Diener:2002if}.

%%%

Additional model systematic uncertainties are attributed to the SM
background MC generators described in section~\ref{ch:sm}, deduced
from the level of agreement observed between the data and simulations
in dedicated samples, as described in appendix~\ref{ch:bgstudies}.

%%%

In the electron and muon channels, the contributions from background
processes modelled using RAPGAP (NC), PYTHIA (photoproduction), GRAPE
(lepton pair) and WABGEN (QED Compton) are attributed an error of
$30\%$, and the CC background contribution, which is modelled using
DJANGO, is attributed an error of $50\%$.
The uncertainties associated with lepton misidentification and the
measurement of missing transverse momentum, as well as the
normalisation of NC DIS and photoproduction processes with at least
two high $P_T$ jets, are included in these errors.

%%%

In the tau channel, the contributions from background processes
modelled using RAPGAP are attributed a $15\%$ model error and from
PYTHIA a $20\%$ model error.
The DJANGO prediction is attributed an error of $20\%$ and the GRAPE
prediction is attributed an error of $30\%$.
In addition to these model uncertainties, an uncertainty on the
description of the jet radius is considered in the tau channel.
The jet radius is varied by $\pm0.012$, corresponding to $10\%$ of the
cut value, resulting in a $17\%$ change in the rate of tau--like jets
selected in the CC--like sample.
All other experimental systematic uncertainties in the tau channel are
included in the model uncertainties.

%%%

In the combination of the electron and muon channels all systematic
errors are treated as fully correlated.
The total error on the SM prediction is determined by adding the MC
statistical error to the effects of all model and experimental
systematic uncertainties in quadrature.

%%%%%%%%%%%%%%%%%%%%%%%%%%%%%%%%%%%%%%%%%%%%%%%%%%%%%%%%%%%%%
\section{Results}
\label{ch:results}
%%%%%%%%%%%%%%%%%%%%%%%%%%%%%%%%%%%%%%%%%%%%%%%%%%%%%%%%%%%%%

%%%%%%%%%%%%%%%%%%%%%%%%%%%%%%%%%%%%%%%%%%%%%%%%%%%%%%%%%%%%%
\subsection{Electron and Muon Channels}
\label{sec:isolepresults}
%%%%%%%%%%%%%%%%%%%%%%%%%%%%%%%%%%%%%%%%%%%%%%%%%%%%%%%%%%%%%

The results of the search for events with an isolated electron or muon
and missing transverse momentum are summarised in
table~\ref{tab:isolepsummary}.
In the full $e^{\pm}p$ data sample $53$ events are observed, in good
agreement with the SM prediction of $54.1\pm7.4$, where $40.4\pm6.3$
events are predicted from signal processes, dominated by single $W$
production.
In the $e^{+}p$ data $40$ events are observed compared to a SM
prediction of $32.3\pm4.4$.
In the $e^{-}p$ data $13$ events are observed compared to a SM
prediction of $21.8\pm3.1$.
Two of the events in the electron channel observed in the previous
analysis~\cite{Andreev:2003pm} are rejected in the present analysis
due to the tighter calorimetric isolation requirement described in
section~\ref{ch:reco}.

%%%

In the complete data set, $39$ electron events are observed, compared
to a SM prediction of $43.1\pm6.0$.
In $17$ of the electron events the lepton charge is measured as
positive, in $9$ events negative and is unmeasured in the remaining
$13$ events.
Similarly for the muon channel, $14$ data events are observed,
compared to a SM prediction of $11.0\pm1.8$.
In $5$ of the muon events the lepton charge is measured as positive,
in $6$ events negative and is unmeasured in the remaining $3$ events.

%%%

\begin{figure}[h]
  \begin{center}
\vspace{-0.4cm}
    \bf{{\large \bf {\boldmath {$e$,~$\mu + P_{T}^{\rm miss}$} events
    at HERA ({\boldmath $e^{\pm}p$},~{\boldmath $474$ pb$^{-1}$})}}}\\
      \includegraphics[width=0.49\textwidth]{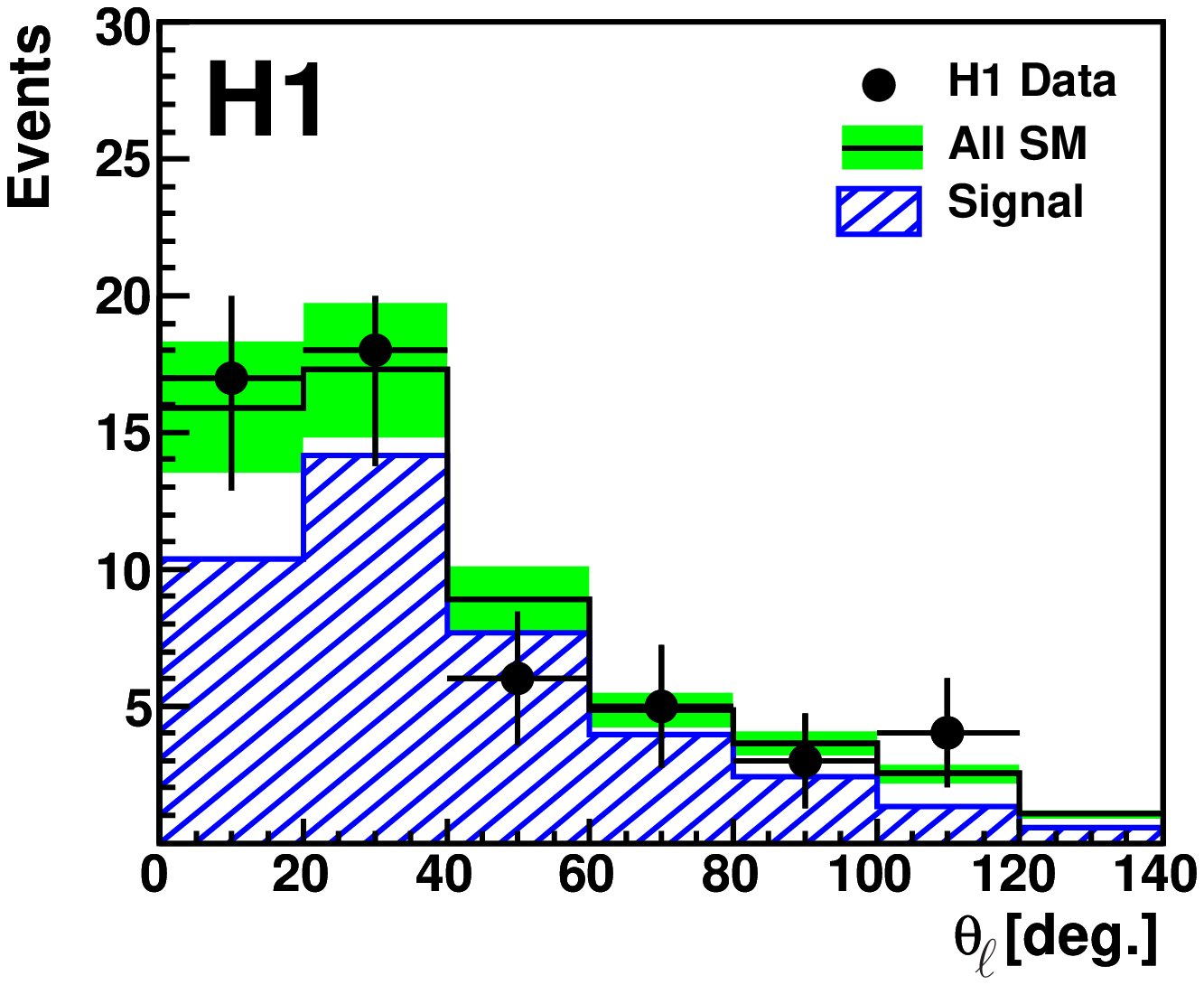}
      \includegraphics[width=0.49\textwidth]{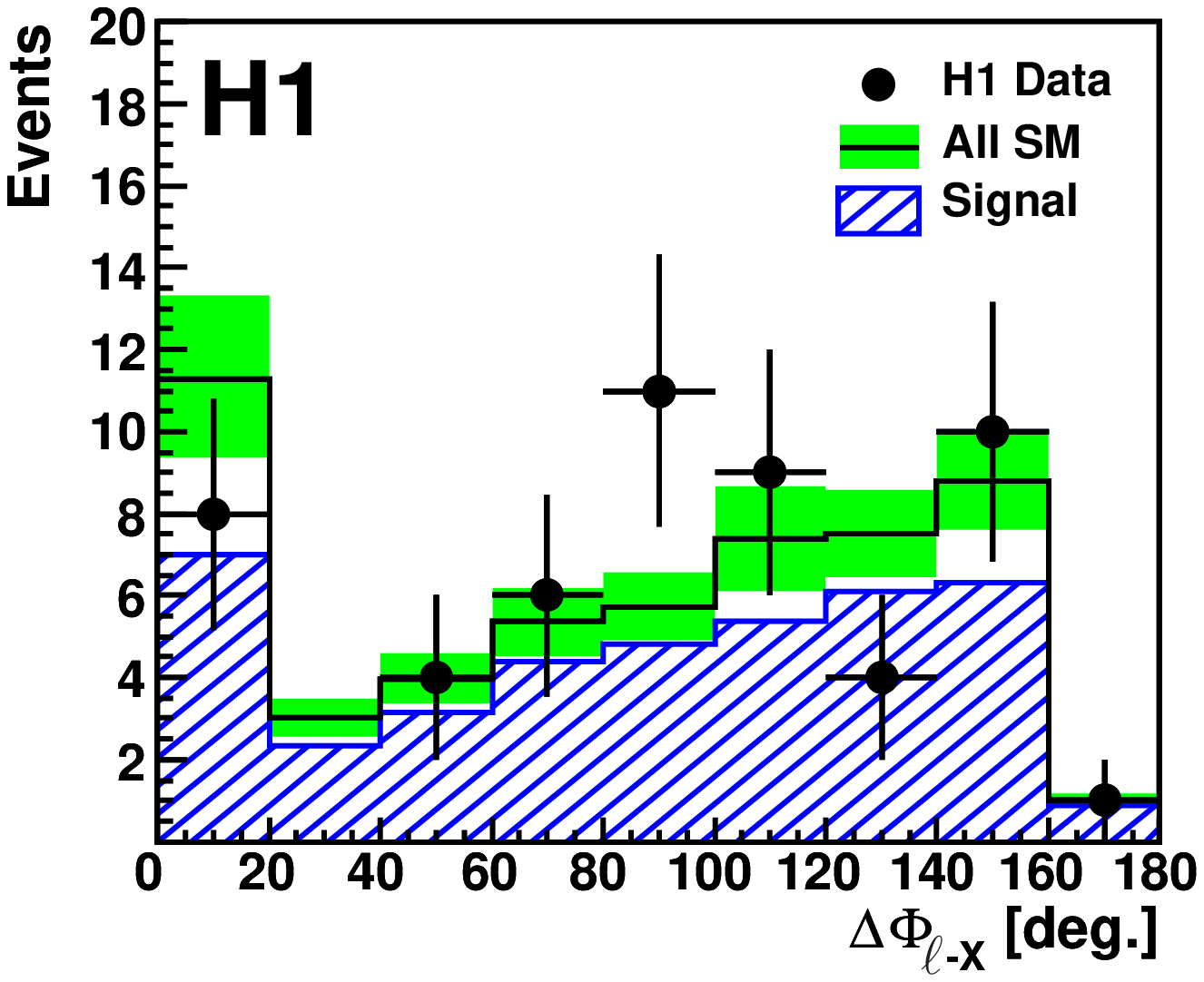}\\
  \vspace{-0.2cm}
      \includegraphics[width=0.49\textwidth]{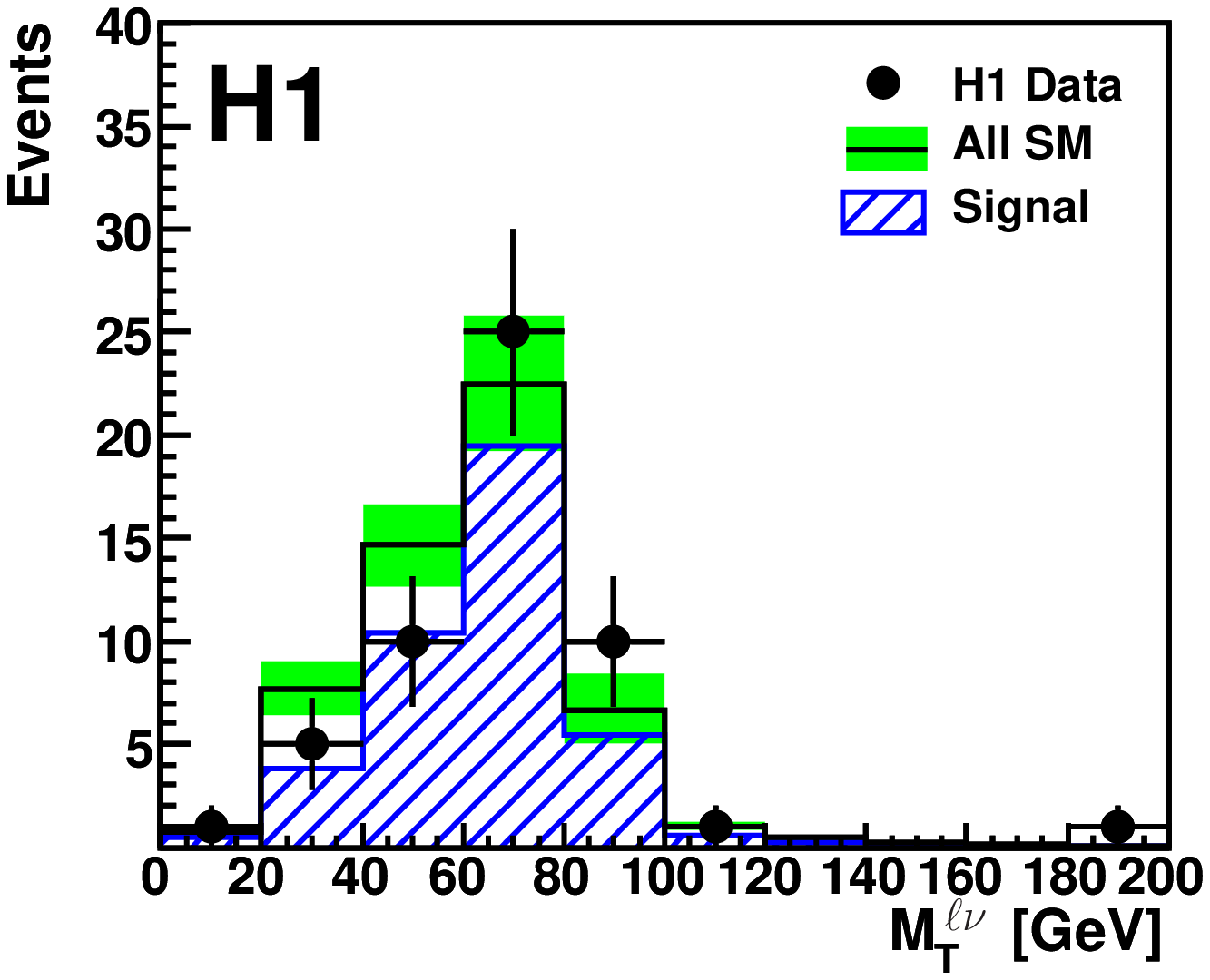}
      \includegraphics[width=0.49\textwidth]{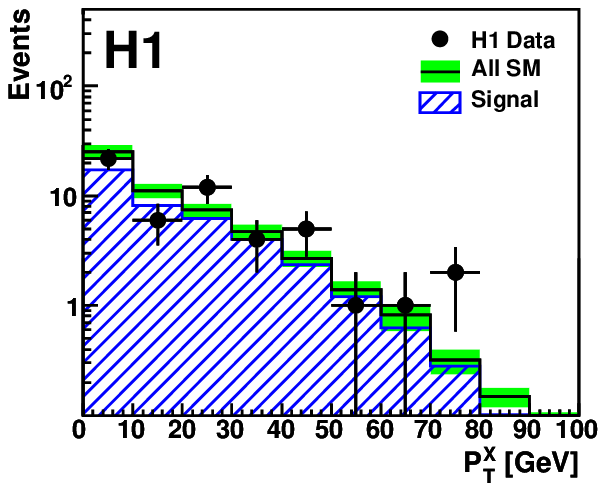}\\
  \vspace{-0.2cm}
      \includegraphics[width=0.49\textwidth]{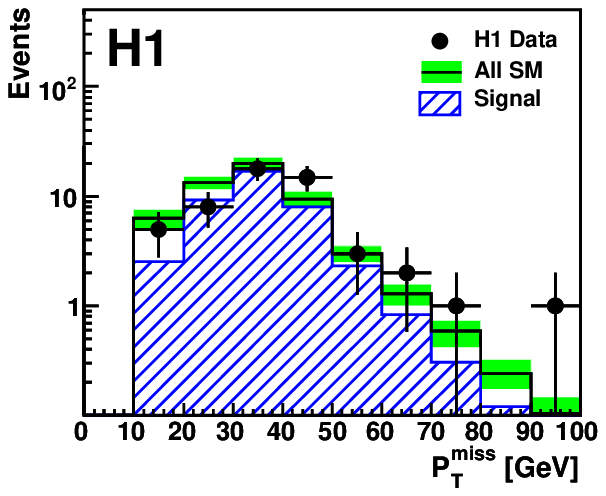}
      \includegraphics[width=0.49\textwidth]{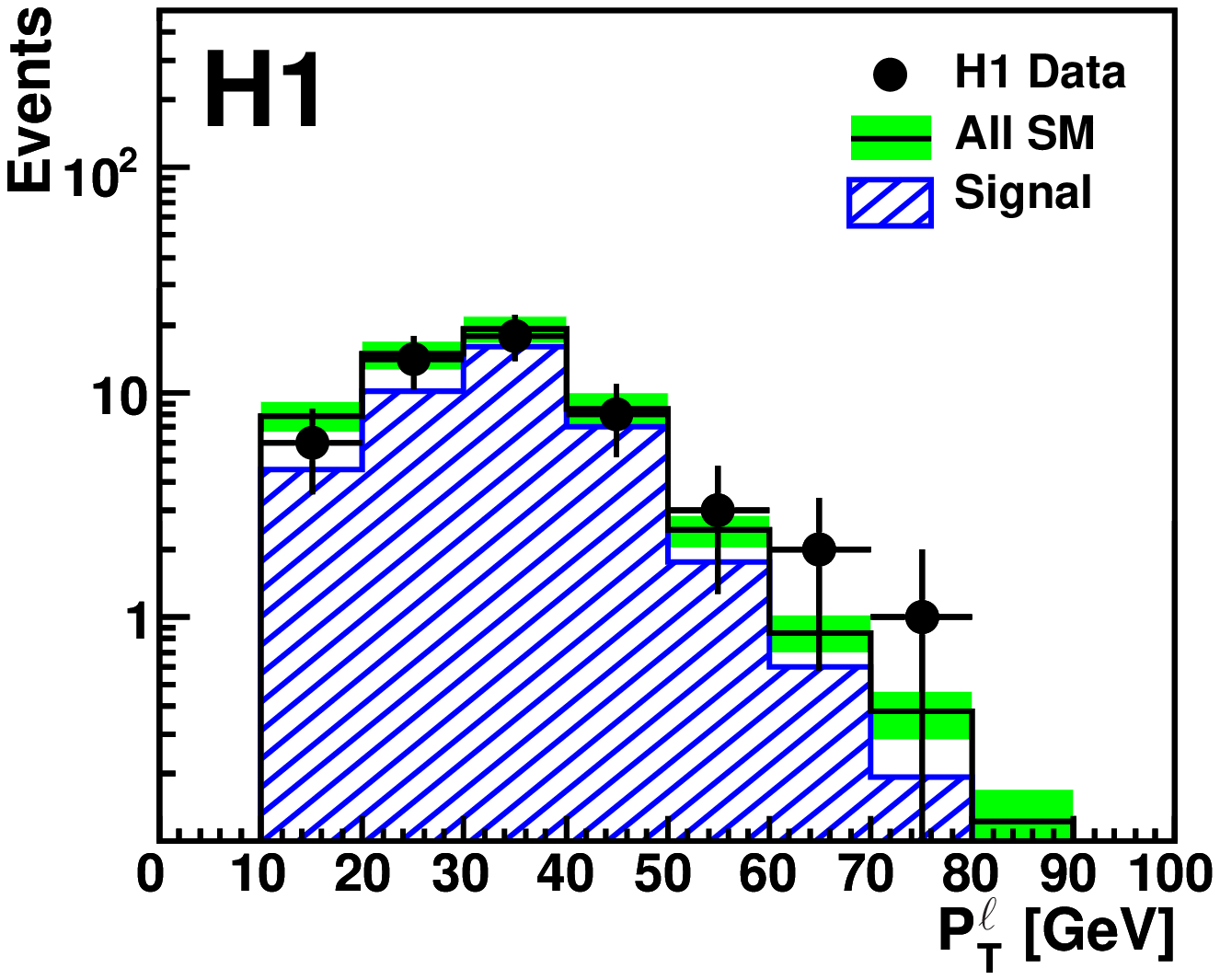}
  \end{center}
  \begin{picture} (0.,0.) 
    \setlength{\unitlength}{1.0cm}
    \put ( 6.3,16.6){(a)} 
    \put (10.2,16.6){(b)} 
    \put ( 6.3,10.3){(c)} 
    \put (14.3,10.3){(d)} 
    \put ( 6.3,4.0){(e)} 
    \put (14.3,4.0){(f)} 
  \end{picture} 
  \vspace{-1cm}
  \caption{Distributions in the combined electron and muon channels
  for the $e^{\pm}p$ data sample.  Shown is the polar angle of the
  lepton $\theta_{\ell}$~(a), the lepton--hadronic system acoplanarity
  $\Delta\phi_{\ell-X}$~(b), the lepton--neutrino transverse mass
  $M_{T}^{\ell\nu}$~(c), the hadronic transverse momentum
  $P_{T}^{X}$~(d), the missing transverse momentum $P_{T}^{\rm
  miss}$~(e) and the transverse momentum of the lepton
  $P_{T}^{\ell}$~(f). The data (points) are compared to the SM
  expectation (open histogram).  The signal component of the SM
  expectation, dominated by real $W$ production, is shown as the
  hatched histogram. The total uncertainty on the SM expectation is
  shown as the shaded band.}
  \label{fig:isolepfinalsample}
\end{figure}

Kinematic distributions of the selected data events are shown in
figure~\ref{fig:isolepfinalsample}.
The events are observed to have generally low values of lepton polar
angle and have a $\Delta \phi_{\ell-X}$ distribution in agreement with
the SM prediction.
The shape of the transverse mass $M_{T}^{\ell\nu}$ distribution shows
a Jacobian peak as expected from single $W$ production.
The observed $P_{T}^{X}$, $P_{T}^{\rm miss}$ and $P_{T}^{\ell}$
distributions are also indicative of $W$ production, where the decay
products of the $W$ peak around $40$~GeV and the hadronic final state
has typically low $P_{T}^{X}$.
In $11$ of the $53$ events in the final sample an additional electron
is observed in the main detector, in agreement with a SM prediction of
$11.7\pm1.5$.
In $W$ production events the additional electron typically corresponds
to the scattered electron, as illustrated in
figure~\ref{fig:feynman}~(a).

%%%

\begin{figure}[h] 

  \begin{center}
  (a)~\includegraphics[width=0.80\textwidth]{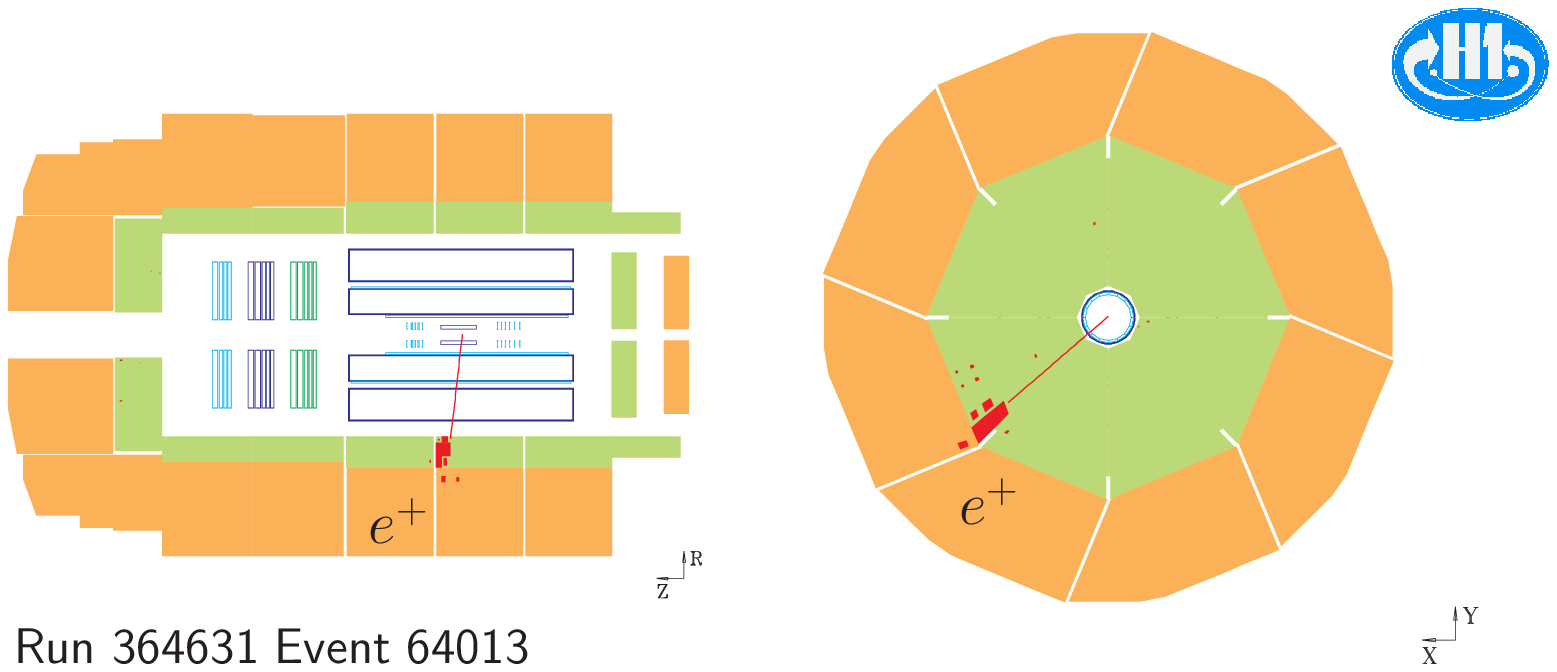}
  \end{center}

  \begin{center}
  (b)~\includegraphics[width=0.80\textwidth]{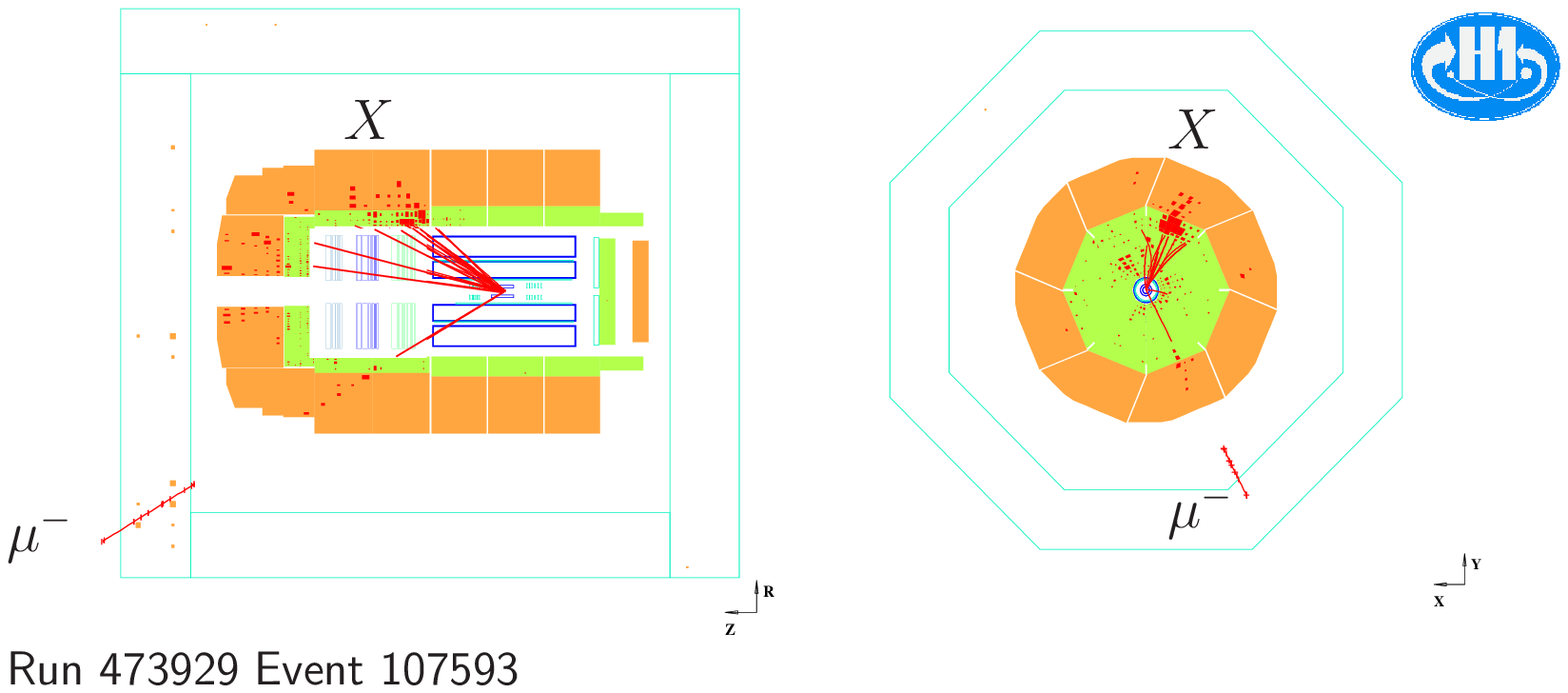}
  \end{center}

  \begin{center}
  (c)~\includegraphics[width=0.80\textwidth]{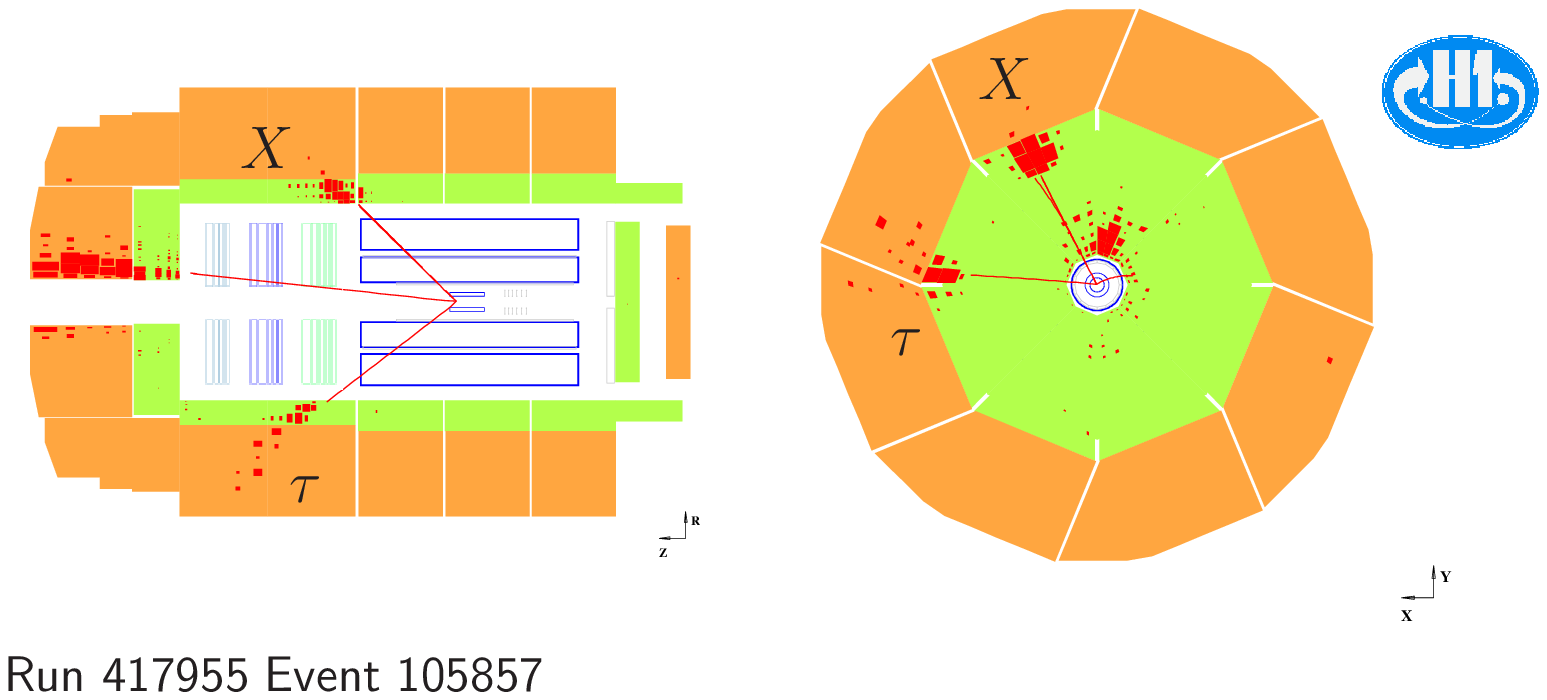}
  \end{center}

  \caption{Displays of isolated lepton events in the final
  sample. (a)~An electron event with missing transverse momentum and
  no visible hadronic final state, typical of low $P_{T}$ single $W$
  production. (b)~An isolated muon event, with missing transverse
  momentum and a prominent hadronic final state $X$. (c)~An event with
  an isolated tau lepton candidate, missing transverse momentum and a
  prominent hadronic final state $X$.}
  \label{fig:eventdisplays}
\end{figure}

Figure~\ref{fig:eventdisplays} displays some examples of data events
selected by the analysis.
Event~(a) features a high $P_{T}$ electron, large missing transverse
momentum and no visible reconstructed hadronic final state, which is
typical of low $P_{T}$ single $W$ production.
Event~(b) features a high $P_{T}$ muon, in addition to a prominent
hadronic jet, resulting in a large $P_{T}^{X}$.
The presence of an undetected particle in the event is particularly
visible in the transverse ($x-y$) plane.

%%%

In the region $P_{T}^{X}>25$~GeV, $18$ events are observed compared to
an expectation of $13.6\pm2.2$, as shown in
table~\ref{tab:isolepsummary}.
Almost all of the high $P_{T}^{X}$ events are seen in the $e^{+}p$
data, where $17$ events are observed compared to an expectation of
$8.0\pm1.3$.
This excess of data events over the SM prediction corresponds to a
$2.4\sigma$ significance, which is lower than previously reported
in~\cite{Andreev:2003pm}.
In the $e^{-}p$ data no excess is seen, where only $1$ event is
observed in the data with $P_{T}^{X}>25$~GeV, compared to a SM
expectation of $5.6\pm0.9$.
The kinematics of the events with $P_T^X>25$ GeV are detailed in
table~\ref{tab:isolepkine}.

%%%

\begin{figure}[h]
  \begin{center}
      \includegraphics[width=0.49\textwidth]{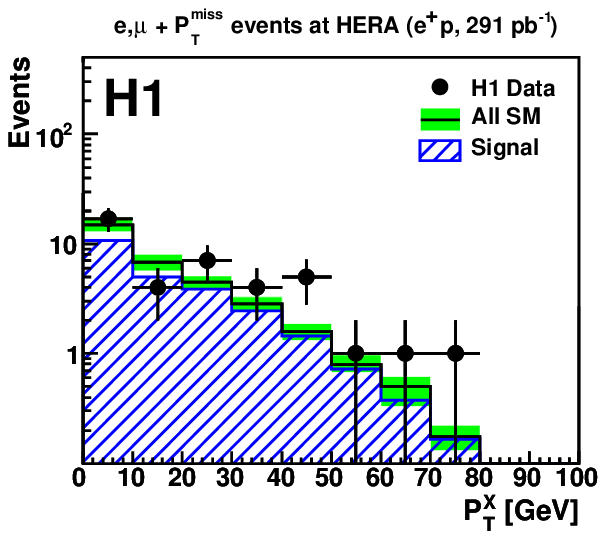}
      \includegraphics[width=0.49\textwidth]{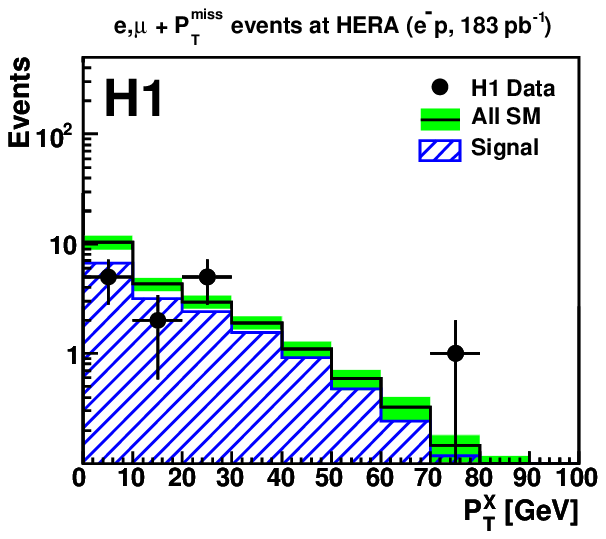}
  \end{center}
  \begin{picture} (0.,0.)
    \setlength{\unitlength}{1.0cm}
    \put ( 6.5,3.8){(a)} 
    \put (14.2,3.8){(b)} 
  \end{picture}  
  \vspace{-0.5cm}
  \caption{The hadronic transverse momentum $P_{T}^{X}$ distributions
  in the combined electron and muon channels for the $e^{+}p$~(a) and
  $e^{-}p$~(b) data samples. The data (points) are compared to the SM
  expectation (open histogram). The signal component of the SM
  expectation, dominated by real $W$ production, is shown as the
  hatched histogram. The total uncertainty on the SM expectation is
  shown as the shaded band.}
 \label{fig:isolepptx}
\end{figure}

The hadronic transverse momentum distribution is shown separately for
$e^{+}p$ and $e^{-}p$ data in figure~\ref{fig:isolepptx}.
In addition to the slight excess of $e^{+}p$ data events visible at
high $P_{T}^{X}$, a slight deficit of data events is visible in the
$e^{-}p$ sample, particularly at low values of $P_{T}^{X}$.
Potential sources of inefficiency in the data due to data quality
requirements, such as trigger conditions, the vertex requirement and
non--$ep$ background suppression, were investigated and ruled out as
the origin of the observed differences between the $e^{+}p$ and
$e^{-}p$ results.
A series of control samples, enriched with SM background physics
processes, are also used to investigate these observations in the
final sample, as discussed in appendix~\ref{ch:bgstudies}.

%%%%%%%%%%%%%%%%%%%%%%%%%%%%%%%%%%%%%%%%%%%%%%%%%%%%%%%%%%%%%
\subsection{Tau Lepton Channel}
\label{sec:isotauresults}
%%%%%%%%%%%%%%%%%%%%%%%%%%%%%%%%%%%%%%%%%%%%%%%%%%%%%%%%%%%%%

\begin{figure}[]
  \begin{center}
    \bf{{\large \bf {\boldmath {$\tau + P_{T}^{\rm miss}$} events at
    HERA ({\boldmath $e^{\pm}p$},~{\boldmath $474$~pb$^{-1}$})}}}\\
    \includegraphics[width=0.49\textwidth]{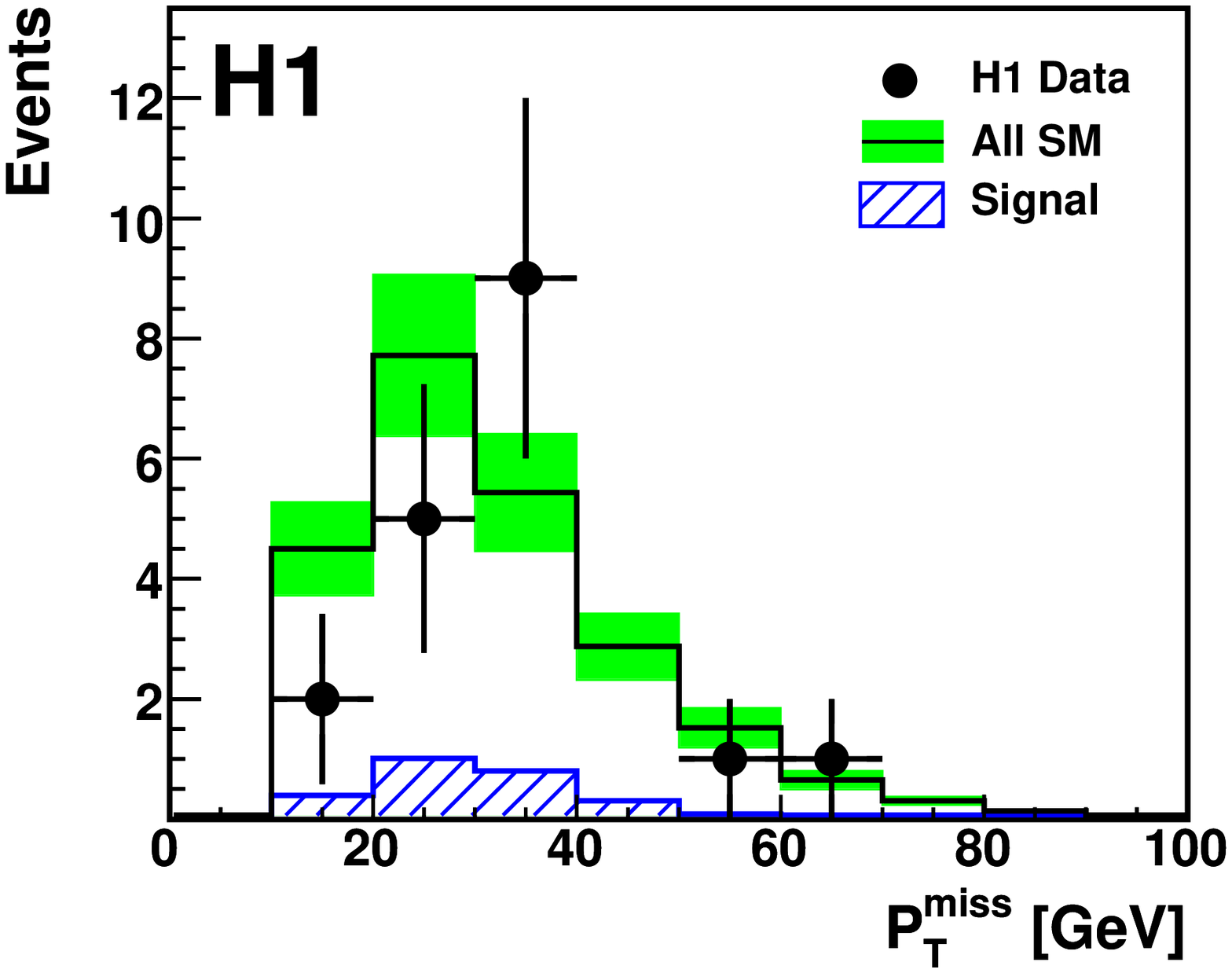}     
    \includegraphics[width=0.49\textwidth]{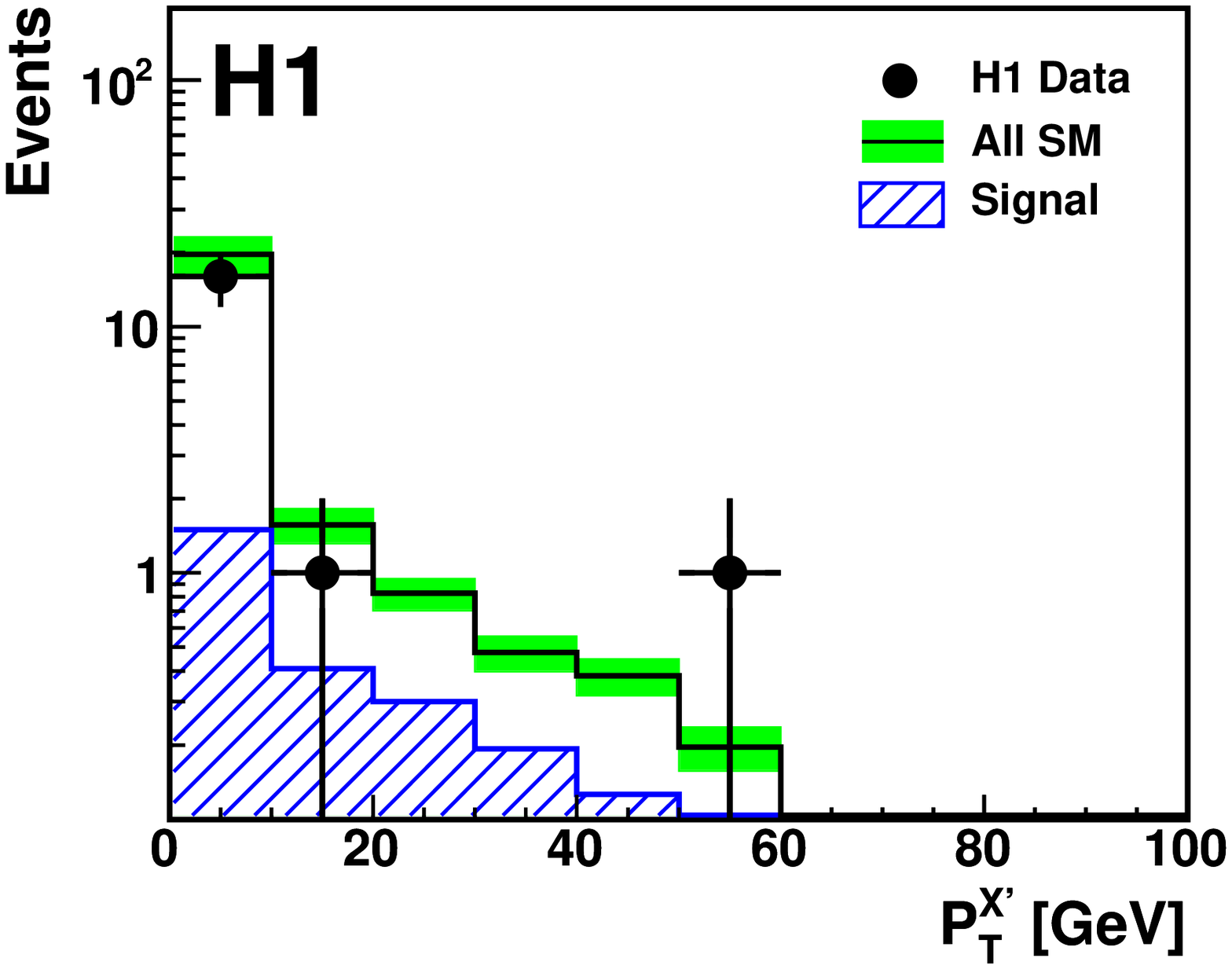}     
    \includegraphics[width=0.49\textwidth]{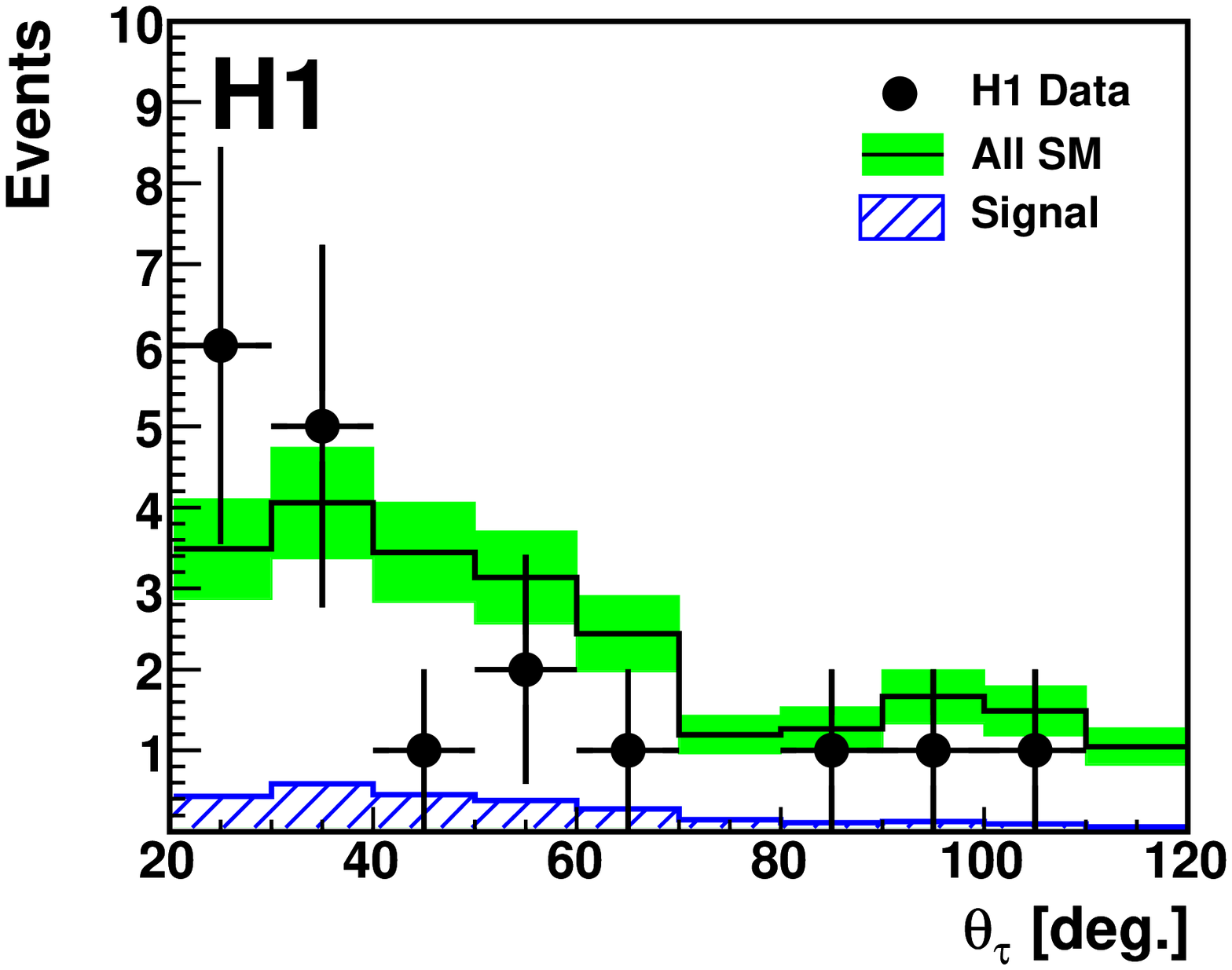}     
    \includegraphics[width=0.49\textwidth]{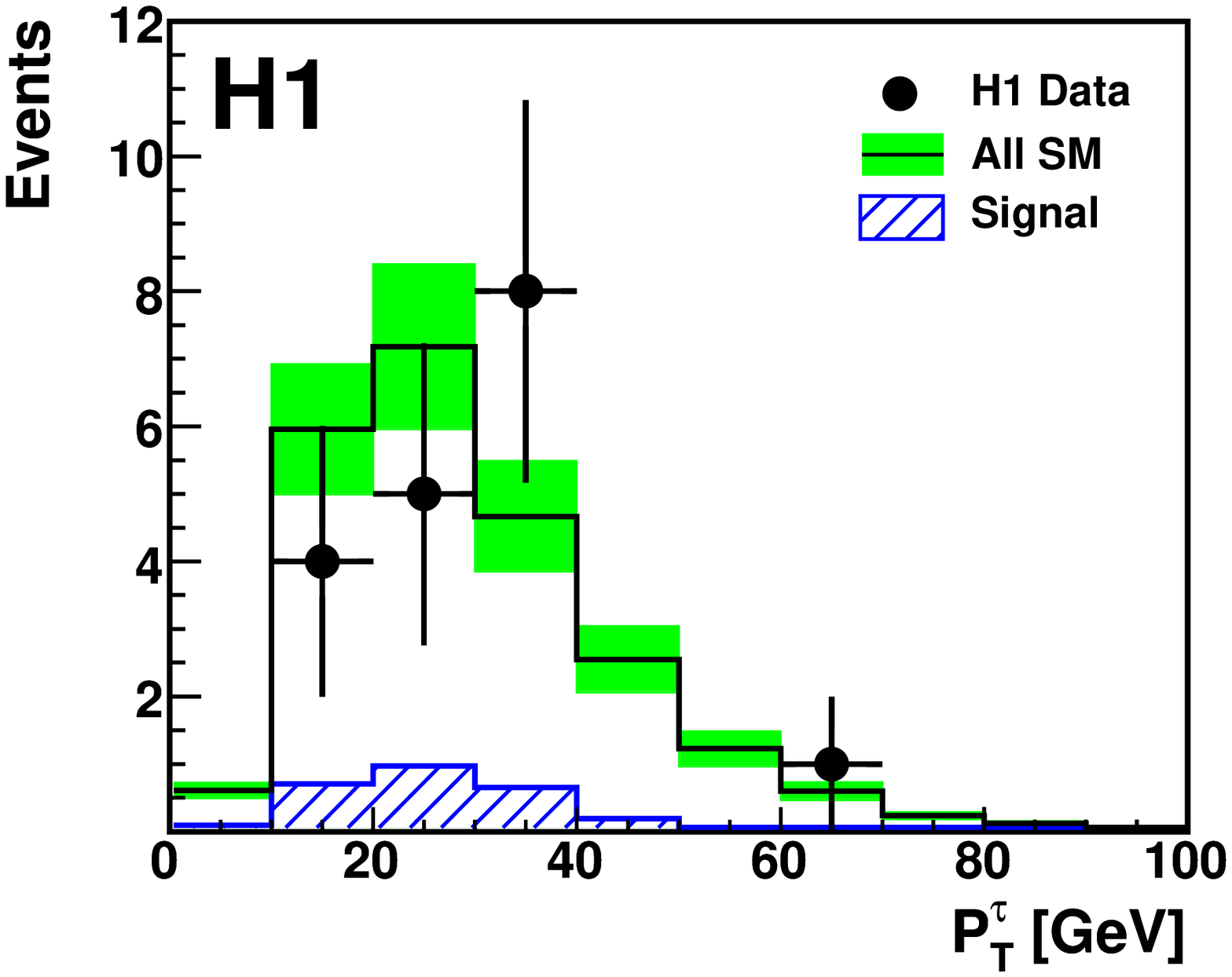}     
  \end{center}
  \begin{picture} (0.,0.)
    \setlength{\unitlength}{1.0cm}
    \put ( 6.5,9.5){(a)} 
    \put (14.4,9.5){(b)} 
    \put ( 6.5,3.7){(c)} 
    \put (14.4,3.7){(d)}  
  \end{picture} 
  \vspace{-0.5cm}

  \caption{Distributions in the tau channel for the $e^{\pm}p$ data
  sample. Shown is the missing transverse momentum
  $P_T^{\mathrm{miss}}$~(a), the hadronic transverse momentum not
  including the tau--jet candidate $P_T^{X'}$~(b), the polar angle of
  the tau--jet candidate $\theta_{\tau}$~(c) and the tau--jet
  candidate transverse momentum $P_{T}^{\tau}$~(d). The data (points)
  are compared to the SM expectation (open histogram). The signal
  component of the SM expectation is shown as the hatched histogram.
  The total uncertainty on the SM expectation is shown as the shaded
  band.}

  \label{fig:isotaufinalsample}
\end{figure}

The results of the search in the tau channel are summarised in 
table~\ref{tab:isotausummary}.
In the final event sample, $18$ events are selected, compared to a SM
expectation of $23.2 \pm 3.8$.
The SM expectation is dominated by charged current background
processes, whereas the signal contribution is only $11\%$.
Distributions of the events in the final sample are shown in
figure~\ref{fig:isotaufinalsample}.
Most of the events are observed at very low $P_{T}^{X'}$.
At $P_{T}^{X'}>25$ GeV one event is observed in the data, compared to
a SM expectation of $1.5\pm 0.2$.
In this region the contribution of single $W$ boson production to the
SM expectation is about $38\%$.
The data event, shown in figure~\ref{fig:eventdisplays}~(c), is
selected in $e^{-}p$ collisions and exhibits
$P_{T}^{\tau}=14.3\pm1.2$~GeV, $P_{T}^{X'}=62\pm5$~GeV and $P_{T}^{\rm
miss}=68\pm6$~GeV.
Of the six events observed in the previous
analysis~\cite{Aktas:2006fc}, three are retained in this selection,
whereas the other three are rejected by the tighter track isolation
described in section~\ref{sec:isotausel}.

%%%

The low efficiency to detect tau--like jets of about~$16\%$, together
with the branching ratio to hadronic one--prong decays of~$49\%$ and
the restricted polar angular range, leads to a significantly lower SM
expectation of $W\rightarrow\tau\nu_{\tau}$ events of only $2.7 \pm
0.4$ events, compared to $30.3 \pm 4.8$ and $10.1 \pm 1.7$ in the
electron and muon channels, respectively.
A comparison and cross check of the electron and muon channel is
impeded by this low efficiency and the high background contamination,
and therefore the tau channel is not included in the cross section
measurements described below.

%%%%%%%%%%%%%%%%%%%%%%%%%%%%%%%%%%%%%%%%%%%%%%%%%%%%%%%%%%%%%
\subsection{Cross Section Measurements}
\label{sec:wxs}
%%%%%%%%%%%%%%%%%%%%%%%%%%%%%%%%%%%%%%%%%%%%%%%%%%%%%%%%%%%%%

A measurement of the visible cross section for the isolated lepton and
missing transverse momentum topology in $ep$ collisions is performed
using the electron and muon channels in the phase space
$5^{\circ}<\theta_{\ell}<140^{\circ}$, $P_{T}^{\ell}>10$~GeV,
$P_{T}^{\rm miss}>12$~GeV and $D(\rm{\ell;jet})>1.0$ at a centre of
mass energy\footnote{Assuming a linear dependence of the cross section
on the proton beam energy.} of $\sqrt{s}=317$~GeV.
The cross section is defined as:

\begin{equation}
\sigma = \frac{N_{\rm obs} - N^{\rm MC}_{\rm bg}}{\mathcal{L}\mathcal{A}},
\label{eq:formulaxsec}
\end{equation}

\noindent
where $N_{\rm obs}$ is the number of observed data events in the complete
$e^{\pm}p$ data set and $N^{\rm MC}_{\rm bg}$ is the MC estimate of the number
of SM background events, due to processes discussed in
section~\ref{ch:sm}.
The total data luminosity is denoted by $\mathcal{L}$.
The acceptance $\mathcal{A}$ is defined as $N^{\rm MC}_{\rm
rec}/N^{\rm MC}_{\rm gen}$, the ratio of reconstructed and generated
signal events, and accounts for detection efficiencies and migrations.
The EPVEC generator is used to calculate the acceptance, which is
predicted to be about the same for $e^{+}p$ and $e^{-}p$ collisions.

%%%

The cross section is measured in several regions of $P_{T}^{X}$ in
both lepton channels, as displayed in
table~\ref{tab:isolepcrosssections}.
For $P_{T}^{X}<12$ GeV, the cross section in the muon channel is
estimated using the electron channel measurement~\cite{deBoer:2007}.
The measured total visible cross section for events with an isolated
lepton and missing transverse momentum is:
\[
\xsisolep = 0.23 \pm 0.05~({\rm stat.}) \pm 0.04~({\rm sys.})~{\rm pb},
\]
\noindent
where the first error is statistical and the second systematic.
The total cross section is in agreement with the SM NLO value of
$0.25 \pm 0.04$~pb.
The results are also found to be in agreement with previously
published numbers~\cite{Andreev:2003pm}.
The measurements in the electron and muon channels are consistent with
each other.

%%%%

The systematic uncertainty includes all of the sources described in
section~\ref{ch:systematics}.
In addition, a $10\%$ model uncertainty is applied to the acceptance,
estimated by comparing two further generators that produce $W$ bosons
with different kinematic distributions from those in EPVEC, namely an
implementation of $W$ boson production within PYTHIA and a flavour
changing neutral current (FCNC) single top generator
ANOTOP~\cite{Aktas:2003yd}.

%%%

\begin{figure}[]
  \begin{center}
    \includegraphics[width=.65\textwidth]{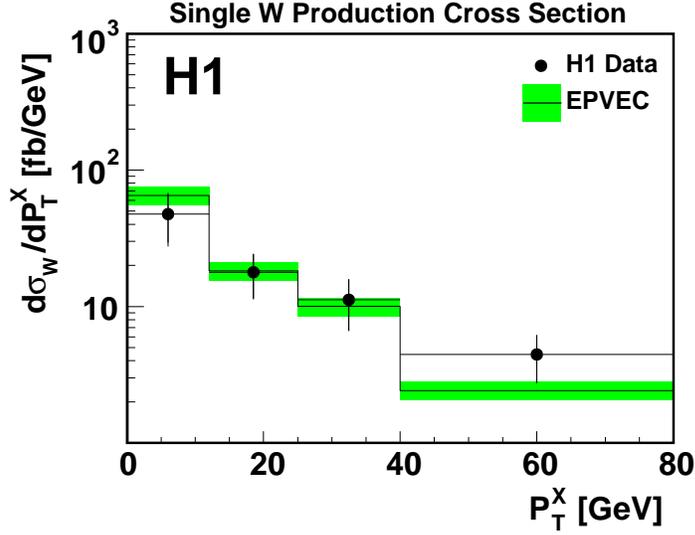}
  \end{center}
  \vspace{-0.5cm}
  \caption{The measured differential single $W$ boson production cross
  section d$\sigma_{W}$/d$P_{T}^{X}$ as a function of the hadronic
  transverse momentum $P_{T}^{X}$ (points). The errors denote the sum
  of the statistical and systematic uncertainties in quadrature. The
  measurement is compared to the EPVEC prediction (open histogram),
  including the theoretical uncertainty of $15\%$ shown as the shaded
  band.}
\label{fig:diffxsec}
\end{figure}

The total single $W$ boson production cross section is also calculated
using equation~\ref{eq:formulaxsec}, where $\mathcal{A}$ is now
defined with respect to the full phase space and the contribution from
$Z$ production illustrated in figure~\ref{fig:feynman}~(d) is
considered as background.
The branching ratio corresponding to the leptonic $W$ boson decay to
any final state with an electron or muon, including the contribution
from leptonic tau--decay~\cite{Yao:2006px}, is now also included in
the calculation.
The total single $W$ boson production cross section at HERA is
measured as:
\[
\xsw= 1.14 \pm 0.25~({\rm stat.}) \pm 0.14~({\rm sys.})~{\rm pb,}
\]
\noindent
in good agreement with the SM expectation of $1.27 \pm 0.19$~pb.
This result is also in good agreement with the ZEUS measurement
presented in~\cite{Chekanov:2008new}.
The differential single $W$ boson production cross section is also
measured as a function of $P_{T}^{X}$, the results of which are
displayed in figure~\ref{fig:diffxsec} and table~\ref{tab:diffxsec},
and is in agreement with the SM prediction.

%%%%%%%%%%%%%%%%%%%%%%%%%%%%%%%%%%%%%%%%%%%%%%%%%%%%%%%%%%%%%
\subsection{Measurement of the {\boldmath $WW\gamma$} Couplings}
\label{sec:wgammagamma}
%%%%%%%%%%%%%%%%%%%%%%%%%%%%%%%%%%%%%%%%%%%%%%%%%%%%%%%%%%%%%

The production of single $W$ bosons at HERA is sensitive to anomalous
triple gauge boson couplings~\cite{Baur:1989gh}, via the process
illustrated in figure~\ref{fig:feynman}~(b).
Due to the large mass of the $Z$ boson, no sensitivity is expected to
the $WWZ$ coupling.
Under the assumption of charge and parity conservation, and Lorentz
and electromagnetic gauge invariance, the $WW\gamma$ vertex can be
parametrised~\cite{Hagiwara:1986vm} using two free coupling
parameters, $\kappa$ and $\lambda$, related to the magnetic dipole
moment $\mu_{W}=e/2M_{W}\left(1+\kappa+\lambda\right)$ and the
electric quadrupole moment
$Q_{W}=-e/M_{W}^{2}\left(\kappa-\lambda\right)$ of the $W$ boson.
In the SM, $\kappa=1$ and $\lambda=0$ at tree level.
In this paper, limits are set on the coupling parameters using the
method of maximum likelihood in a Bayesian approach employing Poisson
statistics.
Instead of $\kappa$, $\Delta\kappa \equiv \kappa-1$ is used, such that
any non--zero value for $\Delta\kappa$ or $\lambda$ represents a
deviation from the SM.

%%%

Since the $P_{T}^{X}$ spectrum is expected to be sensitive to
anomalous values of $\Delta\kappa$ and $\lambda$ \cite{Baur:1989gh},
the likelihood analysis is performed in four bins with lower edges in
$P_{T}^{X}$ of $12,25,40$ and $80$~GeV.
The last bin extends until $120$~GeV and contains no data events, but
for anomalous values of $\lambda$, the expectation in this region
becomes significant and is used to further constrain $\lambda$.
Events with $P_{T}^{X}<12$~GeV are not used since the main SM $W$
boson production diagram, shown in figure~\ref{fig:feynman}~(a),
dominates this region of phase space and no sensitivity to the
$WW\gamma$ vertex parameters is gained from such events.

%%%

\begin{figure}[]
  \begin{center}
   \includegraphics[width=0.49\textwidth]{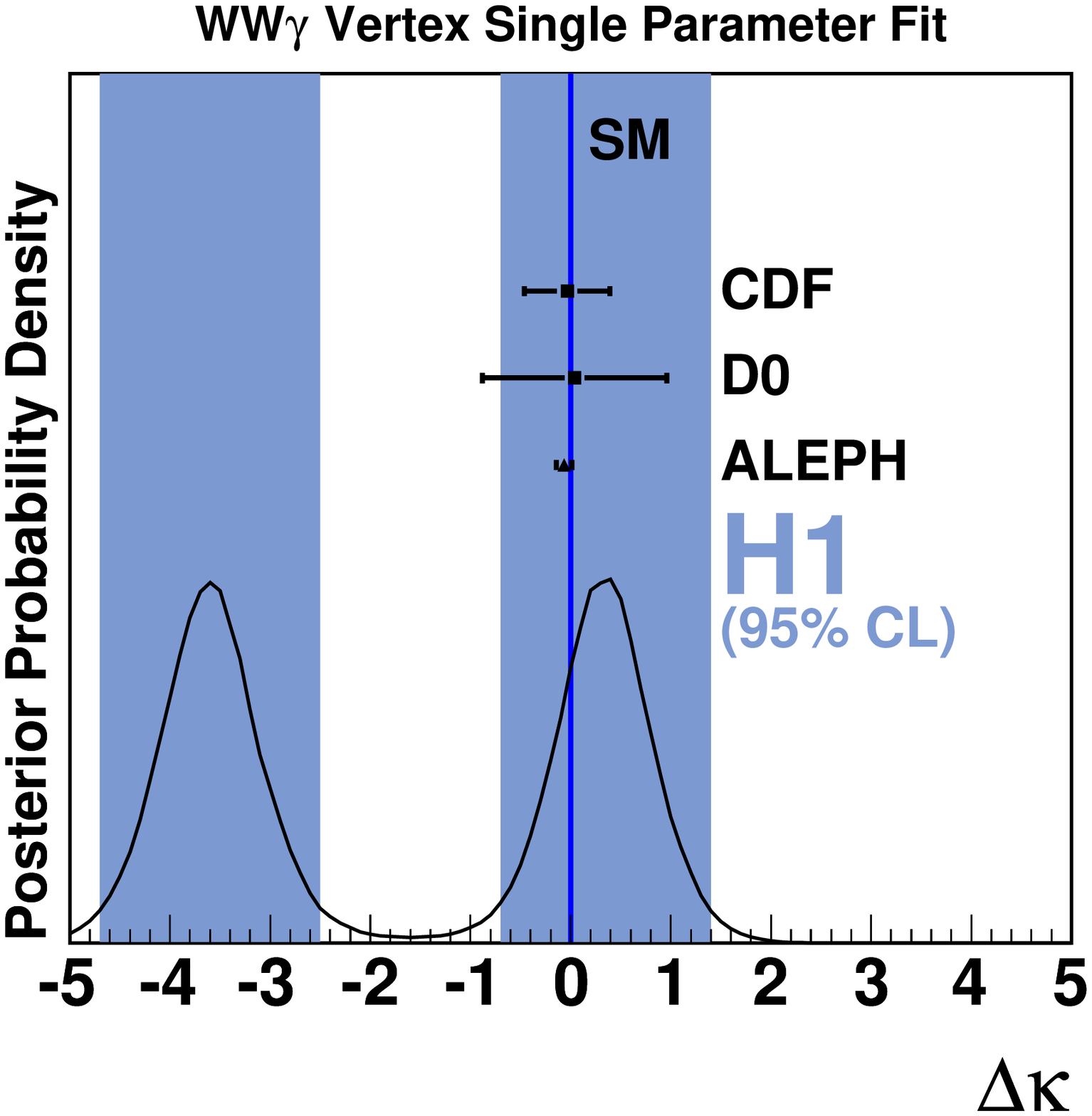}
   \includegraphics[width=0.49\textwidth]{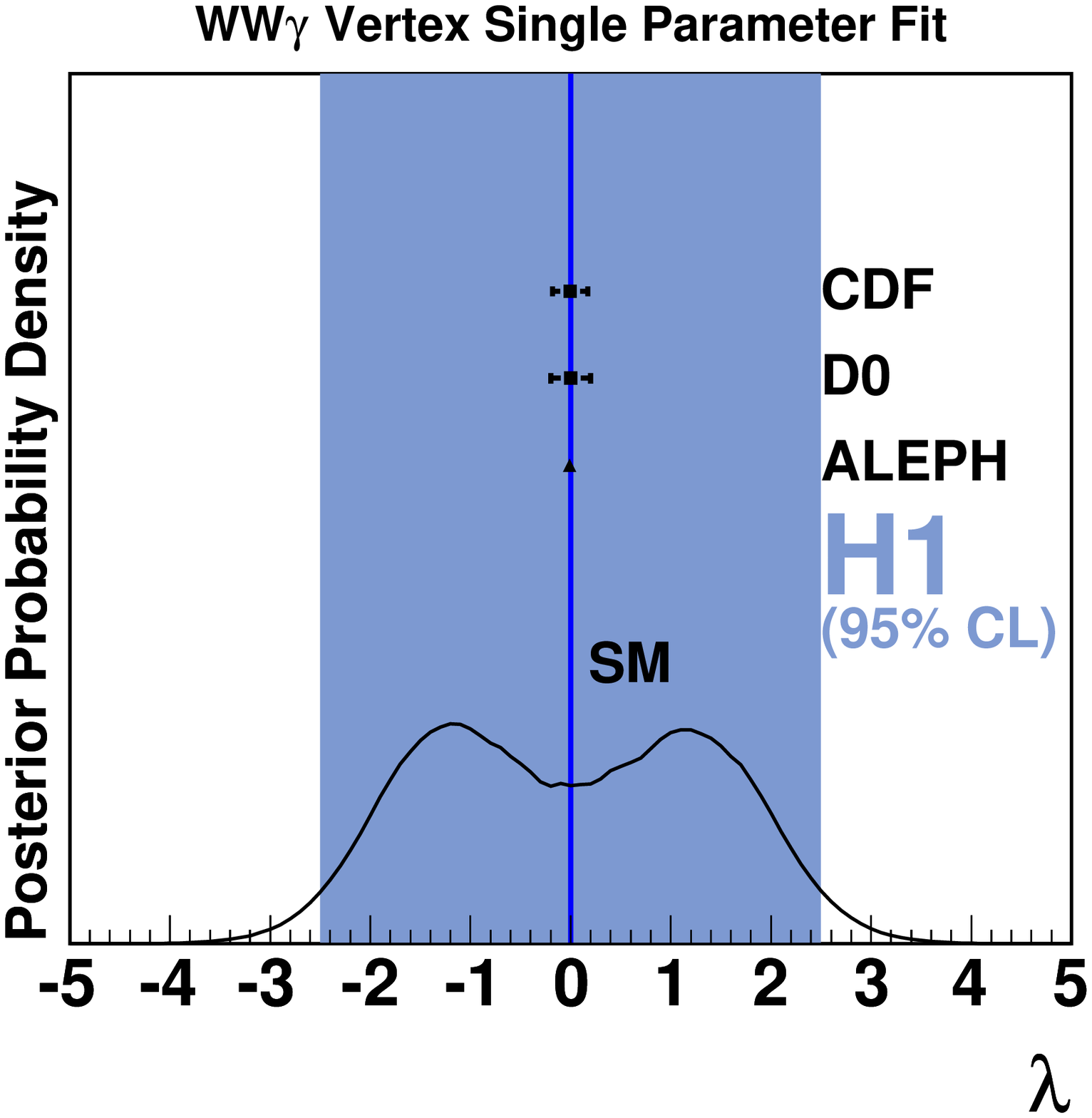}
  \end{center}
  \begin{picture} (0.,0.)
    \setlength{\unitlength}{1.0cm}
    \put ( 6.5,3.0){(a)} 
    \put (14.4,3.0){(b)} 
  \end{picture}
  \vspace{-0.5cm}
  \caption{Probability distributions (in arbitrary units) of the
  single parameter fits to (a)~$\Delta\kappa$ and (b)~$\lambda$. The
  obtained $95\%$~CL limits are shown (shaded areas) in comparison
  with those of D{\O}~\cite{Abazov:2005ni}, CDF~\cite{Aaltonen:2007sd},
  and ALEPH~\cite{Schael:2004tq}. The SM expectation values are also
  indicated.}
 \label{fig:likelihoods}
\end{figure}

The maximum likelihood analysis is performed for each bin in
$P_{T}^{X}$ and the resulting probability distributions are
multiplied.
This is done for $\Delta\kappa$ and $\lambda$ separately, keeping the
other parameter fixed to its SM value.
The resulting probability distributions are shown in
figure~\ref{fig:likelihoods}.
The observed double--peak structure in the probability densities
arises from the quadratic dependence of the cross section to the
coupling parameters.
The following limits are derived at $95\%$ confidence level (CL):

\[
-4.7 < \Delta\kappa < -2.5 {\rm \qquad or \qquad} -0.7 < \Delta\kappa < 1.4,
\]
\[
-2.5 < \lambda < 2.5.
\]

The dominant source of error is the theoretical uncertainty on the
single $W$ boson production cross section.
The limits obtained are in good agreement with the SM prediction.
Since the value of $\Delta\kappa=-1$ is excluded at $95\%$~CL, the
results obtained explicitly demonstrate the presence of a magnetic
coupling of the photon to the $W$ boson, in addition to the coupling
to the electric charge of the $W$ boson.
Given the limits obtained for $\lambda$, the first allowed region
($-4.7~<~\Delta\kappa~<~-2.5$) is excluded by limits derived from loop
contributions to
$\left(g-2\right)_{\mu}$~\cite{Herzog:1984nxWallet:1985mxGrifols:1987dx,Suzuki:1985yh,VanDerBij:1987pk}.
The limits on $\Delta\kappa$, enclosing the SM prediction
$\Delta\kappa=0$, are compatible with those obtained at the
Tevatron~\cite{Abazov:2005ni,Aaltonen:2007sd}.
The most stringent limits were obtained at LEP in single $\gamma$,
single $W$ and $W$ pair
production~\cite{Schael:2004tq,Abdallah:2008sf,Achard:2004ji,Abbiendi:2003mk}.

%%%%%%%%%%%%%%%%%%%%%%%%%%%%%%%%%%%%%%%%%%%%%%%%%%%%%%%%%%%%%
\subsection{Measurement of the {\boldmath $W$} Boson Polarisation Fractions}
\label{sec:wpol}
%%%%%%%%%%%%%%%%%%%%%%%%%%%%%%%%%%%%%%%%%%%%%%%%%%%%%%%%%%%%%

A measurement of the $W$ boson polarisation is performed to further
test the compatibility of the observed results with SM single $W$
production.
For $W$ bosons from the decay of single top quarks, the production of
which is kinematically possible at HERA, the polarisation is expected
to be different from that in the SM.
The \cosths\ distribution in the decay $W\rightarrow e/\mu+\nu$ is
exploited, where $\theta^{*}$ is defined as the angle between the $W$
boson momentum in the lab frame and that of the charged decay lepton
in the $W$ boson rest frame.
The influence of higher order corrections on the \cosths\ distribution
is expected to be small and is neglected.
For the left handed polarisation fraction $F_{-}$, the longitudinal
fraction $F_{0}$ and the right handed fraction $F_{+} \equiv 1 - F_{-}
- F_{0}$, the \cosths\ distributions for $W^{+}$ bosons are
given~\cite{Hagiwara:1986vm} by:

\begin{eqnarray}
\frac{1}{\sigma_{W\rightarrow \ell+\nu }}
\frac{\rm{d} \sigma_{W\rightarrow \ell+\nu }}{\rm{d}cos\,\theta^{*}} =
\frac{3}{4} F_{0} \left( 1 - {\rm cos}^{2}\theta^{*}\right) +  
\frac{3}{8} F_{-} \left( 1 - {\rm cos}\,\theta^{*}\right)^{2} +
\frac{3}{8} F_{+} \left( 1 + {\rm cos}\,\theta^{*}\right)^{2} .
\label{eq:polmodel}
\end{eqnarray}

\noindent
To allow the datasets of both $W$ boson charges to be combined,
\cosths\ is multiplied by the sign of the lepton charge
$q_{\ell}$.
Events in the electron or muon channel originating from W boson to tau
decay, $W\rightarrow\tau\left(\rightarrow
e/\mu+\nu_{e/\mu}\right)+\nu_{\tau}$, are considered background, since
for these events the \cosths\ distributions are not expected to be
described by equation~\ref{eq:polmodel}.

%%%

Events are selected for this measurement starting from the sample
described in section~\ref{sec:isolepsel}.
The $W$ boson four--vector is reconstructed as the sum of the neutrino
and lepton four--vectors using a kinematic constraint to the $W$ mass
if the scattered electron is not observed~\cite{deBoer:2007}.
To ensure a reliable charge measurement only leptons in the central
region $\theta_{\ell} > 20^{\circ}$, and with charge significance
$\sigma_{q} > 1$ are considered.
In the electron channel, the associated track transverse momentum
$P_{T}^{\textrm{track}}$ is required to match the calorimetric
measurement $P_{T}^{e}$ under the condition $1/P_{T}^{\textrm{track}}
- 1/P_{T}^{e} < 0.04$.
Using these requirements the charge misidentification probability is
found to be below $1\%$.
The final sample consists of $21$ electron events and $9$ muon events.
The SM prediction is dominated by single $W$ production, which
contributes $76\%$ to the total expectation.

%%%

\begin{figure}[]
  \begin{center}
    \includegraphics[width=0.47\textwidth]{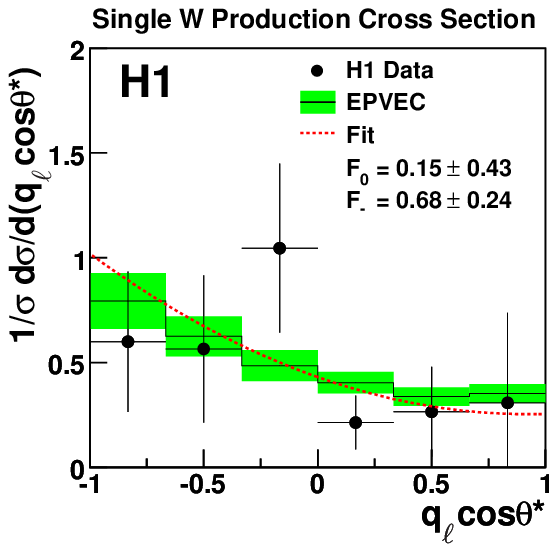}
    \hspace{0.5cm}
    \includegraphics[width=0.45\textwidth]{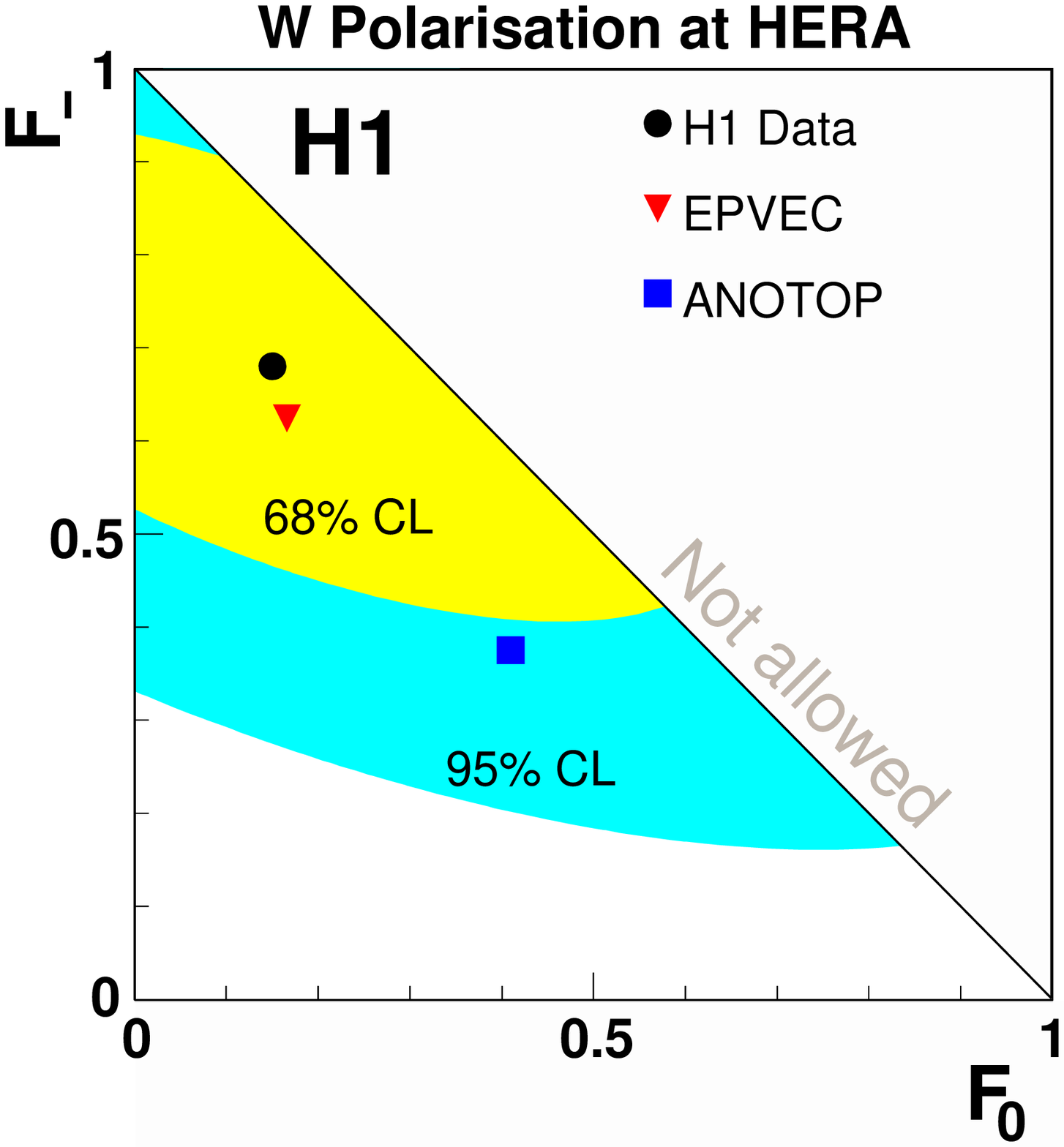}
  \end{center}
  \begin{picture} (0.,0.)
    \setlength{\unitlength}{1.0cm}
    \put ( 6.5,4.5){(a)} 
    \put (14.2,4.5){(b)} 
  \end{picture} 
  \vspace{-0.5cm}
  \caption{(a)~The measured normalised differential cross section
  $1/\sigma$~$d\sigma/d\left(\qcosths\right)$ (points) as a function of
  $\qcosths$ for on--shell $W$ bosons.  The EPVEC prediction is also
  shown (open histogram) with a $15\%$ theoretical uncertainty shown
  by the band. The result of the simultaneous fit of the $W$
  polarisation fractions is shown as the dashed histogram. (b)~The
  plane showing the fit result for the simultaneously extracted left
  handed ($F_{-}$) and longitudinal ($F_{0}$) $W$ boson polarisation
  fractions (point) with the corresponding $68\%$ and $95\%$ CL
  contours. Also shown are the values of predictions from EPVEC
  (triangle) and ANOTOP (square).}
  \label{fig:wpol}
\end{figure}

The \cosths\ distributions of $W$ bosons produced off--shell are not
expected to be described by the polarisation model given in
equation~\ref{eq:polmodel}.
A correction is therefore applied to the measured \qcosths\
distribution~\cite{deBoer:2007}, where the contribution from such
production diagrams is estimated using EPVEC.
The measured \qcosths\ distribution is also corrected for acceptance
and detector effects.
The resulting normalised differential cross section is shown in
figure~\ref{fig:wpol}~(a), with the result of the fit to the
polarisation fractions described in the above equation.
In the fit, $F_{-}$ and $F_{0}$ are simultaneously determined, the
result of which is shown in figure~\ref{fig:wpol}~(b) and found to be
in good agreement with the SM.
To test the stability or possible biases of the result, the fit is
repeated on \cosths\ distributions derived using non-SM polarisation
fractions.
As a further cross check, a fit to the uncorrected \cosths\ distribution
using polarisation templates is performed on detector level.
All results are found to be consistent.

In addition, figure ~\ref{fig:wpol}~(b) shows the expectation obtained
from ANOTOP~\cite{Aktas:2003yd}, where all $W$ bosons are assumed to come
from top decays.
The data are also compatible with this process within the sensitivity
of the measurement, although it can be seen that the SM value is
favoured.

%%%

$F_{-}$ and $F_{0}$ are also extracted in fits where one parameter is
fixed to its SM value, the results of which are presented in
table~\ref{tab:wpol}.
The quoted systematic uncertainties are propagated from the cross
section calculations.
The systematic effects on the shape of the differential cross section
from uncertainties on the background estimates and charge
misidentification are also taken into account.
More details on the analysis can be found in~\cite{deBoer:2007}.

%%%%%%%%%%%%%%%%%%%%%%%%%%%%%%%%%%%%%%%%%%%%%%%%%%%%%%%%%%%%%
\section{Summary}
\label{ch:summary}
%%%%%%%%%%%%%%%%%%%%%%%%%%%%%%%%%%%%%%%%%%%%%%%%%%%%%%%%%%%%%

A search for events with isolated leptons and missing transverse
momentum is performed using the full HERA $e^{\pm}p$ data sample,
corresponding to an integrated luminosity of $474$~pb$^{-1}$.
All three lepton flavours are investigated.
With respect to the previous publication the integrated luminosity of
the data set is increased by about a factor of four, and the signal
efficiency and background suppression have been improved.

%%%

In the combined electron and muon channels $53$ events are observed in
the full $e^{\pm}p$ data sample compared to a SM prediction of
$54.1\pm7.4$, where $40.4\pm6.3$ are expected from signal processes,
dominated by single $W$ production.
Interesting events are observed at high $P_{T}^{X}>25$~GeV in the
$e^{+}p$ data sample, where $17$ events are observed compared to a SM
prediction of $8.0\pm1.3$, including $7.0\pm1.2$ expected from signal
processes.
This excess of data events over the SM prediction in the $e^{+}p$ data
corresponds to a $2.4\sigma$ significance, which is lower than previously
reported in~\cite{Andreev:2003pm}.
No such excess is observed in the $e^{-}p$ data.

%%%

In the search for tau leptons and missing transverse momentum, $18$
events are observed in agreement with the SM expectation of $23.2 \pm
3.8$, which is dominated by CC background.
The signal contribution is small at around $11\%$.
No excess of events is observed in the data at large hadronic
transverse momentum.

%%%

From the events observed in the electron and muon channels the single
$W$ production cross section has been measured as~$1.14 \pm 0.25~({\rm
stat.}) \pm 0.14~({\rm sys.})$~pb, compared to a SM expectation of
$1.27 \pm 0.19$~pb.
The measured cross section is used to derive single parameter limits
on the $WW\gamma$ gauge coupling parameters $\Delta\kappa$ and
$\lambda$ at $95\%$ CL.
The $W$ polarisation fractions are measured for the first time at HERA
and found to be consistent with the SM.

%%%%%%%%%%%%%%%%%%%%%%%%%%%%%%%%%%%%%%%%%%%%%%%%%%%%%%%%%%%%%
\section*{Acknowledgements}
%%%%%%%%%%%%%%%%%%%%%%%%%%%%%%%%%%%%%%%%%%%%%%%%%%%%%%%%%%%%%

We are grateful to the HERA machine group whose outstanding efforts
have made this experiment possible. We thank the engineers and
technicians for their work in constructing and maintaining the H1
detector, our funding agencies for financial support, the DESY
technical staff for continual assistance and the DESY directorate for
support and for the hospitality which they extend to the non--DESY
members of the collaboration.

%%%%%%%%%%%%%%%%%%%%%%%%%%%%%%%%%%%%%%%%%%%%%%%%%%%%%%%%%%%%%
\clearpage
%%%%%%%%%%%%%%%%%%%%%%%%%%%%%%%%%%%%%%%%%%%%%%%%%%%%%%%%%%%%%

\section*{Appendix}

%%%%%%%%%%%%%%%%%%%%%%%%%%%%%%%%%%%%%%%%%%%%%%%%%%%%%%%%%%%%%
\appendix
%%%%%%%%%%%%%%%%%%%%%%%%%%%%%%%%%%%%%%%%%%%%%%%%%%%%%%%%%%%%%
\section{Background Studies}
\label{ch:bgstudies}
%%%%%%%%%%%%%%%%%%%%%%%%%%%%%%%%%%%%%%%%%%%%%%%%%%%%%%%%%%%%%

The description of the SM background and the potential influence of
systematic effects arising from the event selection are investigated
using a set of control samples, where individual background processes
are statistically enriched.
In particular, the origin of features of the final sample, such as the
difference in event yield in the $e^{+}p$ and $e^{-}p$ data have been
examined.
The control samples are defined by removing cuts from the event
selections summarised in tables~\ref{tab:isolepcutstable} and
\ref{tab:isotaucutstable}.
The most conclusive background enriched samples are described in the
following.

%%%%%%%%%%%%%%%%%%%%%%%%%%%%%%%%%%%%%%%%%%%%%%%%%%%%%%%%%%%%%
\subsection{Electron Channel Control Samples}
\label{sec:isoeleccontrol}
%%%%%%%%%%%%%%%%%%%%%%%%%%%%%%%%%%%%%%%%%%%%%%%%%%%%%%%%%%%%%

The SM background in the electron channel is studied by enriching the
final sample with NC and CC events.

%%%

\begin{figure}[]
  \begin{center}  
    \textsf{Electron Channel, NC Enriched}\\
    \includegraphics[width=0.49\textwidth]{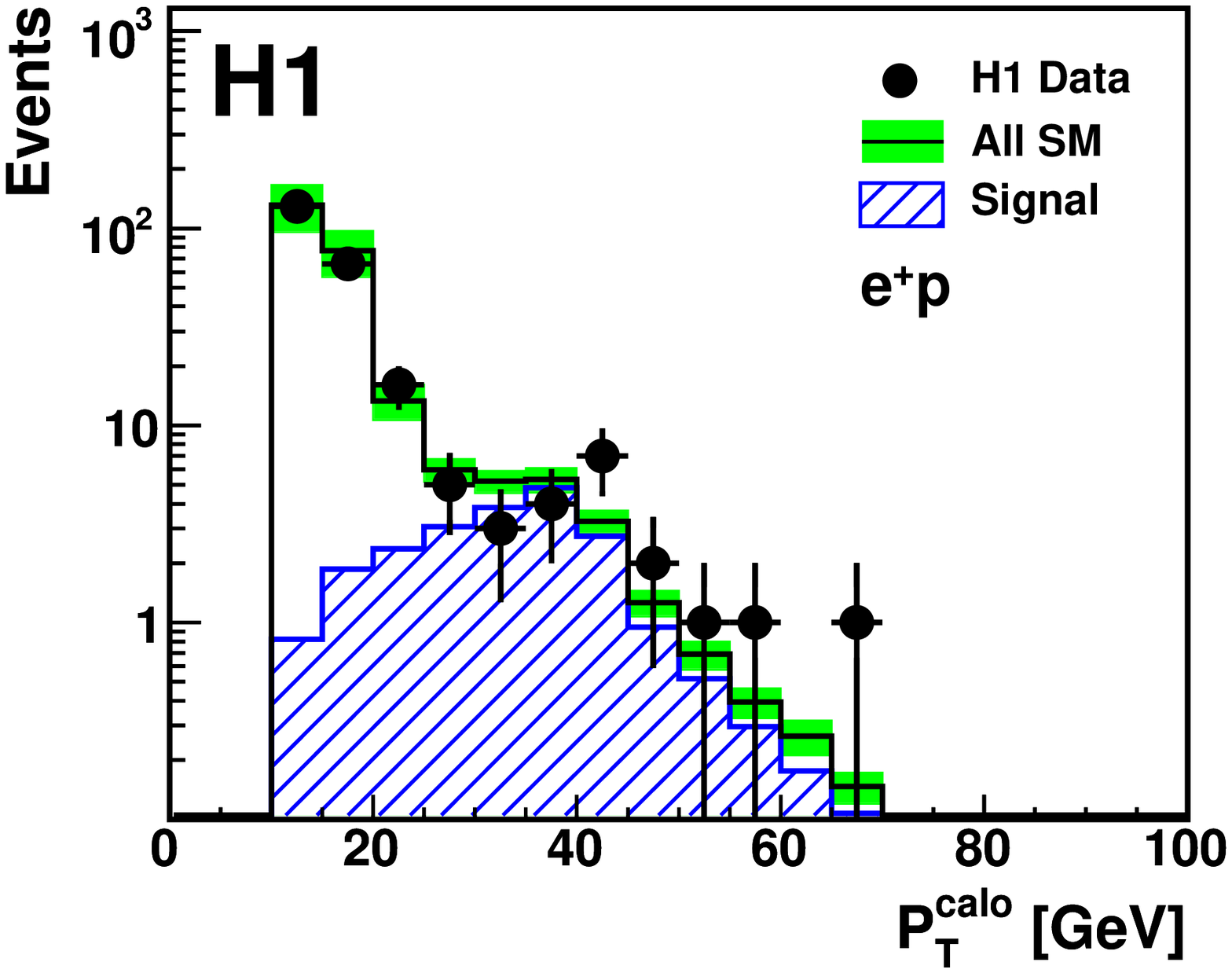}     
    \includegraphics[width=0.49\textwidth]{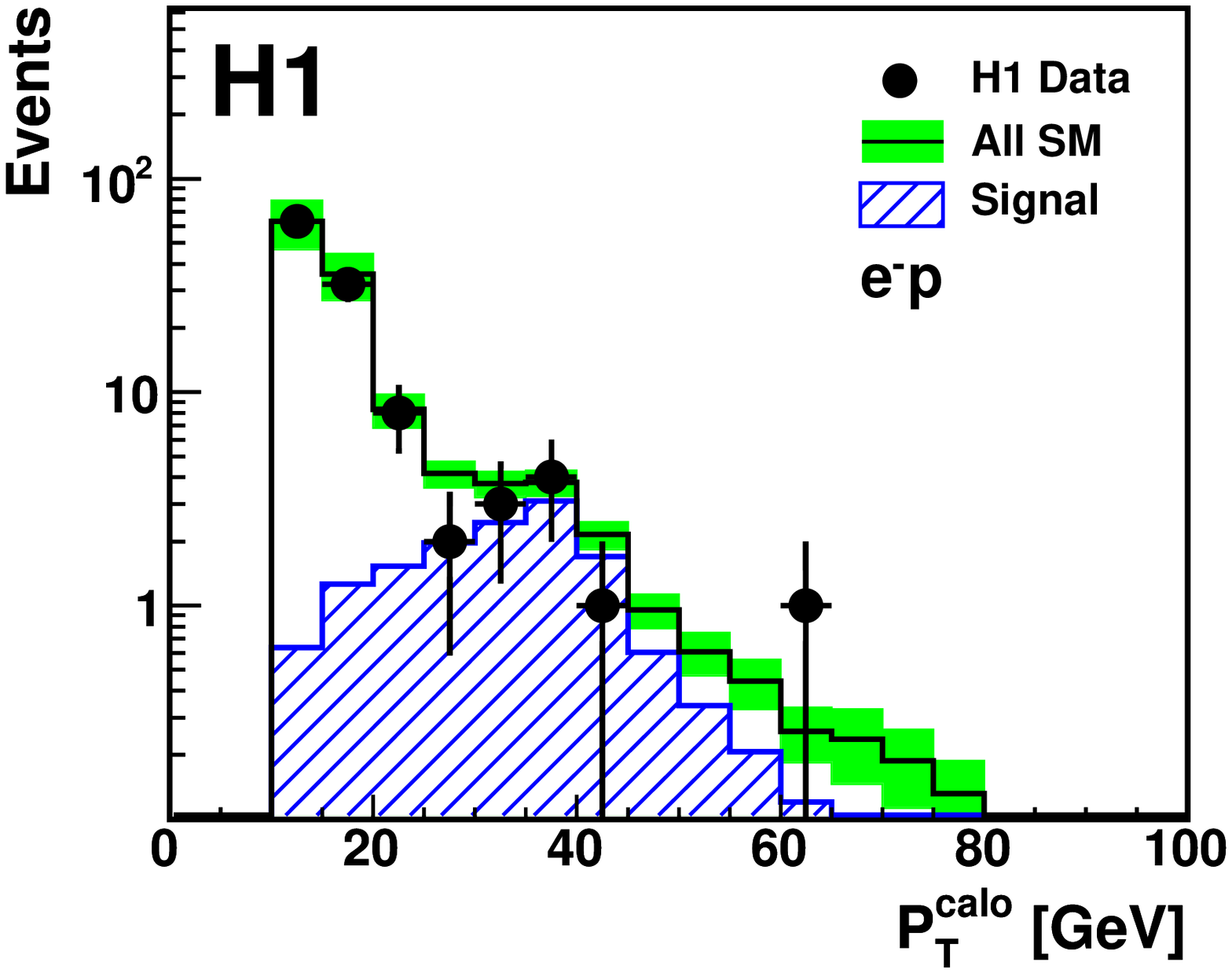}     
    \includegraphics[width=0.49\textwidth]{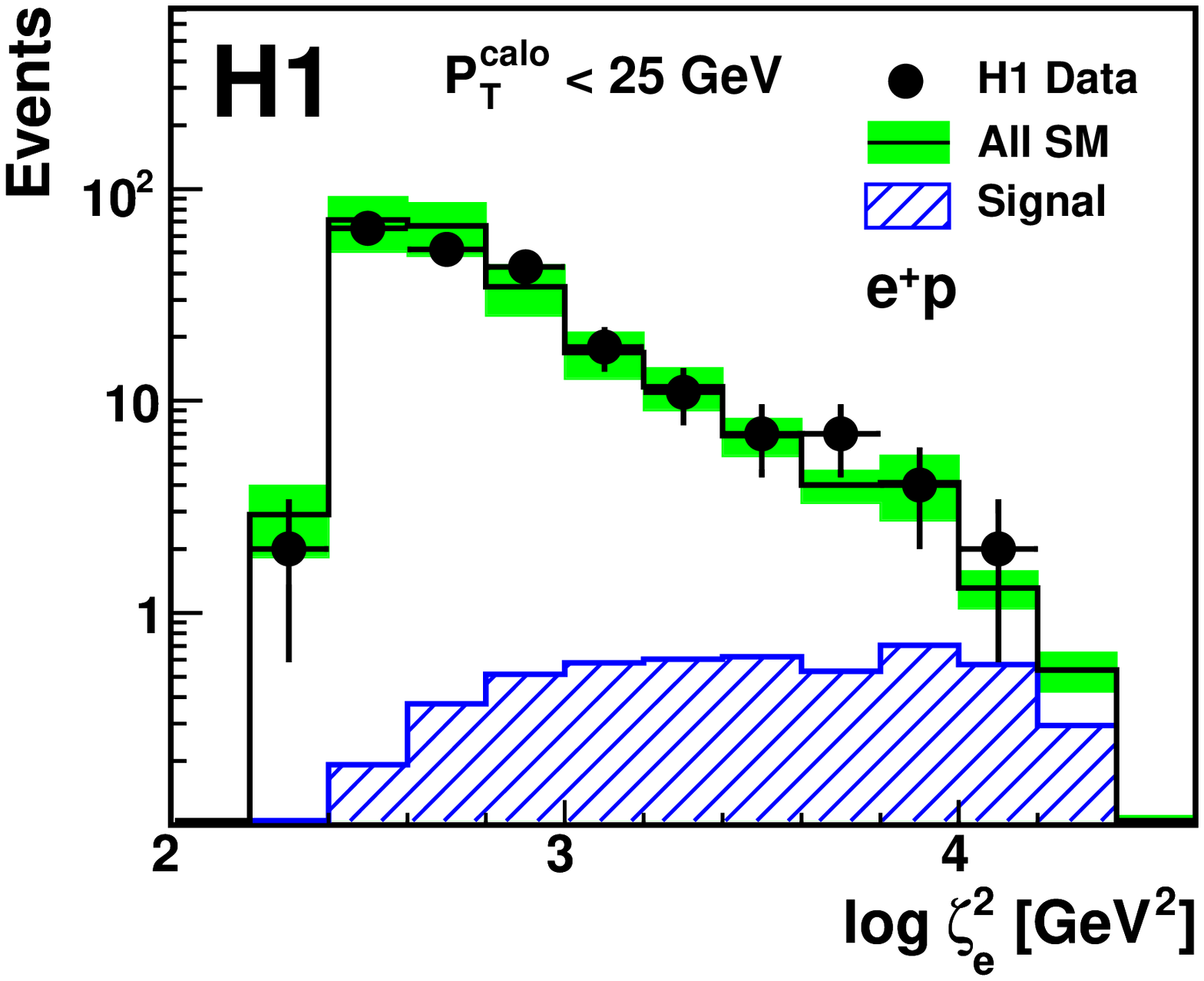}     
    \includegraphics[width=0.49\textwidth]{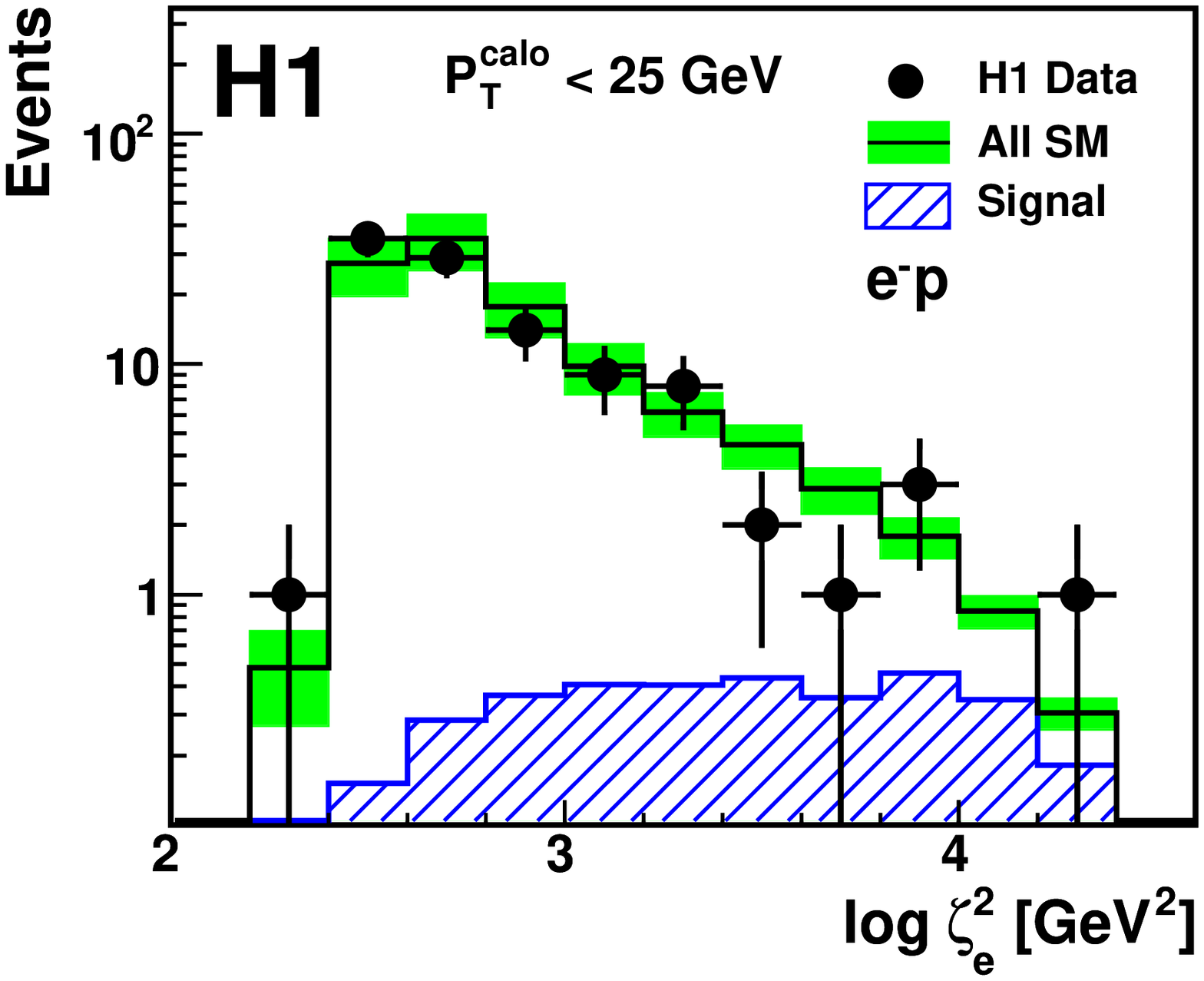}     
  \end{center}
  \begin{picture} (0.,0.)
    \setlength{\unitlength}{1.0cm}
    \put ( 6.5, 9.5){(a)} 
    \put (14.2, 9.5){(b)} 
    \put ( 6.5, 3.7){(c)} 
    \put (14.2, 3.7){(d)}    
  \end{picture} 
  \vspace{-0.5cm}
  \caption{Distributions of data (points) selected in the NC enriched
  sample in the electron channel. Shown are $P_{T}^{\rm calo}$ and
  $\zeta^{2}_{e}$, the latter for events with $P_{T}^{\rm
  calo}<25$~GeV in the $e^{+}p$~(a,~c) and $e^{-}p$~(b,~d) data.  The
  total uncertainty on the SM expectation (open histogram) is shown as
  the shaded band. The signal component is shown as the hatched
  histogram. The final sample populates the region $P_{T}^{\rm calo}>
  25$~GeV or $\zeta^{2}_{e}>5000$~GeV.}
  \label{fig:elecncsample}	
\end{figure}

NC background, concentrated at low $Q^{2}_{e}$, may enter the final
sample due to a mismeasurement of energy which gives rise to a large
value of $P_{T}^{\rm calo}$ and a small value of $V_{\rm ap}/V_{\rm
p}$.
The requirements on $\zeta^{2}_{e}$, $P_{T}^{\rm calo}$ and $V_{\rm
ap}/V_{\rm p}$ are designed to suppress NC that enters due to this
effect.
Figure~\ref{fig:elecncsample} shows a NC enriched electron sample that 
is formed by removing these requirements from the final selection.
The NC contribution to the SM expectation is about $95\%$ in this
sample.
Figures~\ref{fig:elecncsample}~(a) and (b) show the $P_{T}^{\rm calo}$
distribution, and (c) and (d) the $\zeta^{2}_{e}$ distribution under
the condition $P_{T}^{\rm calo}<25$~GeV, in the $e^{+}p$ data and in
the $e^{-}p$ data, respectively.
It can be seen that the part of the sample populated by the NC
background at low $P_{T}^{\rm calo}$ and low $\zeta^{2}_{e}$ is well
described by the simulation for both the $e^{+}p$ data and the
$e^{-}p$ data.
Events in the final sample appear at $P_{T}^{\rm calo}>25$~GeV or
$\zeta^{2}_{e}>5000$~GeV.

%%%

\begin{figure}[]
  \begin{center}
    \textsf{Electron Channel, CC Enriched}\\
    \includegraphics[width=0.49\textwidth]{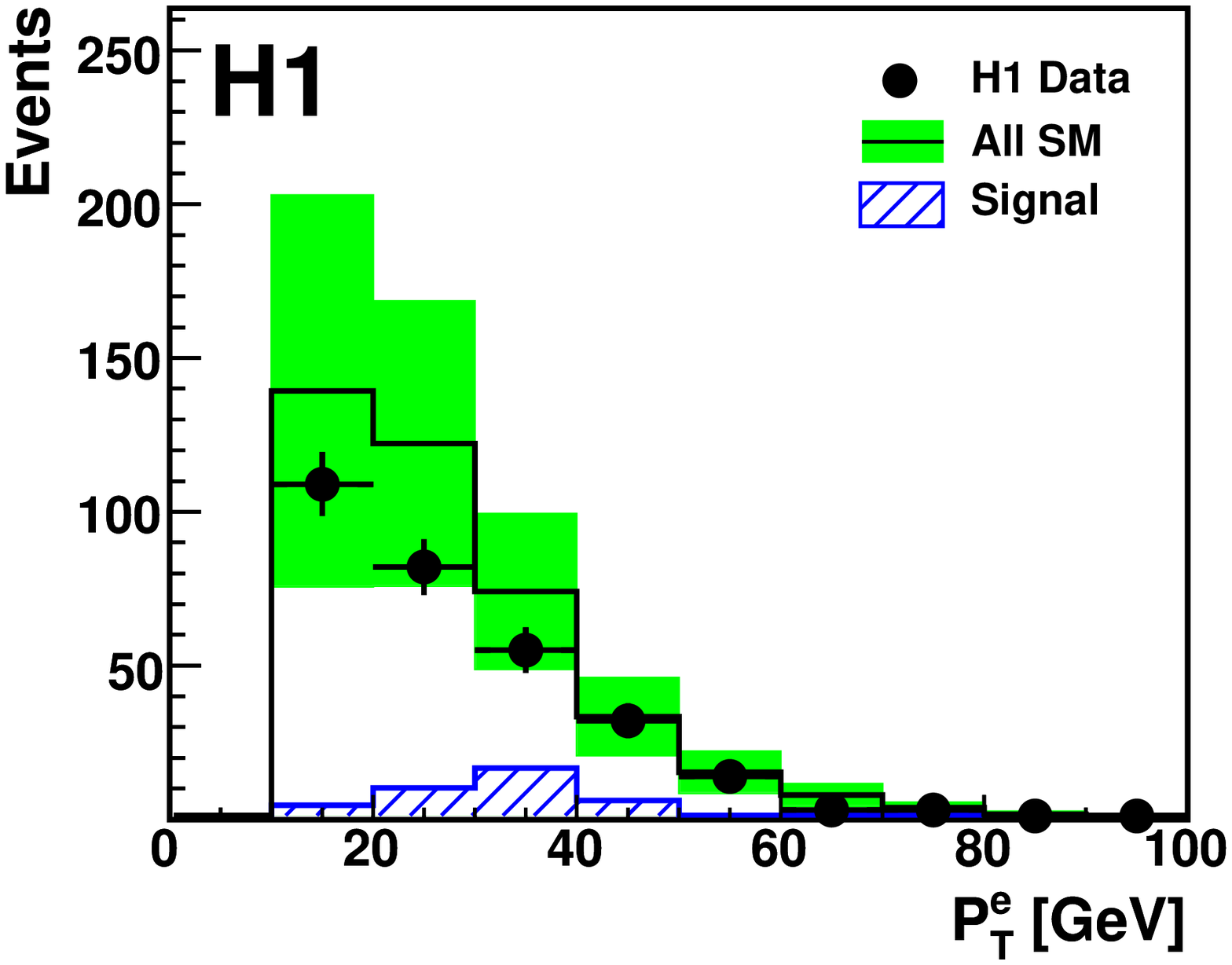}     
    \includegraphics[width=0.49\textwidth]{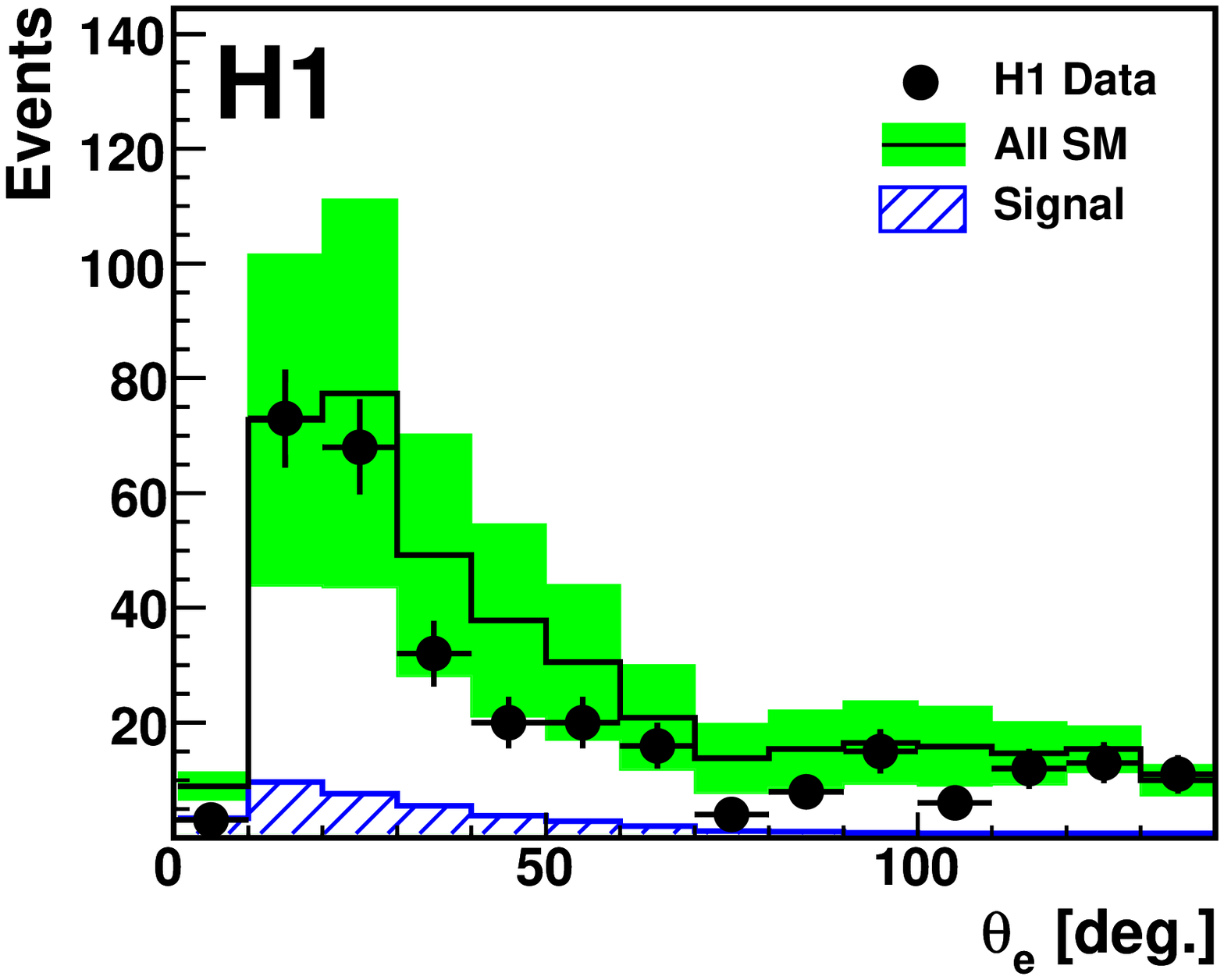}     
    \includegraphics[width=0.49\textwidth]{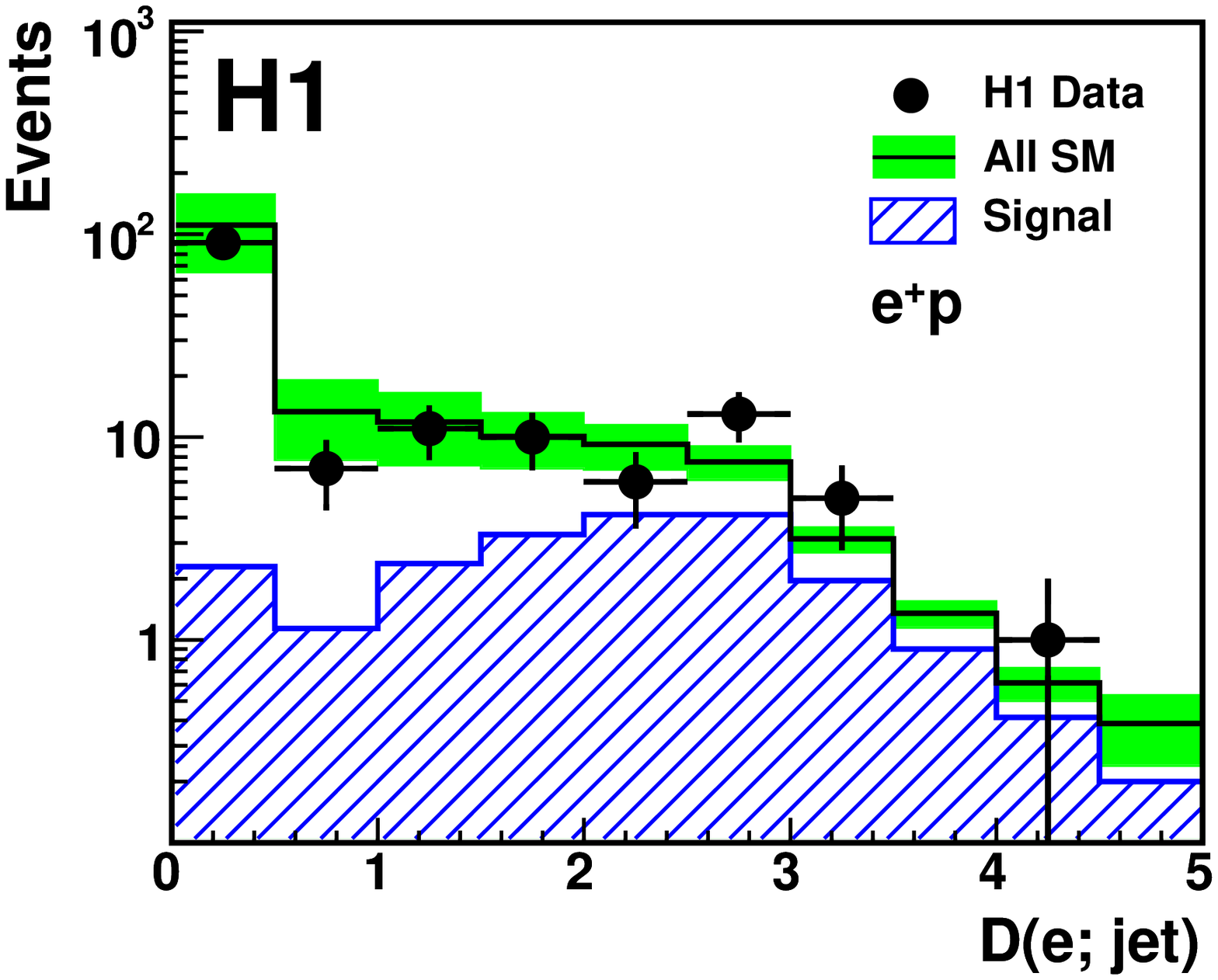}     
    \includegraphics[width=0.49\textwidth]{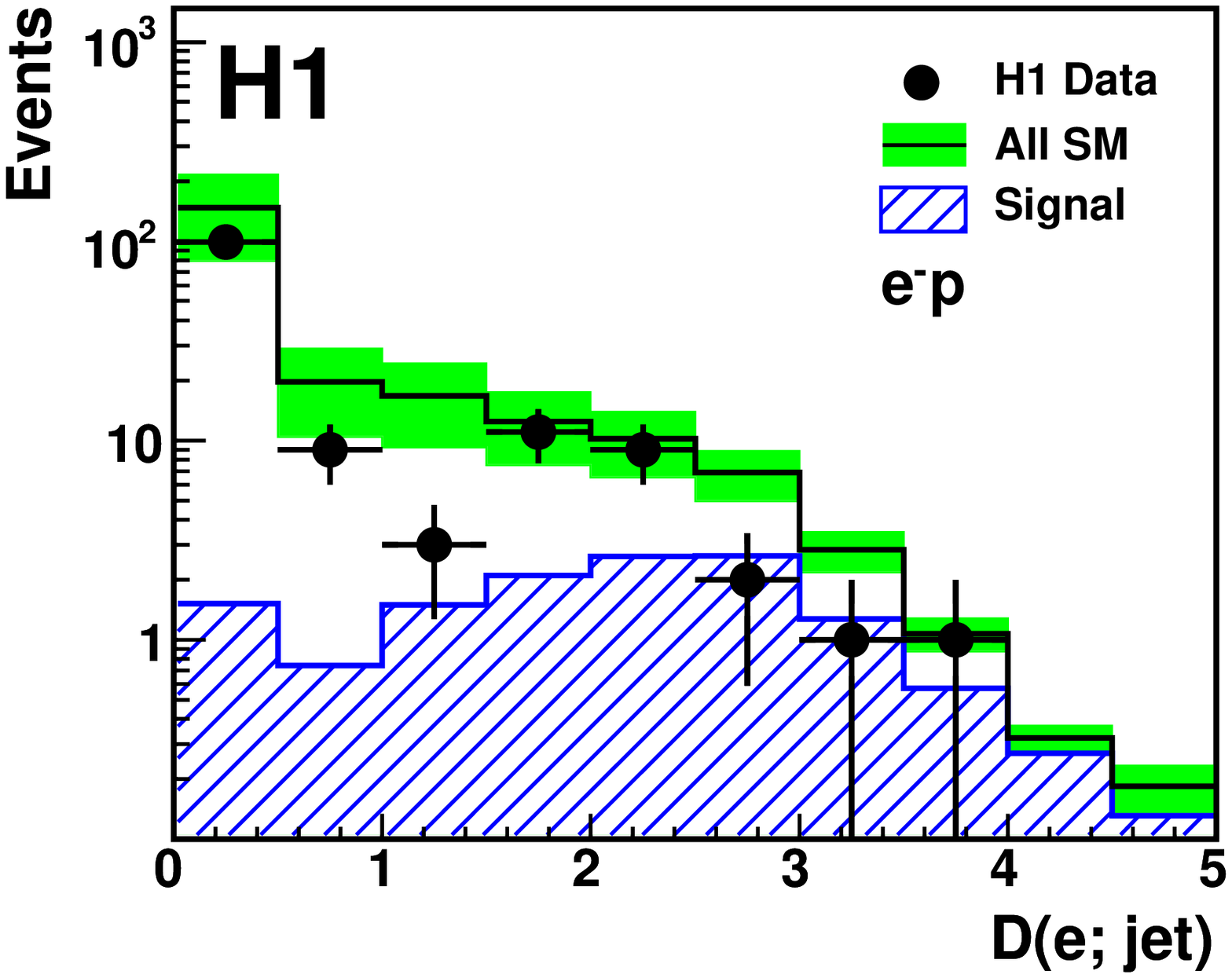}     
  \end{center}
  \begin{picture} (0.,0.)
    \setlength{\unitlength}{1.0cm}
    \put ( 6.5,  9.5){(a)} 
    \put (14.2,  9.5){(b)} 
    \put ( 6.5,  3.7){(c)} 
    \put (14.2,  3.7){(d)} 
  \end{picture} 
  \vspace{-0.5cm}
  \caption{Distributions of data (points) selected in the CC enriched
  sample in the electron channel. Shown is the transverse momentum
  $P_{T}^{e}$~(a) and polar angle $\theta_{e}$~(b) of reconstructed
  electrons for all data $e^{\pm}p$, and their distance to jets
  $D(e; \rm{jet}) > 1.0$ in the $e^{+}p$~(c) and $e^{-}p$~(d) samples
  individually. The total uncertainty on the SM expectation (open
  histogram) is shown as the shaded band. The signal component is
  shown as the hatched histogram. The final sample populates the region
  $D(e; \rm{jet}) > 1.0$.}
  \label{fig:elecccsample}
\end{figure}

CC background may enter the final sample due to hadrons or photons
misidentified as isolated electrons.
Such electron candidates appear predominantly close to the jet and
generally do not fulfill the strict isolation and tracking criteria
applied to electrons in the final sample of this analysis.
Figure~\ref{fig:elecccsample} shows a CC enriched electron sample
where the cuts on $D(e;{\rm jet})$ and $D(e;{\rm track})$ are removed
and only an associated NV track with a DCA~$<12$~cm is required
as a veto against neutral particles.
The electron candidates are still required to be isolated in the
calorimeter, as described in section~\ref{ch:reco}.
The tracking requirements on the electron are relaxed so that the
overall CC contribution to the SM expectation is about $80\%$ in this
selection.
$P_{T}^{e}$ and $\theta_{e}$ distributions of electron candidates in
this sample are shown in figures~\ref{fig:elecccsample}~(a) and (b),
respectively.
The model uncertainty on this type of CC background is $50\%$, as
derived from this control sample, such that the rate of isolated
electrons over the whole range is described within the total
uncertainties.
Figures~\ref{fig:elecccsample}~(c) and (d) show the $D(e\rm{;jet})$
distribution in the $e^{+}p$ and $e^{-}p$ data, respectively. The CC
background peaks close to the jet and is described within the total
error in both data sets.

%%%%%%%%%%%%%%%%%%%%%%%%%%%%%%%%%%%%%%%%%%%%%%%%%%%%%%%%%%%%%
\subsection{Muon Channel Control Samples}
\label{sec:isomuoncontrol}
%%%%%%%%%%%%%%%%%%%%%%%%%%%%%%%%%%%%%%%%%%%%%%%%%%%%%%%%%%%%%

Studies of the SM background in the muon channel are performed by
enriching the final sample with lepton pair and CC events.

%%%

\begin{figure}[]
  \begin{center}
    \textsf{Muon Channel, Lepton Pair Enriched}\\
    \includegraphics[width=0.49\textwidth]{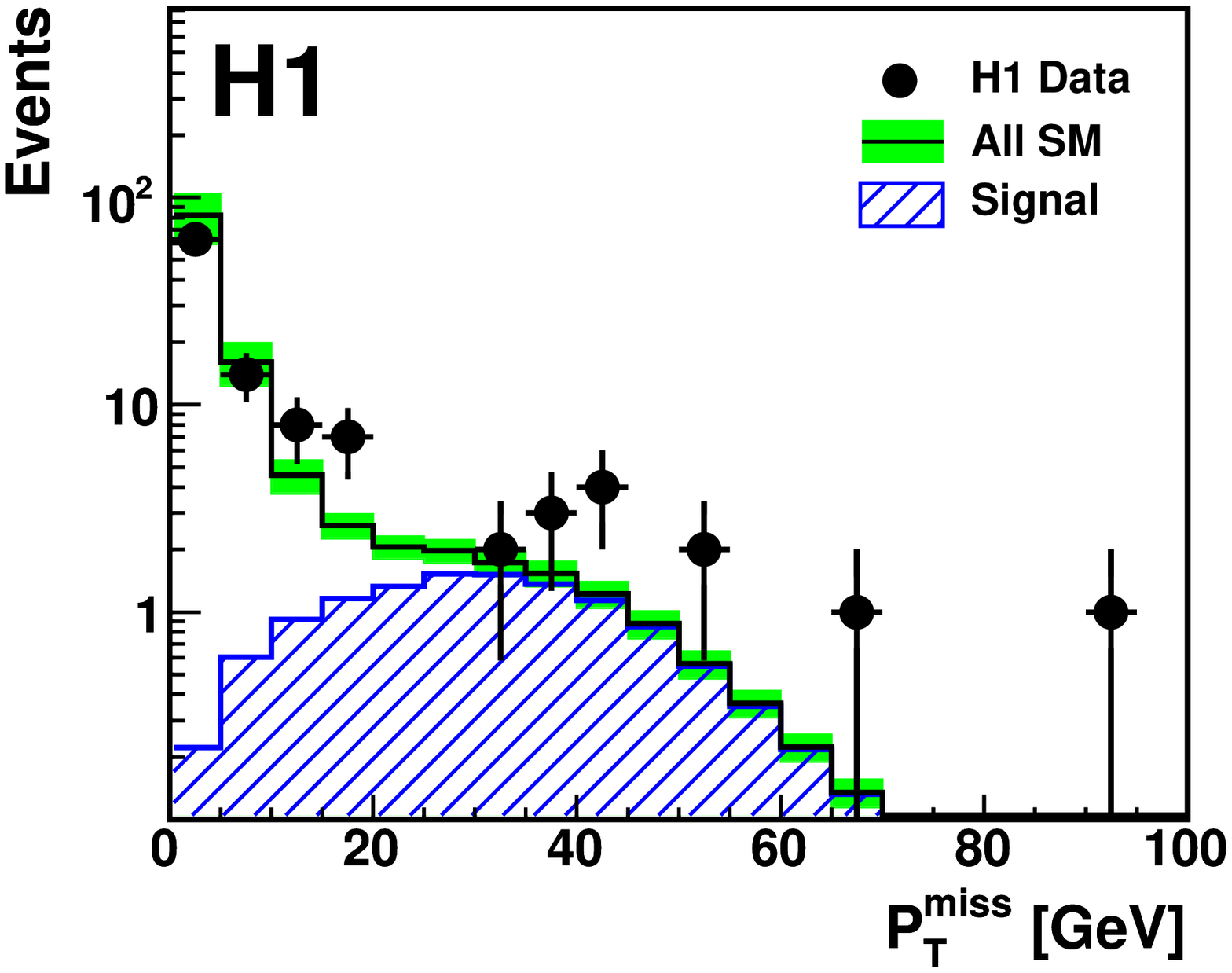}     
    \includegraphics[width=0.49\textwidth]{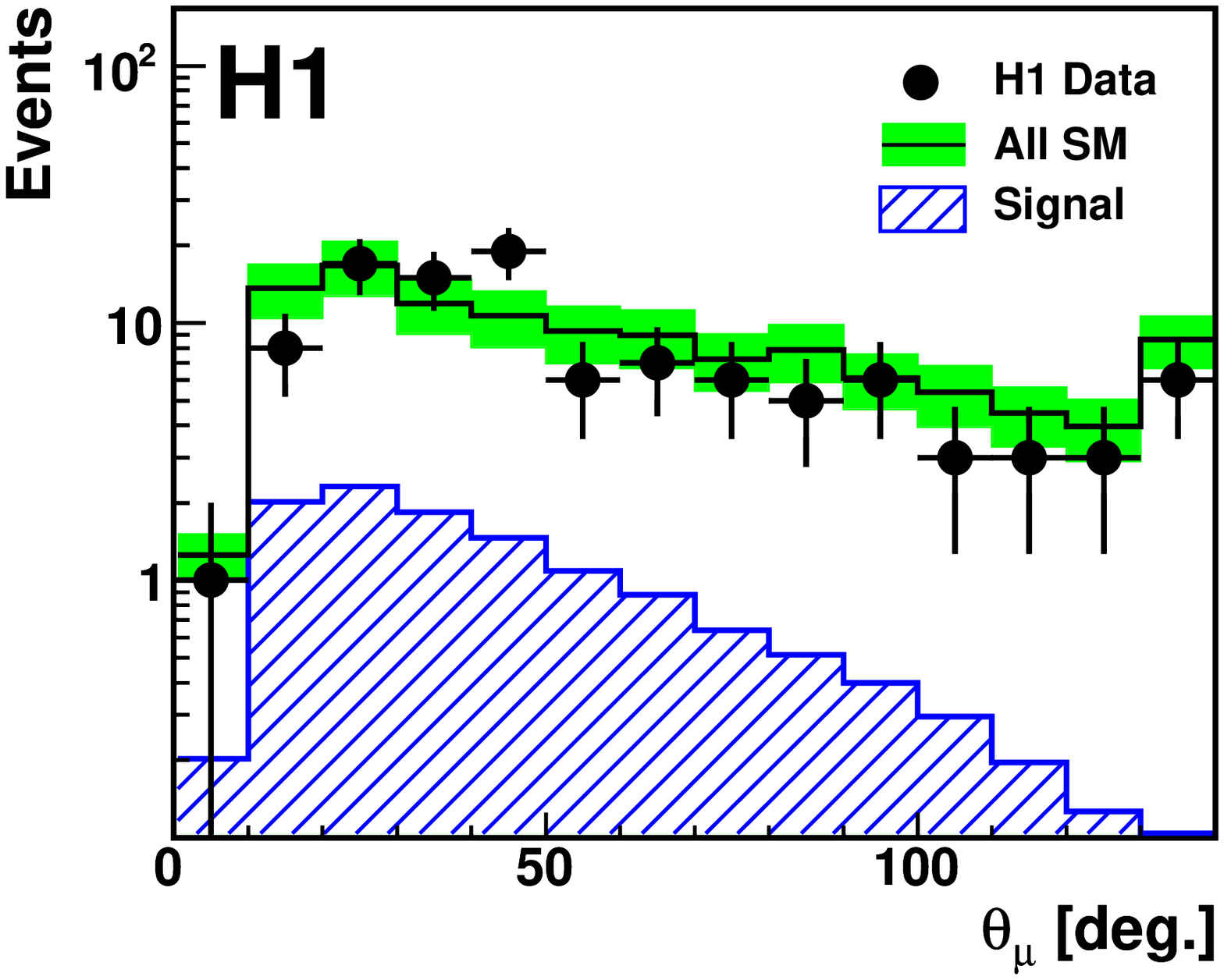}     
    \includegraphics[width=0.49\textwidth]{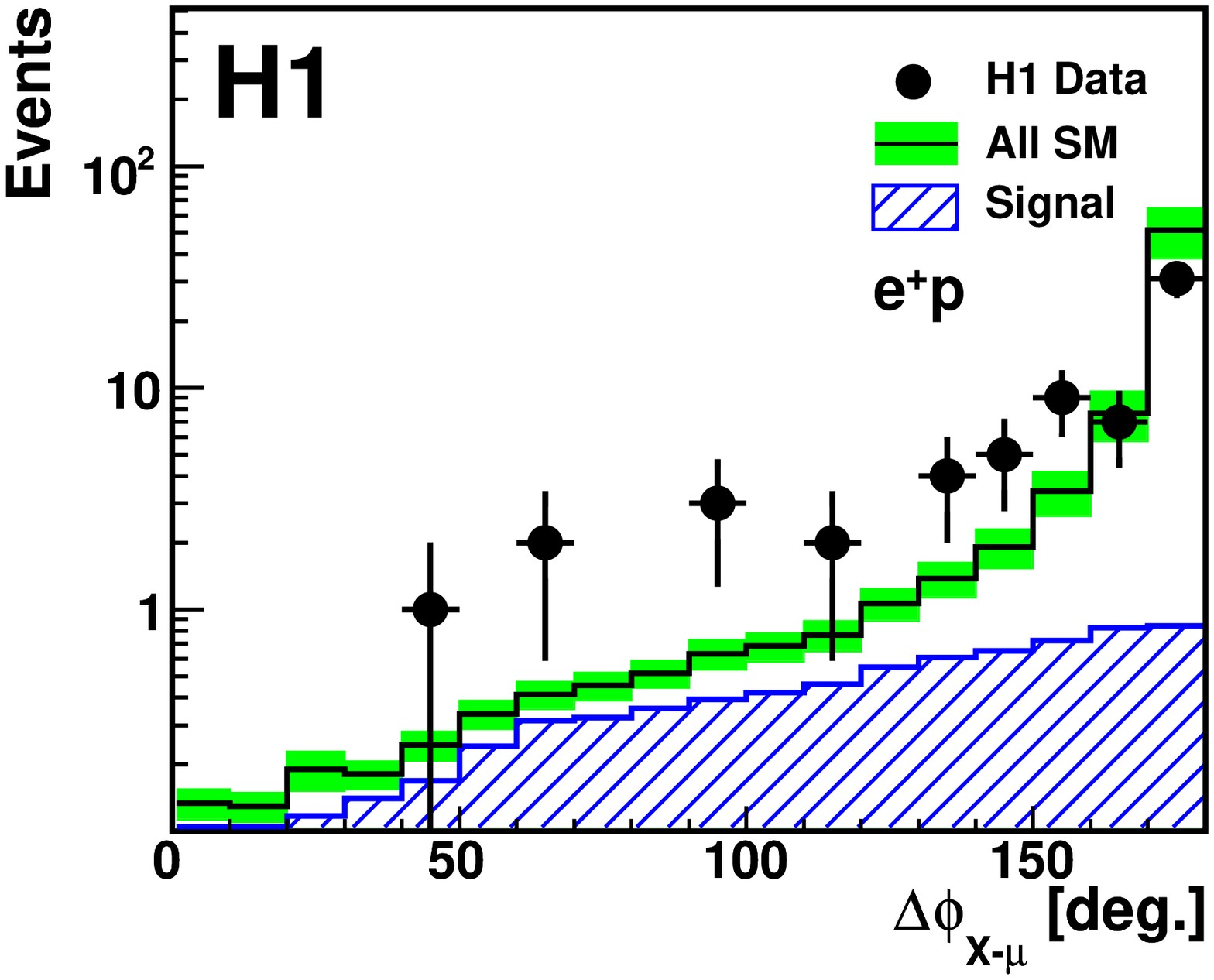}     
    \includegraphics[width=0.49\textwidth]{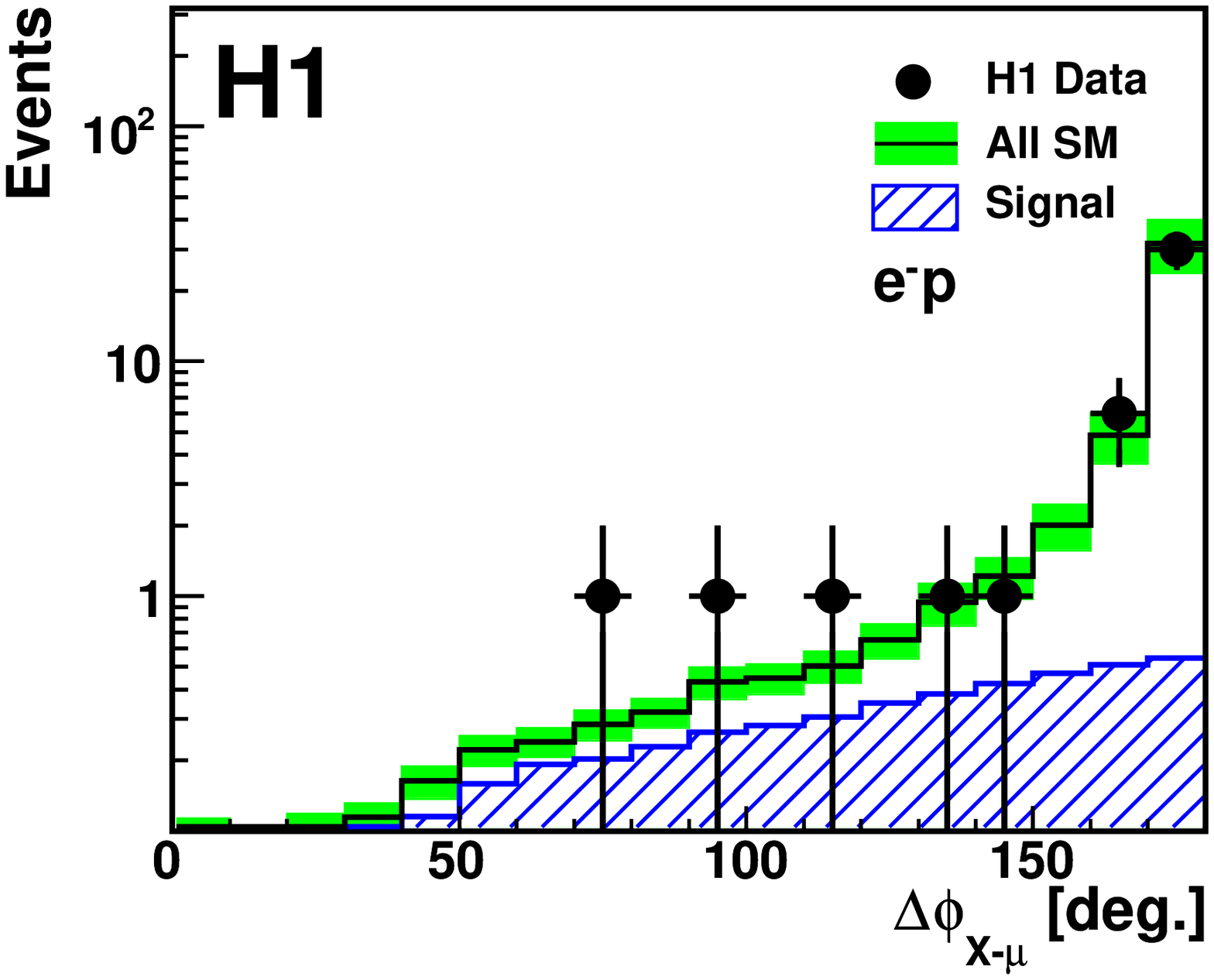}     
  \end{center}
  \begin{picture} (0.,0.)
    \setlength{\unitlength}{1.0cm}
    \put ( 6.4,10.4){(a)} 
    \put (14.3,8.7){(b)} 
    \put ( 1.5,2.8){(c)} 
    \put ( 9.6,2.8){(d)} 
  \end{picture}  
  \vspace{-0.5cm}
  \caption{Distribution of data (points) selected in the lepton pair
  enriched sample in the muon channel. Shown is the missing transverse
  momentum $P_{T}^{\mathrm{miss}}$ (a) and polar angle $\theta_{\mu}$
  (b) of muons for all data $e^{\pm}p$, and the acoplanarity
  $\Delta\phi_{\mu-X}$ in the $e^{+}p$ (c) and $e^{-}p$ (d) samples
  individually. The total uncertainty on the SM expectation (open
  histogram) is shown as the shaded band. The signal component is
  shown as the hatched histogram. The final sample populates the
  region $\Delta\phi_{\mu-X}<170^{\circ}$.} 
  \label{fig:muonlpsample} 
\end{figure}

Muons in pair production events typically have asymmetric transverse
momenta.
If one of the muons is produced at a very low polar angle, it may
escape detection due to geometrical acceptance.
In this case the detected muon tends to balance the hadrons in the 
transverse plane.
Figure~\ref{fig:muonlpsample} shows a muon pair enriched sample of
events that is formed by removing the requirements on $P_{T}^{\rm
miss}$ and $\Delta\phi_{\mu-X}$ from the final muon selection.
The requirement that there be only one muon in the event is also
removed.
The overall lepton pair contribution to the SM expectation is about
$90\%$ in this sample.
The $P_{T}^{\rm miss}$ distribution in this sample is shown in
figure~\ref{fig:muonlpsample}~(a), and demonstrates that the
resolution tail is described within the errors.
The $\theta_{\mu}$ distribution shown in
figure~\ref{fig:muonlpsample}~(b) follows the prediction from the SM
expectation.
The $\Delta\phi_{\mu-X}$ distribution is shown separately for the
$e^{+}p$ data and $e^{-}p$ data in \ref{fig:muonlpsample}~(c) and (d),
respectively, and is described within the errors in the region
populated by lepton pair production.
Events in the final sample are visible in the region
$\Delta\phi_{\mu-X}<170^{\circ}$, where the muon pair prediction is
low.
Most of these events are at values of $\Delta\phi_{\mu-X}$ that
deviate from $180^{\circ}$ significantly.

%%%

\begin{figure}[]
  \begin{center}
    \textsf{Muon Channel, CC Enriched}\\
    \includegraphics[width=0.49\textwidth]{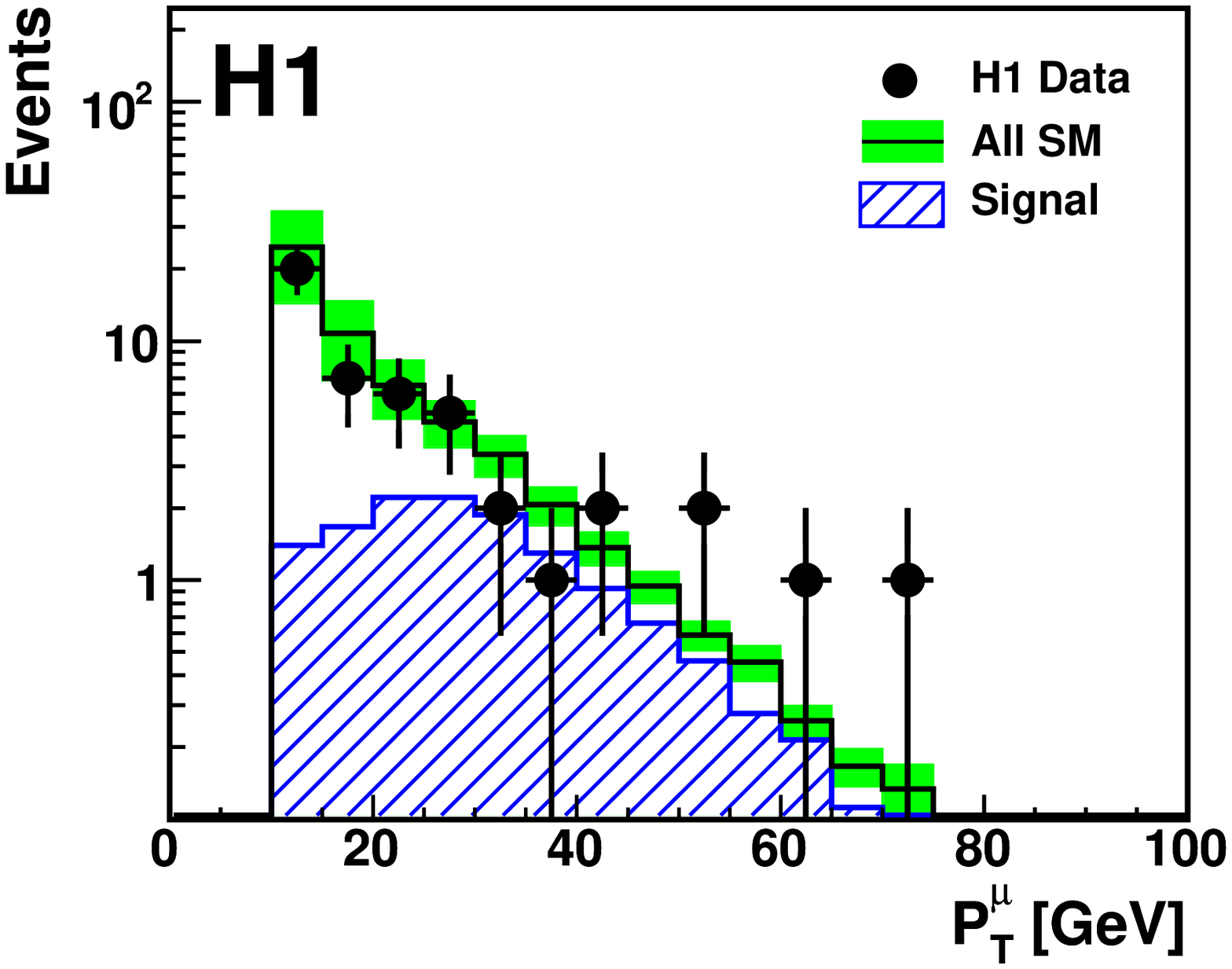}     
    \includegraphics[width=0.49\textwidth]{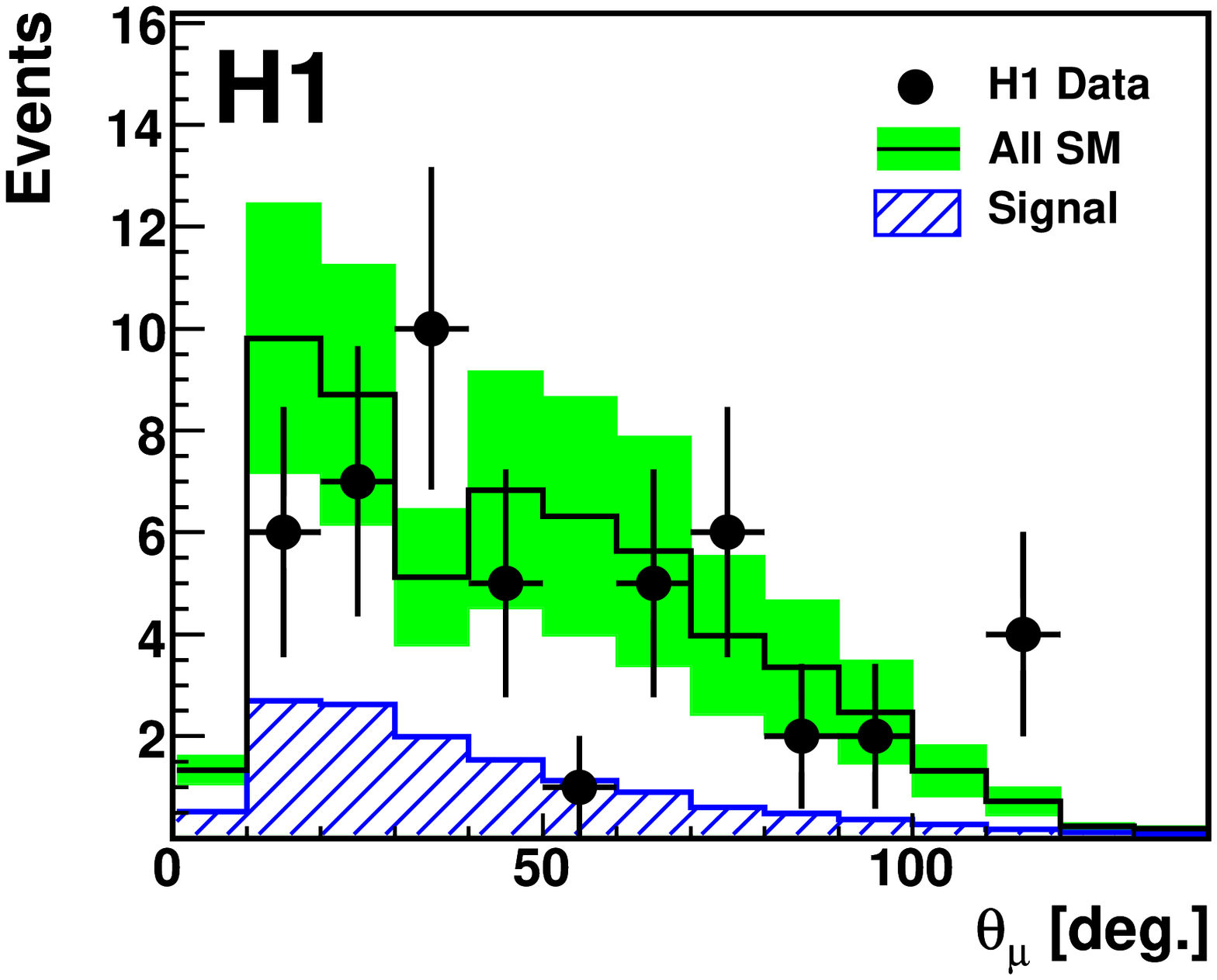}     
    \includegraphics[width=0.49\textwidth]{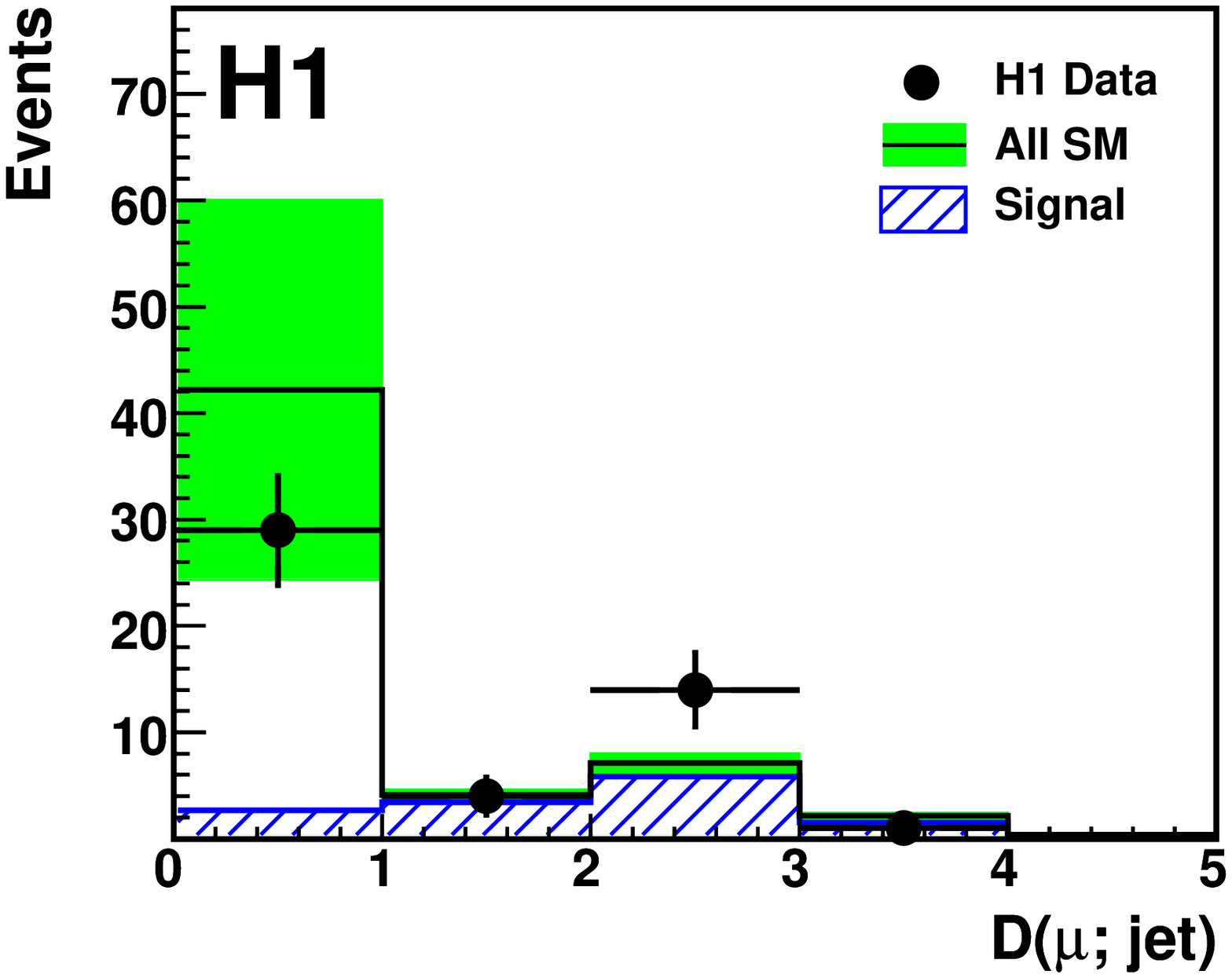}     
    \includegraphics[width=0.49\textwidth]{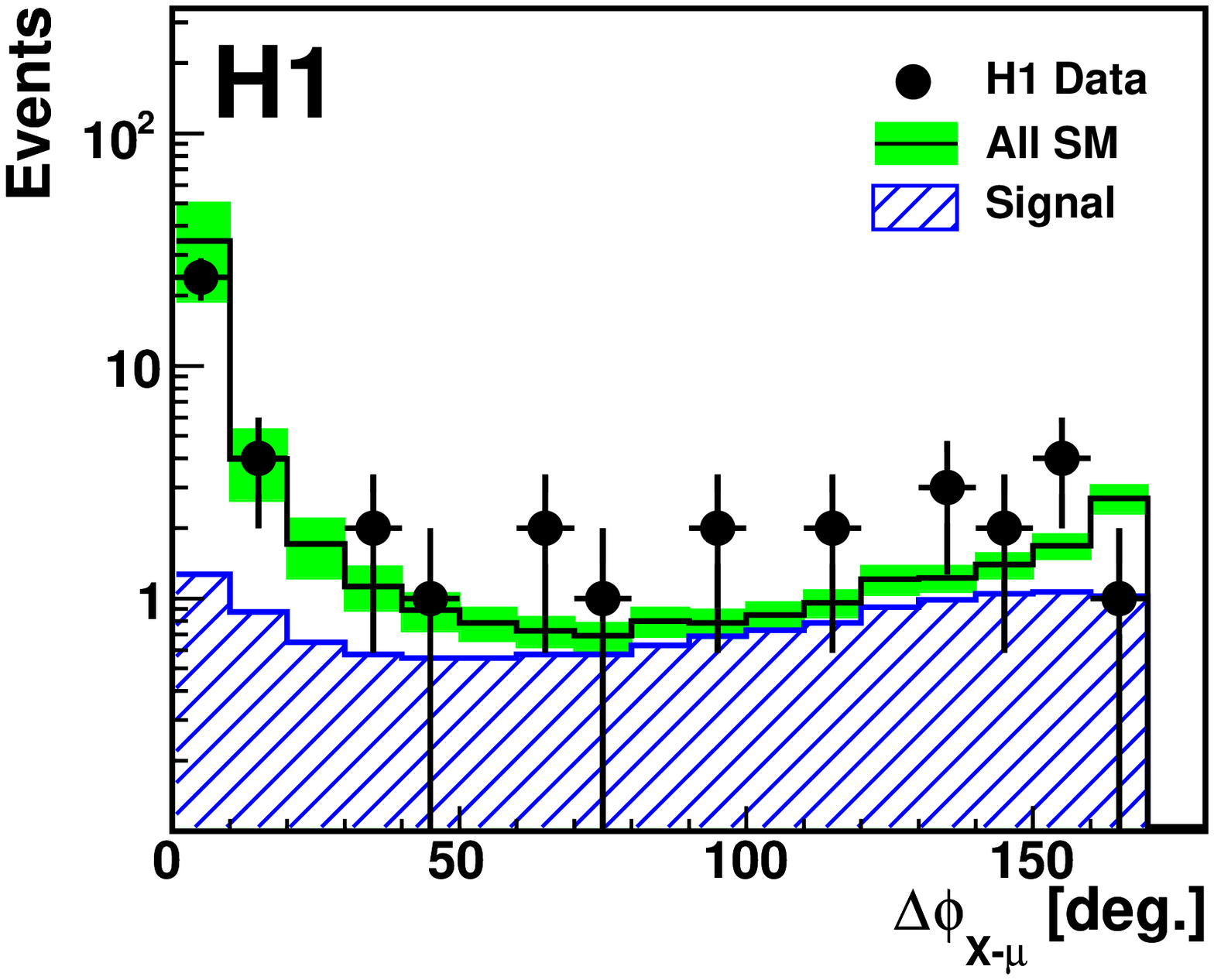}     
  \end{center}
  \begin{picture} (0.,0.)
    \setlength{\unitlength}{1.0cm}
    \put ( 6.0,  9.5){(a)} 
    \put (14.3,  9.5){(b)} 
    \put ( 6.0,  3.7){(c)}
    \put (14.0,  4.6){(d)} 
  \end{picture} 
  \vspace{-0.5cm}
  \caption{Distribution of data (points) selected in the CC enriched
  sample in the muon channel. Shown is the transverse momentum
  $P_{T}^{\mu}$~(a), polar angle ${\theta}_{\mu}$~(b) of identified
  muons, the distance to jets $D(\mu; \rm{jet})$~(c) and the
  acoplanarity $\Delta\phi_{\mu-X}$~(d) for all data. The total
  uncertainty on the SM expectation (open histogram) is shown as the
  shaded band. The signal component is shown as the hatched
  histogram. The final sample populates the region $D(\mu;
  \rm{jet})>1.0$.}
  \label{fig:muonccsample}
\end{figure}

Isolated muons can also appear in CC events, either due to the
presence of real muons produced in the fringe of the jet decay
cascade, or due to misidentification of muons as a combination of
tracks from charged hadrons and noise in the muon systems.
Figure~\ref{fig:muonccsample} shows a CC enriched event sample in the
muon channel that is formed by removing all isolation requirements on
muon candidates, in particular cuts on $D(\rm{\mu;jet})$ and
$D(\rm{\mu;track})$.
In this sample, the contribution from CC to the total expectation is
about $64\%$, with the remainder dominated by signal processes.
Figures~\ref{fig:muonccsample}~(a) and (b) show that $P_T^{\mu}$ and
$\theta_{\mu}$, respectively, of identified muons in this sample
follow the distribution predicted by the simulation.
The rate of identified muons close to the jet is described within the
total uncertainties.
The $ D(\rm{\mu;jet})$ distribution is shown in
figure~\ref{fig:muonccsample}~(c).
Events in the final sample are distributed across the
$\Delta\phi_{\mu-X}$ distribution, shown in
figure~\ref{fig:muonccsample}~(d), away from the region of the phase
space dominated by CC processes.

%%%%%%%%%%%%%%%%%%%%%%%%%%%%%%%%%%%%%%%%%%%%%%%%%%%%%%%%%%%%%
\subsection{Tau Channel Control Samples}
\label{sec:isotaucontrol}
%%%%%%%%%%%%%%%%%%%%%%%%%%%%%%%%%%%%%%%%%%%%%%%%%%%%%%%%%%%%%

\begin{figure}[] 
  \begin{center}
    \textsf{Tau Channel, Control Samples}\\
    \includegraphics[width=0.49\textwidth]{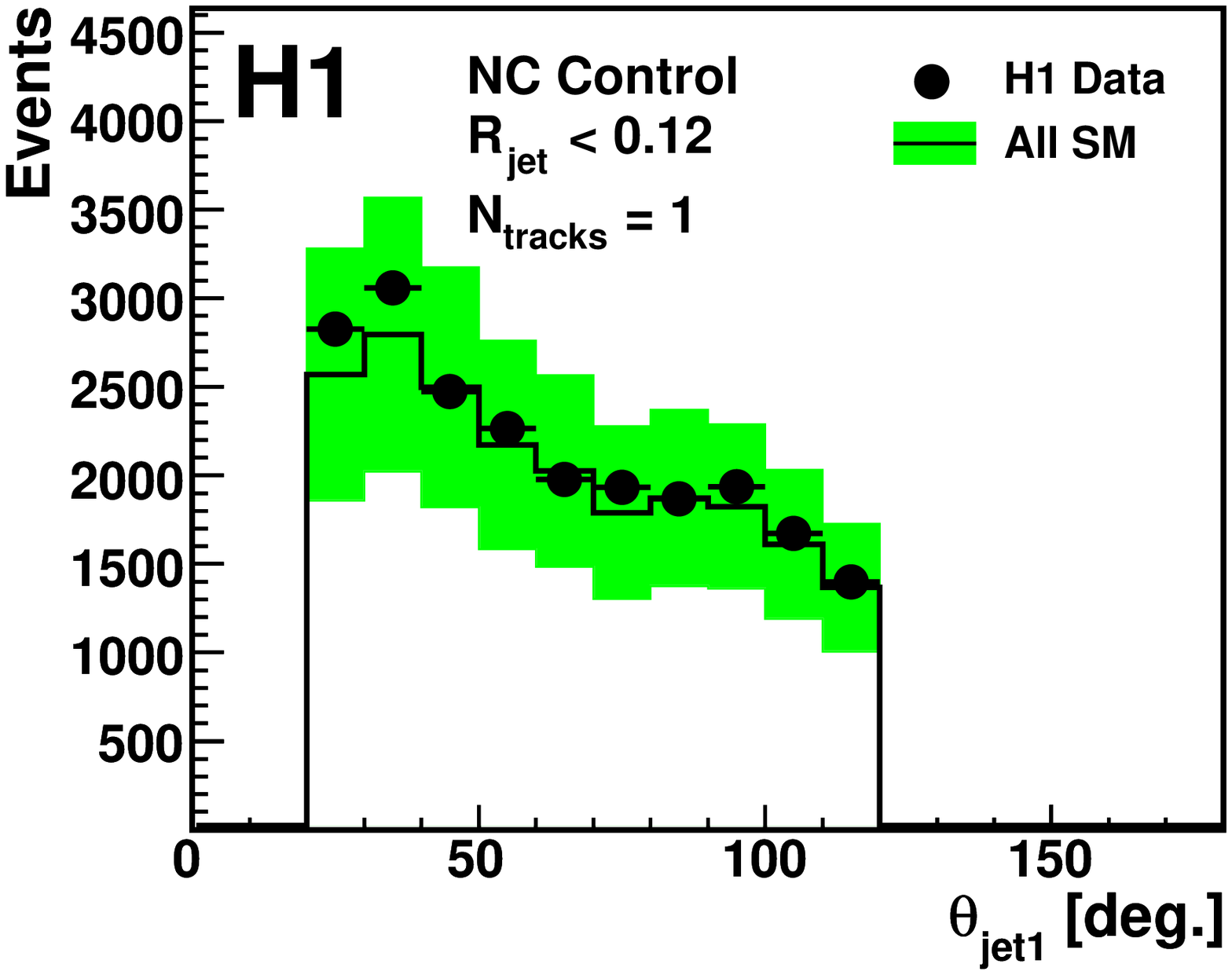}     
    \includegraphics[width=0.49\textwidth]{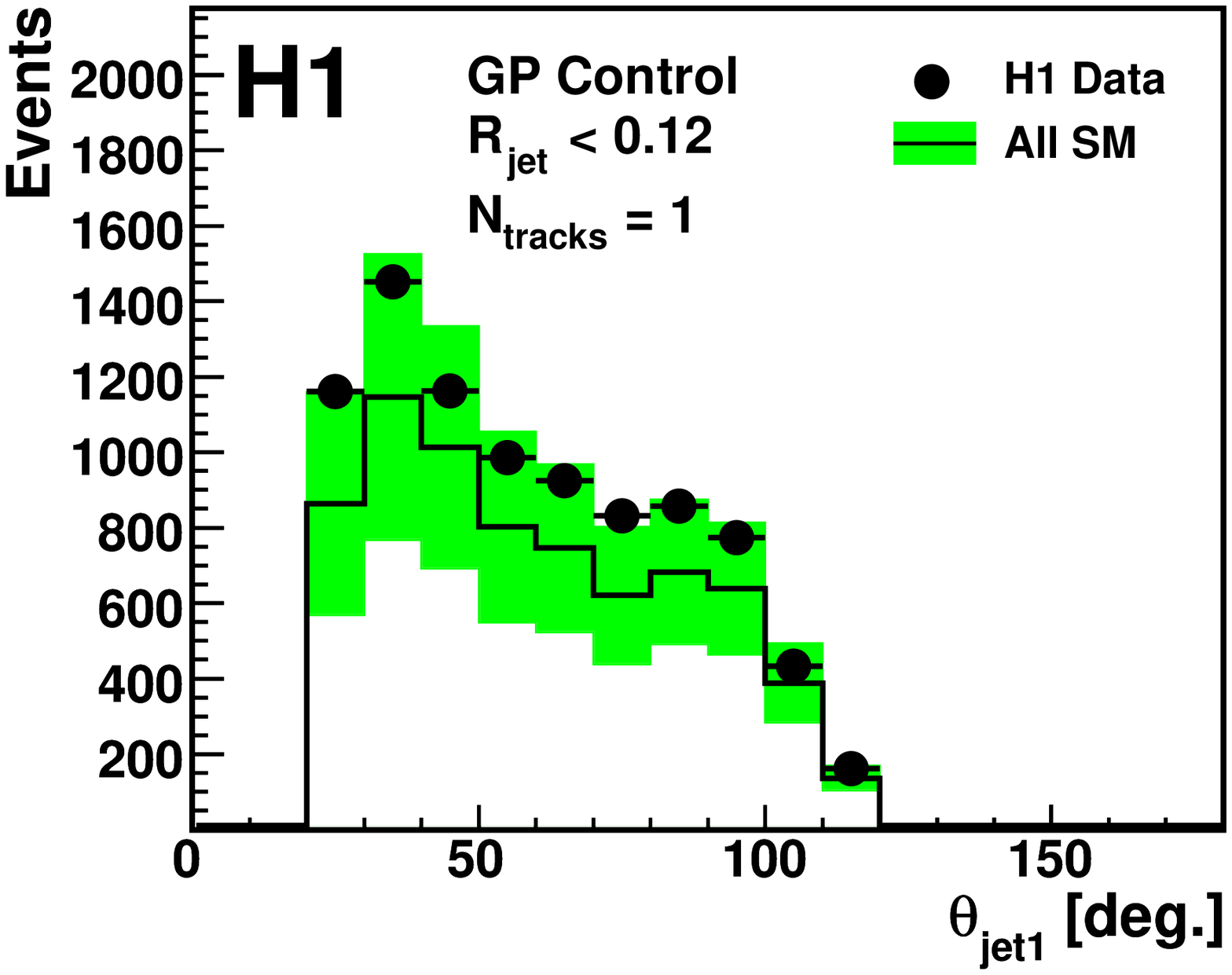}     
    \includegraphics[width=0.49\textwidth]{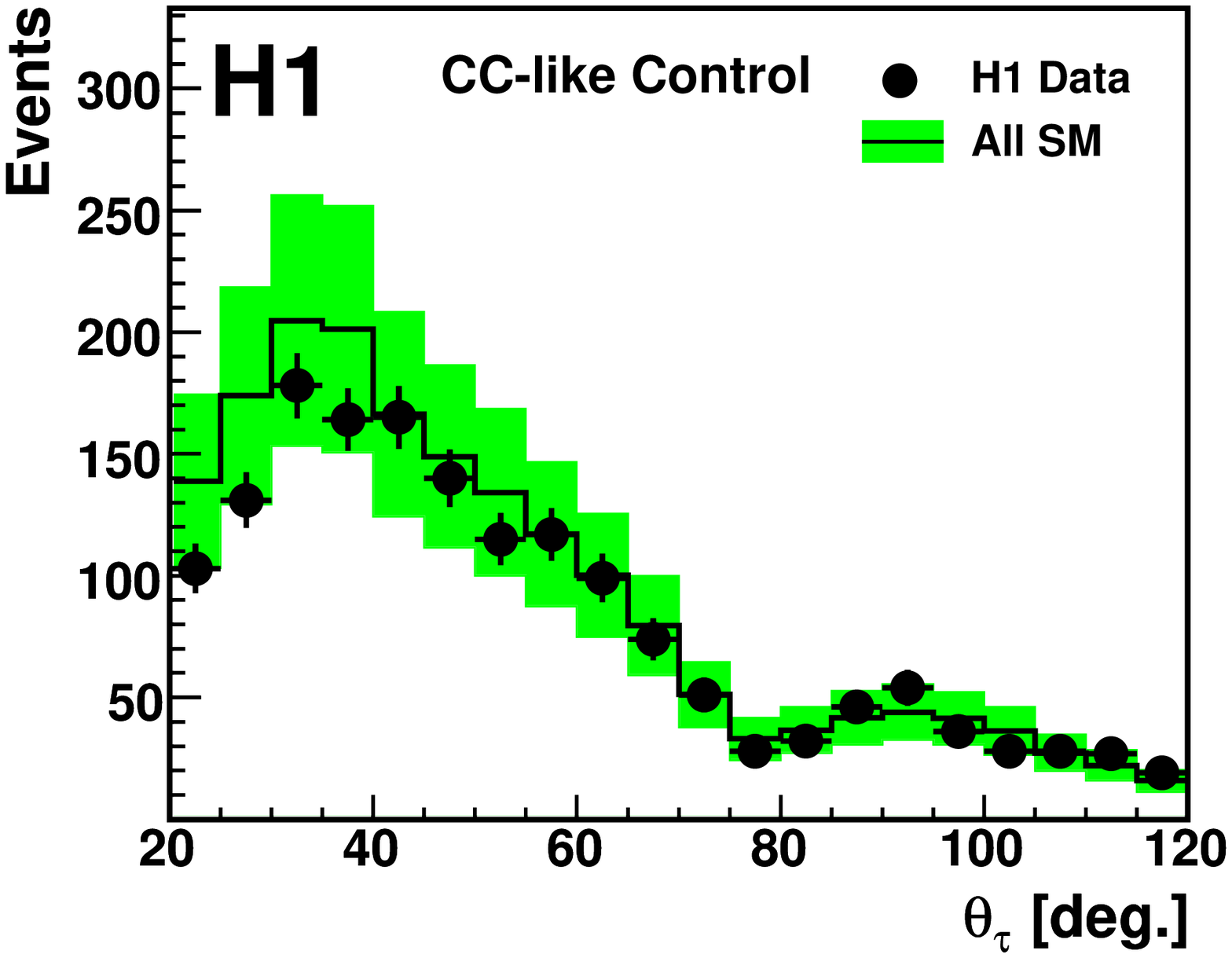}     
    \includegraphics[width=0.49\textwidth]{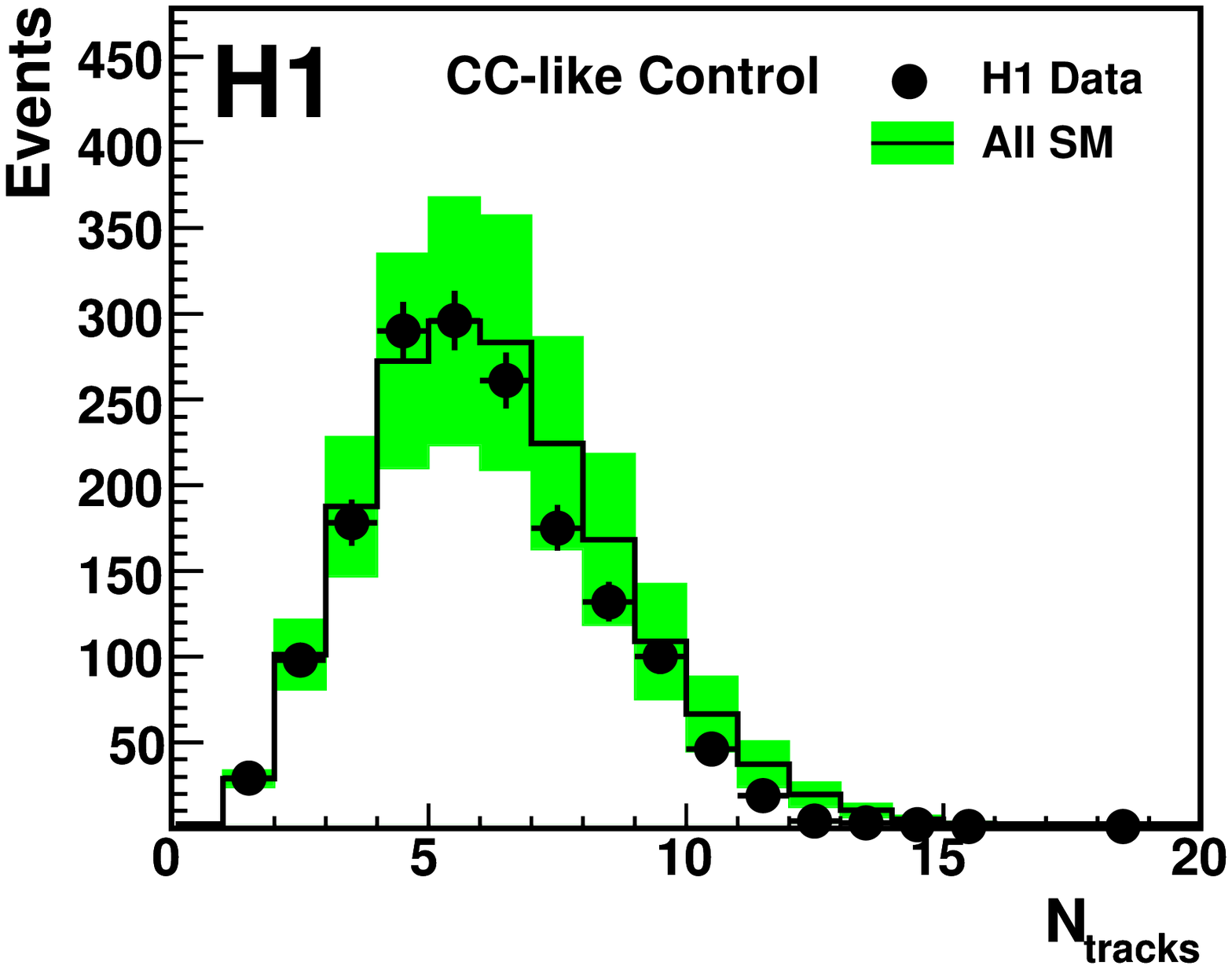}     
  \end{center}   
  \begin{picture} (0.,0.)
    \setlength{\unitlength}{1.0cm}
    \put ( 6.0,9.5){(a)} 
    \put (14.0,9.5){(b)} 
    \put ( 6.0,3.7){(c)} 
    \put (14.0,3.7){(d)} 
  \end{picture} 
  \vspace{-0.5cm} 
  \caption{Control distributions of data (points) in the tau
  channel. Shown is the polar angle distribution $\theta_{\rm jet}$ of
  tau-like jets in an inclusive NC control sample~(a), in a
  photoproduction control sample~(b), in the CC-like sample~(c), and
  the multiplicity of $P_{T} > 5$~GeV tracks associated to the jet in
  the CC-like sample~(d).  The total uncertainty on the SM expectation
  (open histogram) is shown as the shaded band. }
  \label{fig:isotaustudysamples}
\end{figure}

The description of narrow jets with at least one central high $P_T$
track as background to tau jets is checked by investigating tau--like
jets satisfying the requirements given in
table~\ref{tab:isotaucutstable}, in the same $P_{T}$ and polar angle
region as tau jets in the final sample.
Such tau--like jets are selected in an event sample of inclusive NC
events defined by an isolated electron in the phase space $P_{T}^{e} >
10$~GeV, $\theta_{e}<140^{\circ}$ and $D(e;{\rm jet}) > 1.0$.
The polar angle distribution of the tau--like jet with the highest
$P_{T}$ is shown in figure~\ref{fig:isotaustudysamples}~(a).
To control any influence of lost or mismeasured energy on the hadronic
final state and therefore on tau--like jets, $P_{T}^{\rm miss}>12$~GeV
is required in the inclusive NC event sample.
A sample of dijets in photoproduction is
also selected, mainly by requiring $V_{\rm ap}/V_{\rm p}>0.2$,
$\delta_{\rm miss}>10$~GeV and rejecting NC events with a scattered
electron, in addition to requiring two jets with $P_{T}^{\rm
jet1,jet2}>20,15$~GeV.
The polar angle distribution of the tau--like jet with the highest
$P_{T}$ is shown in figure~\ref{fig:isotaustudysamples}~(b).
The data are found to agree with the MC simulation within the 
total uncertainties in all of the above samples.

%%%  

The total background in the final sample is controlled using the
selection steps described in table~\ref{tab:isotaucutstable}.
Figure~\ref{fig:isotaustudysamples} shows the polar angle~(c) and
track multiplicity~(d) distributions of tau--like jets in the CC--like
sample, where all the isolation criteria of
table~\ref{tab:isotaucutstable} are applied with the exception of the
track isolation requirements.
Good agreement between the data and the SM prediction is observed
in these samples.

%%%%%%%%%%%%%%%%%%%%%%%%%%%%%%%%%%%%%%%%%%%%%%%%%%%%%%%%%%%%%%

%%%%%%%%%%%%%%%%%%%%%%%%%%%%%%%%%%%%%%%%%%%%%%%%%%%%%%%%%%%%%%%%%%%%%
\clearpage
%%%%%%%%%%%%%%%%%%%%%%%%%%%%%%%%%%%%%%%%%%%%%%%%%%%%%%%%%%%%%%%%%%%%%

%% Figures
%%%%%%%%%%%%%%%%%%%%%%%%%%%%%%%%%%%%%%%%%%%%%%%%%%%%%%%%%%%%%%%%%%%%%

%%% Event Rates in Electron and Muon Channels
%%%%%%%%%%%%%%%%%%%%%%%%%%%%%%%%%%%%%%%%%%%%%%%%%%%%%%%%%%%%%%%%%%%%%

\begin{table}[]
\begin{center}
\renewcommand{\arraystretch}{1.45} 

\begin{tabular}{|cc||c|rcl||rcl|rcl|}
\hline
\multirow{2}{10mm}{\LARGE \bf H1} & 1994-2007 $e^{+}p$ & Data & \multicolumn{3}{|c||}{SM} & \multicolumn{3}{|c|}{SM} & \multicolumn{3}{|c|}{Other SM}  \\
 & 291~pb$^{-1}$ & & \multicolumn{3}{|c||}{Expectation} & \multicolumn{3}{|c|}{Signal} & \multicolumn{3}{|c|}{Processes} \\
\hline
\hline
 Electron & Total & $28$ & $25.6$ & $\pm$ & $3.5$ & $18.6$ & $\pm$ & $2.9$ &$6.9$ & $\pm$ & $1.7$ \\
\cline{2-12}
  & $P_{T}^{X}>25$ GeV & $9$ & $4.32$ & $\pm$ & $0.71$ & $3.56$ & $\pm$ & $0.61$ &$0.76$ & $\pm$ & $0.32$ \\
\hline
 Muon & Total & $12$ & $6.7$ & $\pm$ & $1.1$ & $6.1$ & $\pm$ & $1.0$ &$0.55$ & $\pm$ & $0.18$ \\
\cline{2-12}
  & $P_{T}^{X}>25$ GeV & $8$ & $3.70$ & $\pm$ & $0.63$ & $3.43$ & $\pm$ & $0.60$ &$0.28$ & $\pm$ & $0.09$ \\
\hline
 Combined & Total & $40$ & $32.3$ & $\pm$ & $4.4$ & $24.8$ & $\pm$ & $3.9$ &$7.5$ & $\pm$ & $1.8$ \\
\cline{2-12}
  & $P_{T}^{X}>25$ GeV & $17$ & $8.0$ & $\pm$ & $1.3$ & $7.0$ & $\pm$ & $1.2$ &$1.04$ & $\pm$ & $0.37$ \\
\hline

\multicolumn{12}{c}{}\\

\hline
\multirow{2}{10mm}{\LARGE \bf H1} & 1998-2006 $e^{-}p$ & Data & \multicolumn{3}{|c||}{SM} & \multicolumn{3}{|c|}{SM} & \multicolumn{3}{|c|}{Other SM}  \\
 & 183~pb$^{-1}$ & & \multicolumn{3}{|c||}{Expectation} & \multicolumn{3}{|c|}{Signal} & \multicolumn{3}{|c|}{Processes} \\
\hline
\hline
 Electron & Total & $11$ & $17.5$ & $\pm$ & $2.7$ & $11.6$ & $\pm$ & $1.8$ &$5.9$ & $\pm$ & $1.9$ \\
\cline{2-12}
  & $P_{T}^{X}>25$ GeV & $1$ & $3.18$ & $\pm$ & $0.59$ & $2.23$ & $\pm$ & $0.38$ &$0.95$ & $\pm$ & $0.41$ \\
\hline
 Muon & Total & $2$ & $4.29$ & $\pm$ & $0.69$ & $3.96$ & $\pm$ & $0.66$ &$0.33$ & $\pm$ & $0.11$ \\
\cline{2-12}
  & $P_{T}^{X}>25$ GeV & $0$ & $2.40$ & $\pm$ & $0.41$ & $2.22$ & $\pm$ & $0.39$ &$0.19$ & $\pm$ & $0.06$ \\
\hline
 Combined & Total & $13$ & $21.8$ & $\pm$ & $3.1$ & $15.6$ & $\pm$ & $2.4$ &$6.2$ & $\pm$ & $1.9$ \\
\cline{2-12}
  & $P_{T}^{X}>25$ GeV & $1$ & $5.58$ & $\pm$ & $0.91$ & $4.45$ & $\pm$ & $0.75$ &$1.14$ & $\pm$ & $0.44$ \\
\hline

\multicolumn{12}{c}{}\\

\hline
\multirow{2}{10mm}{\LARGE \bf H1} & 1994-2007 $e^{\pm}p$ & Data & \multicolumn{3}{|c||}{SM} & \multicolumn{3}{|c|}{SM} & \multicolumn{3}{|c|}{Other SM}  \\
 & 474~pb$^{-1}$ & & \multicolumn{3}{|c||}{Expectation} & \multicolumn{3}{|c|}{Signal} & \multicolumn{3}{|c|}{Processes} \\
\hline
\hline
 Electron & Total & $39$ & $43.1$ & $\pm$ & $6.0$ & $30.3$ & $\pm$ & $4.8$ &$12.8$ & $\pm$ & $3.4$ \\
\cline{2-12}
  & $P_{T}^{X}>25$ GeV & $10$ & $7.5$ & $\pm$ & $1.3$ & $5.79$ & $\pm$ & $0.99$ &$1.71$ & $\pm$ & $0.72$ \\
\hline
 Muon & Total & $14$ & $11.0$ & $\pm$ & $1.8$ & $10.1$ & $\pm$ & $1.7$ &$0.88$ & $\pm$ & $0.28$ \\
\cline{2-12}
  & $P_{T}^{X}>25$ GeV & $8$ & $6.1$ & $\pm$ & $1.0$ & $5.64$ & $\pm$ & $0.99$ &$0.47$ & $\pm$ & $0.15$ \\
\hline
 Combined & Total & $53$ & $54.1$ & $\pm$ & $7.4$ & $40.4$ & $\pm$ & $6.3$ &$13.7$ & $\pm$ & $3.5$ \\
\cline{2-12}
  & $P_{T}^{X}>25$ GeV & $18$ & $13.6$ & $\pm$ & $2.2$ & $11.4$ & $\pm$ & $1.9$ &$2.18$ & $\pm$ & $0.80$ \\
\hline
\end{tabular}
\vspace{-0.5cm}
\end{center}
   \caption{Summary of the H1 results of searches for events with
   isolated electrons or muons and missing transverse momentum for the
   $e^{+}p$ data (291~pb$^{-1}$, top), $e^{-}p$ data (183~pb$^{-1}$,
   middle) and the full HERA data set (474~pb$^{-1}$, bottom). The
   results are shown for the full selected sample and for the
   subsample at large transverse momentum $P_{T}^{X}>25$~GeV. The
   number of observed events is compared to the SM prediction. The SM
   signal (dominated by single $W$ production) and the total
   background contribution are also shown. The quoted errors contain
   statistical and systematic uncertainties added in quadrature.}
\label{tab:isolepsummary}
\end{table}

%%% Kinematics of PtX>25 GeV events
%%%%%%%%%%%%%%%%%%%%%%%%%%%%%%%%%%%%%%%%%%%%%%%%%%%%%%%%%%%%%%%%%%%%%

\begin{landscape}

\begin{table}
\begin{center}
\renewcommand{\arraystretch}{1.2} 
\begin{xtabular}{|rrl||r|r|r|r|r|}
\hline
\multicolumn{8}{|c|}{\bf H1 Isolated Lepton Events at High {\boldmath $P_{T}^{X}$}}\\
\hline
Run & Event & Lepton $ ^{q(\sigma_q)} $ & $ P_T^{\ell}$ [ GeV ] & $ \theta_{\ell}$ [ $ ^\circ$ ] &
$ P_T^X $ [ GeV ] & $ M_T^{\ell\nu}$ [ GeV ] &  $ P_T^{\mathrm{miss}} $ [ GeV ] \\
\hline
\hline
$ 186729 $ & $ 702 $ & $\mu$ & $ > 42.5 $ & $ 30.0 \pm 0.4$ & $ 75.3 \pm   5.5$  & $ > 33.7 $  & $ > 40.0 $ \\
$ 188108 $ & $ 5066 $ & $\mu$$^{-\;(8.3 \sigma)}$ & $ 40.9 $ $ ^{ +5.6}_{ -4.4}  $ & $ 35.1 \pm 0.4$ & $ 29.4 \pm   2.4$  & $ 79.2 ^{+  8.0}_{- 10.1}$  & $ 43.7 ^{+  3.3}_{-  4.2}$ \\
$ 192227 $ & $ 6208 $ & $\mu$$^{-\;(7.0 \sigma)}$ & $ 73.3 $ $ ^{+12.2}_{ -9.2}  $ & $ 28.6 \pm 0.3$ & $ 63.9 \pm   5.9$  & $ 67.8 ^{+ 19.8}_{- 24.9}$  & $ 19.8 ^{+  5.4}_{-  6.8}$ \\
$ 195308 $ & $ 16793 $ & $\mu$$^{+\;(4.2 \sigma)}$ & $ 60.1 $ $ ^{+18.6}_{-11.5}  $ & $ 30.9 \pm 0.4$ & $ 30.1 \pm   2.6$  & $ 88.7 ^{+ 23.5}_{- 37.0}$  & $ 33.5 ^{+ 10.6}_{- 15.8}$ \\
$ 248207 $ & $ 32134 $ & $e$$^{+\;(15 \sigma)}$ & $ 32.1 \pm   1.3$  & $ 32.2 \pm 0.3$ & $ 42.0 \pm   3.9$  & $ 62.7 \pm   2.3$  & $ 43.4 \pm   2.8$ \\
$ 252020 $ & $ 30485 $ & $e$$^{+\;(40 \sigma)}$ & $ 25.6 \pm   1.2$  & $110.2 \pm 0.3$ & $ 39.1 \pm   3.3$  & $ 48.6 \pm   2.1$  & $ 35.5 \pm   2.5$ \\
$ 266336 $ & $ 4126 $ & $\mu$$^{+\;(26 \sigma)}$ & $ 19.7 $ $  ^{ +0.8}_{ -0.7} $ & $ 67.3 \pm 0.4$ & $ 50.0 \pm   3.8$  & $ 69.8 ^{+  2.4}_{-  2.5}$  & $ 66.6 \pm   3.7$ \\
$ 268338 $ & $ 70014 $ & $e$$^{+\;(1.6 \sigma)}$ & $ 33.8 \pm   1.3$  & $ 29.7 \pm 0.2$ & $ 45.2 \pm   3.2$  & $ 90.3 \pm   3.1$  & $ 67.2 \pm   3.0$ \\
$ 275991 $ & $ 29613 $ & $e$$^{+\;(37 \sigma)}$ & $ 37.8 \pm   1.5$  & $ 41.7 \pm 0.3$ & $ 27.1 \pm   1.8$  & $ 73.3 \pm   2.8$  & $ 40.3 \pm   1.4$ \\
$ 369241 $ & $ 6588 $ & $e$ & $ 29.2 \pm   1.1$  & $ 20.3 \pm 0.2$ & $ 40.5 \pm   4.8$  & $ 74.3 \pm   3.0$  & $ 55.5 \pm   4.2$ \\
$ 385422 $ & $ 76666 $ & $e$$^{+\;(22 \sigma)}$ & $ 28.1 \pm   1.3$  & $ 96.1 \pm 0.3$ & $ 25.9 \pm   2.8$  & $ 63.1 \pm   2.8$  & $ 40.0 \pm   2.3$ \\
$ 389826 $ & $ 2783 $ & $e$$^{-\;(10 \sigma)}$ & $ 62.0 \pm   2.2$  & $ 45.6 \pm 0.3$ & $ 45.3 \pm   4.5$  & $ 79.7 \pm   6.0$  & $ 30.3 \pm   2.1$ \\
$ 391884 $ & $ 49715 $ & $e$ & $ 38.2 \pm   1.4$  & $ 22.7 \pm 0.2$ & $ 32.4 \pm   2.6$  & $ 48.5 \pm   3.0$  & $ 20.1 \pm   0.8$ \\
$ 473929 $ & $ 107593 $ & $\mu$$^{-\;(9.6 \sigma)}$ & $ 53.5 $ $ ^{ +6.2}_{ -5.1}  $ & $ 31.4 \pm 0.4$ & $ 49.1 \pm   4.5$  & $ 80.6 ^{+  8.7}_{- 10.7}$  & $ 40.9 ^{+  2.8}_{-  3.4}$ \\
$ 494115 $ & $ 121996 $ & $\mu$$^{+\;(22 \sigma)}$ & $ 22.6 $ $ ^{ +1.0}_{ -1.0} $ & $ 61.5 \pm 0.4$ & $ 37.0 \pm   3.7$  & $ 45.2 ^{+  1.8}_{-  1.9}$  & $ 35.8  ^{+  3.0}_{-  3.0}$ \\
$ 495399 $ & $ 85500 $ & $\mu$$^{-\;(32 \sigma)}$ & $ 29.4 $ $ ^{ +0.9}_{ -0.8}  $ & $ 62.4 \pm 0.4$ & $ 29.6 \pm   2.8$  & $ 63.1 ^{+  1.7}_{-  1.8}$  & $ 40.3 ^{+  2.0}_{-  2.0}$ \\
$ 498117 $ & $ 316609 $ & $e$$^{+\;(9.8 \sigma)}$ & $ 27.4 \pm   1.1$  & $ 30.7 \pm 0.3$ & $ 26.7 \pm   1.8$  & $ 72.5 \pm   2.5$  & $ 49.9 \pm   2.0$ \\
\hline  
\hline  
$ 433051 $ & $ 64528 $ & $e$$^{-\;(24 \sigma)}$ & $ 26.2 \pm   1.3$  & $ 69.9 \pm 0.3$ & $ 72.9 \pm   5.6$  & $ 71.3 \pm   2.9$  & $ 75.8 \pm   5.2$ \\
\hline  
\end{xtabular}
  \caption{Kinematics of the isolated lepton events at high
  $P_{T}^{X}>25$~GeV. Resolutions on the experimental observables are
  propagated to the reconstructed variables.  The measured charge $q$
  of the lepton and the associated charge significance $\sigma_{q}$ is
  given where appropriate. For muon events where the curvature
  $|\kappa| \propto 1/P_{T}$ measurement is compatible with zero
  within 2$\sigma=|\kappa|/\delta\kappa$, the kinematic variables are
  estimated at $|\kappa|+\delta\kappa$ corresponding to a lower limit
  on $P_{T}$ at the 1$\sigma$ level.  All events in the table are
  observed in $e^{+}p$ collisions with exception of the event in the
  last row, which was observed in $e^{-}p$ collisions.}
  \label{tab:isolepkine}
\end{center}
\end{table}
\end{landscape}

%%% Event Rates in the Tau Channel
%%%%%%%%%%%%%%%%%%%%%%%%%%%%%%%%%%%%%%%%%%%%%%%%%%%%%%%%%%%%%%%%%%%%%

\begin{table}[]
\renewcommand{\arraystretch}{1.5} 
\begin{center}

\begin{tabular}{|cc||c|rcl||rcl|rcl|}
\hline
\multirow{2}{10mm}{\LARGE \bf H1} & Tau Channel & Data & \multicolumn{3}{|c||}{SM} & \multicolumn{3}{|c|}{SM} & \multicolumn{3}{|c|}{Other SM} \\
 & & & \multicolumn{3}{|c||}{Expectation} & \multicolumn{3}{|c|}{Signal} & \multicolumn{3}{|c|}{Processes} \\
\hline
\hline
 1994-2007 $e^{+}p$ & Total & $9$ & $12.3$ & $\pm$ & $2.0$ & $1.66$ & $\pm$ & $0.25$ & $10.6$ & $\pm$ & $1.8$ \\
\cline{2-12}
 291~pb$^{-1}$ & $P_{T}^{X}>25$ GeV & $0$ & $0.82$ & $\pm$ & $0.12$ & $0.38$ & $\pm$ & $0.06$ & $0.44$ & $\pm$ & $0.06$ \\
\hline
\hline
 1999-2006 $e^{-}p$ & Total & $9$ & $11.0$ & $\pm$ & $1.9$ & $1.00$ & $\pm$ & $0.15$ & $10.0$ & $\pm$ & $1.8$ \\
\cline{2-12}
 183~pb$^{-1}$ & $P_{T}^{X}>25$ GeV & $1$ & $0.68$ & $\pm$ & $0.11$ & $0.21$ & $\pm$ & $0.03$ & $0.47$ & $\pm$ & $0.07$ \\
\hline
\hline
 1994-2007 $e^{\pm}p$ & Total & $18$ & $23.2$ & $\pm$ & $3.8$ & $2.66$ & $\pm$ & $0.40$ & $20.6$ & $\pm$ & $3.4$ \\
\cline{2-12}
 474~pb$^{-1}$ & $P_{T}^{X}>25$ GeV & $1$ & $1.50$ & $\pm$ & $0.21$ & $0.59$ & $\pm$ & $0.09$ & $0.91$ & $\pm$ & $0.12$ \\
\hline
\end{tabular}
\end{center}
  \caption{Summary of the H1 results of the search for events with tau
  leptons and missing transverse momentum for the $e^{+}p$ data (291
  pb$^{-1}$), $e^{-}p$ data (183 pb$^{-1}$) and the full HERA data set
  (474 pb$^{-1}$).  The results are shown for the full selected sample
  and for the subsample at $P_{T}^{X}>25$~GeV. The number of observed
  events is compared to the SM prediction. The SM signal
  ($W\rightarrow\tau\nu_{\tau}$) and the background contributions are
  also shown. The quoted errors contain statistical and systematic
  uncertainties added in quadrature.}
\label{tab:isotausummary}
\end{table}

%%% Isolated Lepton Cross Section
%%%%%%%%%%%%%%%%%%%%%%%%%%%%%%%%%%%%%%%%%%%%%%%%%%%%%%%%%%%%%%%%%%%%%

\begin{table}[]
\renewcommand{\arraystretch}{1.5}
\begin{center}
\begin{tabular}{| c  c | r @{$\,\pm\,$} c @{$\,\pm\,$} l | c |}
\hline 
\multicolumn{6}{|c|}{\bf H1 Isolated Lepton and {\boldmath $P_{T}^{\rm miss}$} Cross Section} \\
\hline
 & & Measured & stat. & sys. [fb] & SM NLO [fb]\\
\hline\hline 
 Electron & $P_{T}^{X}\leq 12$~GeV &   $63$ &  $22$ & $13$ &  ~~$84$ $\pm$ $13$ \\
\cline{2-6}
 & $P_{T}^{X}>12$~GeV &   $54$ &  $17$ & ~~$9$ &  ~~$49$ $\pm$ ~~$7$ \\
\hline
 Muon &     $P_{T}^{X}>12$~GeV &   $56$ &  $16$ & ~~$7$ &  ~~$44$ $\pm$ ~~$7$ \\
\hline 
 Combined & $P_{T}^{X}\leq 25$~GeV &  $164$ & $45$ & $32$ & $207$ $\pm$ $31$ \\
\cline{2-6}
 & $P_{T}^{X}>25$~GeV   &  $64$ & $18$ & $10$ & ~~$47$ $\pm$ ~~$7$ \\
\cline{2-6}
 & Total &   $228$ &  $48$ & $39$ &  $253$ $\pm$ $38$  \\
\hline 
\end{tabular}
  \caption{The measured cross sections with statistical (stat.) and
  systematic (sys.) errors for events with an isolated high energy
  electron or muon with missing transverse momentum, in different
  kinematic regions of the hadronic transverse momentum
  $P_{T}^{X}$. The cross sections are determined for the kinematic
  region: $5^{\circ}<\theta_{\ell}<140^{\circ}$,~$P_{T}^{\ell}>
  10$~GeV,~$P_{T}^{\textrm{miss}}>12$~GeV~and~$D(\rm \ell;jet)>1.0$ at
  a centre of mass energy of $\sqrt{s}=317$~GeV. Also shown are the
  signal expectations including the theoretical uncertainties for the
  Standard Model where the dominant contribution $ep\rightarrow eWX$
  is calculated at next to leading order (SM NLO).}
\label{tab:isolepcrosssections}
\end{center}
\end{table}

%%% Differential Single W Cross Section
%%%%%%%%%%%%%%%%%%%%%%%%%%%%%%%%%%%%%%%%%%%%%%%%%%%%%%%%%%%%%%%%%%%%%

\begin{table}[]
\renewcommand{\arraystretch}{1.5}
\begin{center}
\begin{tabular}{| r @{$\,-\,$} l | r @{$\,\pm\,$} r @{$\,\pm\,$} l | c | }
 \hline
 \multicolumn{6}{|c|}{ \textbf{ H1 Differential Single {\boldmath $W$} Production Cross Section at {\boldmath $\sqrt{s}=317$} GeV} } \\
\hline 
 \multicolumn{2}{|c|}{$P_{T}^{X}$~[GeV] } & Measured & stat. & sys. [fb / GeV] & SM NLO [fb / GeV]\\
\hline
\hline
   $0$ & $12$  &  $47.6$ &   $18.2$ &   $8.9$ &  $65.2$ $\pm$   $9.8$ \\
  $12$ & $25$  &  $17.4$ &    $6.3$ &   $1.7$ &  $18.3$ $\pm$   $2.7$ \\
  $25$ & $40$  &  $11.2$ &    $4.5$ &   $1.0$ &  $10.1$ $\pm$   $1.5$ \\
  $40$ & $80$  &   $4.5$ &    $1.7$ &   $0.6$ & $~~2.4$ $\pm$   $0.4$ \\
\hline
\end{tabular}
\vspace{0.5cm}
  \caption{The differential single $W$ boson production cross section
  d$\sigma_{W}$/d$P_{T}^{X}$ with statistical (stat.) and systematic
  (sys.) errors, derived from the electron and muon channels as a
  function of the hadronic transverse momentum $P_{T}^{X}$. Also shown
  are the expectations including the theoretical uncertainties for the
  Standard Model calculated at next to leading order (SM NLO). In the
  phase space where $P_{T}^{X}<12$~GeV, the electron measurement is
  used to estimate the muon cross section under the assumption of
  lepton universality.}
\label{tab:diffxsec}
\end{center}
\end{table}

%%% W Polarisation fractions
%%%%%%%%%%%%%%%%%%%%%%%%%%%%%%%%%%%%%%%%%%%%%%%%%%%%%%%%%%%%%%%%%%%%%

\begin{table}[]
\renewcommand{\arraystretch}{1.5}
\begin{center}
\begin{tabular}{ | l | r @{$\,\pm\,$} r @{$\,\pm\,$} r | c | }
\hline
\multicolumn{5}{|c|}{ \textbf{ H1 {\boldmath $W$} Polarisation Fractions } } \\
\hline
Fraction & Measured & stat. & sys. & EPVEC \\
\hline\hline
$F{-} $ & $0.67$ & $0.19$ & $0.08$ & $0.62$ \\
$F{0} $ & $0.21$ & $0.35$ & $0.10$ & $0.17$ \\
\hline
\end{tabular}
\vspace{0.5cm}
  \caption{The polarisation fractions $F_{-}$ and $F_{0}$ obtained
  from one parameter fits, with statistical (stat.) and systematic
  (sys.) errors. The central values are obtained by fixing one
  parameter to its SM prediction and fitting the other. The SM
  predictions from EPVEC are also shown.}
\label{tab:wpol}
\end{center}
\end{table}

%% Figures
%%%%%%%%%%%%%%%%%%%%%%%%%%%%%%%%%%%%%%%%%%%%%%%%%%%%%%%%%%%%%%%%%%%%%

%%% Feynman Diagrams
%%%%%%%%%%%%%%%%%%%%%%%%%%%%%%%%%%%%%%%%%%%%%%%%%%%%%%%%%%%%%%%%%%%%%

%%% Final Sample Plots
%%%%%%%%%%%%%%%%%%%%%%%%%%%%%%%%%%%%%%%%%%%%%%%%%%%%%%%%%%%%%%%%%%%%%

%%% Event Displays
%%%%%%%%%%%%%%%%%%%%%%%%%%%%%%%%%%%%%%%%%%%%%%%%%%%%%%%%%%%%%%%%%%%%%

%%% PtX e+p and e-p
%%%%%%%%%%%%%%%%%%%%%%%%%%%%%%%%%%%%%%%%%%%%%%%%%%%%%%%%%%%%%%%%%%%%%

%%% Tau Channel Distributions
%%%%%%%%%%%%%%%%%%%%%%%%%%%%%%%%%%%%%%%%%%%%%%%%%%%%%%%%%%%%%%%%%%%%%

%%% Differential W boson production cross section as a fn of PTX
%%%%%%%%%%%%%%%%%%%%%%%%%%%%%%%%%%%%%%%%%%%%%%%%%%%%%%%%%%%%%%%%%%%%%

%%% WWgamma Vertex Parameter Likelihood distributions
%%%%%%%%%%%%%%%%%%%%%%%%%%%%%%%%%%%%%%%%%%%%%%%%%%%%%%%%%%%%%%%%%%%%%

%%% Polarisation figures
%%%%%%%%%%%%%%%%%%%%%%%%%%%%%%%%%%%%%%%%%%%%%%%%%%%%%%%%%%%%%%%%%%%%%

%%% Elec NC Control Plots
%%%%%%%%%%%%%%%%%%%%%%%%%%%%%%%%%%%%%%%%%%%%%%%%%%%%%%%%%%%%%%%%%%%%%

%%% Elec CC Control Plots
%%%%%%%%%%%%%%%%%%%%%%%%%%%%%%%%%%%%%%%%%%%%%%%%%%%%%%%%%%%%%%%%%%%%%

%%% Muon LP Control Plots
%%%%%%%%%%%%%%%%%%%%%%%%%%%%%%%%%%%%%%%%%%%%%%%%%%%%%%%%%%%%%%%%%%%%%

%%% Muon CC Control Plots
%%%%%%%%%%%%%%%%%%%%%%%%%%%%%%%%%%%%%%%%%%%%%%%%%%%%%%%%%%%%%%%%%%%%%

%%% Tau Control Plots
%%%%%%%%%%%%%%%%%%%%%%%%%%%%%%%%%%%%%%%%%%%%%%%%%%%%%%%%%%%%%%%%%%%%%

\end{document}